\documentclass[iop,apj,numberedappendix,appendixfloats]{emulateapj}

\pdfoutput=1
\usepackage{epstopdf}
\usepackage{amsmath}
\usepackage{longtable}
\usepackage{multirow}

\usepackage{appendix}
\usepackage{color}
\newcommand{\redpen}[1]{{\bf\textcolor{red}{#1}}}

\newcommand{\about}{$\sim\!\!$~}

\newif\iflongtable
\longtabletrue
\newif\ifcheck
\checkfalse
\checktrue                                                                     

\ifcheck
\newcommand{\REF}[1]{\redpen{REF\##1}}
\else
\newcommand{\REF}[1]{}
\fi

\slugcomment{\today}
\def\reff@jnl#1{{\rm#1\/}}
\def\pasa{\reff@jnl{PASA}}% 

\iflongtable
\LongTables
\fi

\newcommand{\gps}{\ensuremath{g_{\rm P1}}}
\newcommand{\rps}{\ensuremath{r_{\rm P1}}}
\newcommand{\ips}{\ensuremath{i_{\rm P1}}}
\newcommand{\zps}{\ensuremath{z_{\rm P1}}}
\newcommand{\yps}{\ensuremath{y_{\rm P1}}}

\newcommand{\grizy}{\ensuremath{grizy_{\rm P1}}}
\newcommand{\griz}{\ensuremath{griz_{\rm P1}}}
\newcommand{\PS}{\protect \hbox {Pan-STARRS1}}
\newcommand{\photpipe}{\protect \hbox {\it photpipe}}
\newcommand{\transphot}{\protect \hbox {\it transphot}}

\newcommand{\Tonryphot}{T12b}
\newcommand{\Scolnicsys}{S14}

\newcommand{\OM}{\Omega_{\rm M}}

\newcommand{\OL}{\Omega_{\Lambda}}
\newcommand{\Ok}{\Omega_{\rm k}}
\newcommand{\SINSINHFUN}{{\cal S}_k}  % F(x) = sin(x), sinh(x) or x

\newcommand{\Pfit}{ {\cal P}_{\rm fit} }

\newcommand{\DL}{d_L}

\newcommand{\SNIAPSall}{146}

\newcommand{\SNIAlowztot}{222}

\newcommand{\SNIAPSused}{113}
\newcommand{\SNIAlowzused}{197}
\newcommand{\SNIAused}{310}
\newcommand{\SNIAlowzall}{497}
\newcommand{\wSNstat} {\ensuremath{-1.010^{+0.360}_{-0.206}}}
\newcommand{\wCstat} {\ensuremath{-1.131^{+0.049}_{-0.049}}}
\newcommand{\fomCstat} {\ensuremath{0.284^{+0.010}_{-0.010}}}
\newcommand{\wCsys} {\ensuremath{-1.166^{+0.072}_{-0.069}}}
\newcommand{\fomCsys} {\ensuremath{0.280^{+0.013}_{-0.012}}}
\newcommand{\wCwsys}{\ensuremath{-1.124^{+0.083}_{-0.065}}}
\newcommand{\wSNsys} {\ensuremath{-1.120^{+0.450}_{-0.357}}}
\newcommand{\wSNstasys} {\ensuremath{-1.120^{+0.360}_{-0.206}\textrm{(Stat)} ^{+0.269}_{-0.291}\textrm{(Sys)}}}
\newcommand{\fomSNsys} {\ensuremath{0.256^{+0.201}_{-0.174}}}
\newcommand{\fomSNstat} {\ensuremath{0.223^{+0.209}_{-0.221}}}

\newcommand{\snOMstatf}{\ensuremath{0.242^{+0.039}_{-0.041}}}
\newcommand{\snOMsysf}{\ensuremath{0.226^{+0.057}_{-0.061}}}
\newcommand{\snOLstatf}{\ensuremath{0.758^{+0.039}_{-0.041}}}
\newcommand{\snOLsysf}{\ensuremath{0.774^{+0.057}_{-0.061}}}
\newcommand{\cOMstat}{\ensuremath{0.274^{+0.018}_{-0.015}}}
\newcommand{\cOLstat}{\ensuremath{0.722^{+0.014}_{-0.014}}}
\newcommand{\cOMsys}{\ensuremath{0.308^{+0.033}_{-0.030}}}
\newcommand{\cOLsys}{\ensuremath{0.693^{+0.024}_{-0.025}}}

\newcommand{\accel}{\ensuremath{99.999}}

\shorttitle{Cosmology with PS1 SN~Ia}
\shortauthors{Rest, Scolnic et al.}
\begin{document}
\title{Cosmological Constraints from Measurements of Type Ia Supernovae discovered during the first 1.5 years of the Pan-STARRS1 Survey}

\def\stsci{1}
\def\jhu{2}
\def\cfa{3}
\def\illast{4}
\def\illphy{5}
\def\hawaii{6}
\def\harvard{7}
\def\qub{8}
\def\mpia{9}
\def\hubble{10}
\def\inafNapoli{11}
\def\waterlooA{12}
\def\waterlooB{13}
\def\ucsantacruz{14}
\def\inafPadova{15}
\def\lco{16}
\def\ucsb{17}
\def\pitt{18}
\def\durham{19}
\def\princeton{20}

\author{
% dscolnic@pha.jhu.edu,rfoley@cfa.harvard.edu,mhuber@pha.jhu.edu,rchornock@cfa.harvard.edu, gnarayan@noao.edu,jt@ifa.hawaii.edu,eberger@cfa.harvard.edu,asoderbe@cfa.harvard.edu,stubbs@physics.harvard.edu,ariess@stsci.edu,rkirshner@cfa.harvard.edu,S.Smartt@qub.ac.uk,eschlafly@gmail.com,rodney@jhu.edu
A.~Rest\altaffilmark{\stsci},
D.~Scolnic\altaffilmark{\jhu},
R.~J.~Foley\altaffilmark{\cfa,\illast,\illphy},
M.~E.~Huber\altaffilmark{\hawaii},
R.~Chornock\altaffilmark{\cfa},
G.~Narayan\altaffilmark{\harvard},
J.~L.~Tonry\altaffilmark{\hawaii},
E.~Berger\altaffilmark{\cfa},
A.~M.~Soderberg\altaffilmark{\cfa},
C.~W.~Stubbs\altaffilmark{\harvard,\cfa},
A.~Riess\altaffilmark{\jhu},
R.~P.~Kirshner\altaffilmark{\cfa,\harvard},
S.~J.~Smartt\altaffilmark{\qub}, 
E.~Schlafly\altaffilmark{\mpia},
%D. Finkbeiner\altaffilmark{\cfa},
S.~Rodney\altaffilmark{\jhu,\hubble},
%
% camille.leibler@gmail.com,mike.hudson@uwaterloo.ca,mariateresa.botticella@oapd.inaf.it,andrea.pastorello@oapd.inaf.it,djbrout@gmail.com,pchallis@cfa.harvard.edu,iczekala@cfa.harvard.edu,mrdrout@gmail.com,r.kotak@qub.ac.uk,rlunnan@cfa.harvard.edu,gmarion@cfa.harvard.edu,mmccrum04@qub.ac.uk,dmilisav@cfa.harvard.edu,nsanders@cfa.harvard.edu,k.w.smith@qub.ac.uk,estaffo5@jhu.edu,dthilker@pha.jhu.edu,svalenti@lcogt.net,wmwv@pitt.edu
M.~T.~Botticella\altaffilmark{\inafNapoli}, 
D.~Brout\altaffilmark{\jhu},
P.~Challis\altaffilmark{\cfa},
I.~Czekala\altaffilmark{\cfa},
M.~Drout\altaffilmark{\cfa}, 
M.~J. Hudson\altaffilmark{\waterlooA,\waterlooB}, 
R.~Kotak\altaffilmark{\qub}, 
C.~Leibler\altaffilmark{\ucsantacruz},
R.~Lunnan\altaffilmark{\cfa}, 
G.~H. Marion\altaffilmark{\cfa}, 
M.~McCrum\altaffilmark{\qub},
D.~Milisavljevic\altaffilmark{\cfa},
A.~Pastorello\altaffilmark{\inafPadova},
N.~E.~Sanders\altaffilmark{\cfa},
K.~Smith\altaffilmark{\qub},
E.~Stafford\altaffilmark{\jhu},
D.~Thilker\altaffilmark{\jhu},
S.~Valenti\altaffilmark{\lco,\ucsb},
W.~M.~Wood-Vasey\altaffilmark{\pitt},
Z.~Zheng\altaffilmark{\jhu},
% Builders
% kaiser@IfA.Hawaii.Edu,heather@ifa.hawaii.edu,denneau@ifa.hawaii.edu,chambers@ifa.hawaii.edu,eugene@ifa.hawaii.edu,price@astro.princeton.edu,bills@ifa.hawaii.edu,rjw@IfA.Hawaii.Edu,kud@IfA.Hawaii.Edu,watersc1@IfA.Hawaii.Edu,hodapp@hawaii.edu,nigel.metcalfe@durham.ac.uk,p.w.draper@durham.ac.uk
W.~S.~Burgett\altaffilmark{\hawaii},
K.~C.~Chambers\altaffilmark{\hawaii}, 
L.~Denneau\altaffilmark{\hawaii},
P.~W.~Draper\altaffilmark{\durham},
H.~Flewelling\altaffilmark{\hawaii},
K.~W.~Hodapp\altaffilmark{\hawaii}, 
N.~Kaiser\altaffilmark{\hawaii}, 
R.-P.~Kudritzki\altaffilmark{\hawaii}, 
E.~A.~Magnier\altaffilmark{\hawaii}, 
N.~Metcalfe\altaffilmark{\durham},
P.~A.~Price\altaffilmark{\princeton},
W.~Sweeney\altaffilmark{\hawaii},
R.~Wainscoat\altaffilmark{\hawaii},
C.~Waters\altaffilmark{\hawaii}.
}

% The ordering here should be sequential, matching the sequence in the list of authors:
\altaffiltext{\stsci}{Space Telescope Science Institute, 3700 San Martin Drive, Baltimore, MD 21218, USA}
\altaffiltext{\jhu}{Department of Physics and Astronomy, Johns Hopkins University, 3400 North Charles Street, Baltimore, MD 21218, USA}
\altaffiltext{\cfa}{Harvard-Smithsonian Center for Astrophysics, 60 Garden Street, Cambridge, MA 02138, USA}
\altaffiltext{\illast}{Astronomy Department, University of Illinois at Urbana-Champaign, 1002 West Green Street, Urbana, IL 61801, USA}
\altaffiltext{\illphy}{Department of Physics, University of Illinois Urbana-Champaign, 1110 W. Green Street, Urbana, IL 61801, USA}
\altaffiltext{\hawaii}{Institute for Astronomy, University of Hawaii, 2680 Woodlawn Drive, Honolulu, HI 96822, USA}
\altaffiltext{\harvard}{Department of Physics, Harvard University, 17 Oxford Street, Cambridge MA 02138}
\altaffiltext{\qub}{Astrophysics Research Centre, School of Mathematics and Physics, Queens University Belfast, Belfast, BT71NN, UK}
\altaffiltext{\mpia}{Max Planck Institute for Astronomy, K\:onigstuhl 17, D-69117 Heidelberg, Germany}
\altaffiltext{\hubble}{Hubble Postdoctoral Fellow}
\altaffiltext{\inafNapoli}{INAF - Osservatorio Astronomico di Capodimonte, Salita Moiariello 16, 80131 Napoli, Italy}
\altaffiltext{\waterlooA}{University of Waterloo, 200 University Ave W  Waterloo, ON N2L 3G1, Canada}
\altaffiltext{\waterlooB}{Perimeter Institute for Theoretical Physics, 31 Caroline St N, Waterloo, Ontario, N2L 2Y5, Canada}
\altaffiltext{\ucsantacruz}{Department of Astronomy \& Astrophysics, University of California, Santa Cruz, CA 95060, USA}
\altaffiltext{\princeton}{Department of Astrophysical Sciences, Princeton University, Princeton, NJ 08544, USA}
\altaffiltext{\inafPadova}{INAF - Osservatorio Astronomico di Padova, Vicolo dell'Osservatorio 5, 35122 Padova, Italy}
\altaffiltext{\lco}{Las Cumbres Observatory Global Telescope Network, Inc., Santa Barbara, CA 93117, USA} 
\altaffiltext{\ucsb}{Department of Physics, University of California Santa Barbara, Santa Barbara, CA 93106-9530, USA}
\altaffiltext{\pitt}{PITT PACC, Department of Physics and Astronomy, University of Pittsburgh, Pittsburgh, PA 15260, USA.}
\altaffiltext{\durham}{Department of Physics, University of Durham Science Laboratories, South Road Durham DH1 3LE, UK}

\begin{abstract}
We present \griz\ light curves of \SNIAPSall\ spectroscopically
confirmed Type Ia Supernovae ($0.03 < z <0.65$) discovered during the
first 1.5 years of the \PS\ Medium Deep Survey.  The \PS\ natural
photometric system is determined by a combination of on-site
measurements of the instrument response function and observations of
spectrophotometric standard stars. We find that the systematic
uncertainties in the photometric system are currently 1.2\% without
accounting for the uncertainty in the {\it HST} Calspec definition of
the AB system. A Hubble diagram is constructed with a subset of
\SNIAPSused\ out of \SNIAPSall\ SNe~Ia that pass our light curve
quality cuts. The cosmological fit to \SNIAused\ SNe~Ia (\SNIAPSused\
PS1 SNe~Ia + \SNIAlowztot\ light curves from \SNIAlowzused\ low-z
SNe~Ia), using only SNe and assuming a constant dark energy equation
of state and flatness, yields $w=\wSNstasys$.  When combined with
BAO+CMB({\it Planck})+$H_0$, the analysis yields $\OM=\fomCsys$ and
$w=\wCsys$ including all identified systematics
\citep{Scolnic13_sys}. The value of $w$ is inconsistent with the
cosmological constant value of $-1$ at the 2.3$\sigma$ level.  Tension
endures after removing either the BAO or the $H_0$ constraint, though
it is strongest when including the $H_0$ constraint.  If we include
{\it WMAP9} CMB constraints instead of those from {\it Planck}, we
find $w=\wCwsys$, which diminishes the discord to $<2\sigma$.  We
cannot conclude whether the tension with flat $\Lambda$CDM is a
feature of dark energy, new physics, or a combination of chance and
systematic errors.  The full \PS\ supernova sample with \about3 times
as many SNe should provide more conclusive results.
\end{abstract}


\section{Introduction}
\label{sec:intro}

We have used the Medium Deep Field (MDF) survey of the \PS\ Science
program as the source for detecting thousands of transient
events. These include supernovae (SN) and other transients of unusual
types \citep{ Gezari10, Botticella10, Chomiuk11, Narayan11, Lee12,
Gezari12, Sanders12, Berger12,Lunnan13, Chornock13,
Sanders13,Berger13}. In parallel, we are also exploiting the \PS\
3$\pi$ survey for detection of brighter transients at lower redshifts
\citep{Pastorello10,Valenti12, Inserra13}.  Here, we describe \SNIAPSall\ Type
Ia supernovae (SNe~Ia) from the first year and a half of observations
that we use to measure the history of cosmic expansion to help
constrain the properties of dark energy.  This report explains our
observing strategy, photometric reductions, and cosmological analysis.
The systematic errors in the photometry are discussed at length in
this paper, while the companion paper by \cite{Scolnic13_sys}
(hereafter \Scolnicsys) focuses on the systematic uncertainties in the
cosmological analysis.

SN~Ia have been proven to be reliable standard candles at cosmological
distances.  The Supernova Cosmology Project and the High-Z Supernova
Team discovered SNe~Ia at redshifts from 0.3
to 1 that provided the first evidence for cosmic acceleration
\citep{Riess98:lambda,Perlmutter99}.  This result in combination with
measurements of the baryon acoustic peak in the large-scale
correlation function of galaxies \citep[e.g.,][]{Blake11, Anderson12}
and the power spectrum of fluctuations in the cosmic microwave background (CMB)
\citep[e.g.,][]{Hinshaw12, Planck13} indicates that we live in a flat,
accelerating Universe composed of baryons, dark matter, and dark
energy.

SN~Ia are more-or-less homogeneous thermonuclear explosions of white
dwarfs. The discovery by \cite{Phillips93} that the shape of the
supernova light curve is related to the SN Ia luminosity, opened the
door for their use as precise cosmic standard candles.  In 1996, the
Cal\'an/Tololo group published light curves of 29 SN in 4 colors
\citep{Hamuy96:lc}.  This data set was large enough to develop
reliable ways to use the SN light curves to determine the intrinsic
luminosity of SN Ia and to measure the luminosity distance with an
precision of \about 10\% to each object (e.g., $\Delta m_{15}$,
\citealt{Hamuy96:temp, Hamuy96:lum, Phillips99}, and MLCS,
\citealt{Riess96}). Since then, new approaches and new algorithms have
improved light-curve fitting so that well-observed SN~Ia have
individual luminosity distances good to 5\% (e.g., MLCS2K2,
\citealt{Jha07}; Stretch, \citealt{Goldhaber01}; SALT,
\citealt{Guy07}; SiFTO, \cite{Conley08_SIFTO}; and BayeSN,
\citealt{Mandel09, Mandel11}).

With more accurate methods to determine distances of SN~Ia, it is now
possible to probe the nature of dark energy by constraining its
equation of state, characterized by the parameter $w=P/(\rho c^2)$,
where $P$ is its pressure and $\rho$ is its density.  Evidence for
cosmic acceleration and constraints on the dark energy are derived
from the combination of low redshift samples and high redshift
samples. In the last decade, many groups have worked on assembling
large sets of low-redshift SNe (e.g., CfA1-CfA4,
\cite{Riess99,Jha06,Hicken09,Hicken09b,Hicken12}; CSP,
\cite{Contreras10,Folatelli10,Stritzinger11}; LOSS,
\cite{Ganeshalingam13}).  In total, the currently published low-z SN
sample comprises more than 500 SNe~Ia. There have been three main
surveys probing the higher redshift range, ESSENCE
\citep{Miknaitis07,Wood-Vasey07}, SNLS \citep{Conley11,Sullivan11},
and SDSS \citep{Frieman08, Kessler09_SDSS}.  These surveys have
overlapping redshift ranges from $0.1 \la z \la 0.4$ for SDSS, $0.2
\la z \la 0.7$ for ESSENCE, and $0.3 \la z \la 1.1$ for SNLS. For a
more complete review of SN~Ia cosmology from these surveys, see
\cite{Kirshner10}.  

Increasing the number of SN alone will not improve the limits on $w$
unless we decrease the systematic errors as well.  The photometric
calibration is currently the most significant source of systematic
bias \citep{Sullivan11}, since different sets of SN with light curves
from different telescope/detector systems are compared to each
other. There are significant efforts underway to improve and
homogenize the photometric calibration of previous \citep[e.g., SNLS
and SDSS,][]{Betoule12} and on-going wide-field transient surveys like
PS1 \citep{PS1_system}, PTF \citep{Rau09}, and the Dark Energy Survey
\citep[DES,][]{Flaugher12}.  Another technical issue is how the light
curve fitters deal with the degeneracy between intrinsic luminosity,
extinction, and intrinsic color.  For example, \cite{Scolnic13_color}
show that the distance residuals can be decreased by assuming an intrinsic
SN~Ia color dispersion as a prior, which also is consistent with a
Milky Way-like reddening law.

In the future, the full \PS\ SN sample will be uniquely suited to better
constrain the nature of dark energy by covering a very wide redshift
range ($0.03<z<0.65$) with a single instrument. Although the \PS\
sample presented in this paper does not have enough low-redshift SN~Ia
for a stand-alone cosmological analysis, it is sufficient for us to
investigate the best approach to combining data from separate
photometric systems.  In this paper, we use the \SNIAPSall\ SN~Ia from
the first 1.5 years (2009 September to 2011 May) of \PS\ to constrain
the cosmological parameters along with a joint constraint by other
cosmological probes.  We emphasize the reduction of systematic
uncertainties that affect the measurement of dark energy properties.
The number of SN~Ia that pass all light-curve quality cuts
(\SNIAPSused) is almost half of the largest published individual
high-redshift sample (248, \cite{Conley11,Sullivan11}). Our sample
allows us to identify where we have the largest systematic errors and
to develop remedies for them.  Future analysis of the full data set
will build on our current analysis, and improve today's largest
systematic uncertainties: the photometric calibration, and flaws in
light-curve fitting techniques.

This paper is organized as follows. In \S\ref{sec:PS1}, we
introduce the technical aspects of the \PS\ system. We describe the
transient alert system and the spectroscopic follow-up in
\S\ref{sec:alerts} and \S\ref{sec:specfollowup}, respectively.
The SN~Ia light curves are presented in \S\ref{sec:transphot}, and the
photometric calibration is presented in \S\ref{sec:photcal}.  We
discuss how we determine distances from the SN~Ia light curves in
\S\ref{sec:lcfit}.  In \S\ref{sec:results}, we determine
the cosmological parameters by combining our sample of \SNIAPSused\
high-quality SN~Ia with the low-redshift sample and constraints from
other cosmological probes.

\section{\PS\ Survey}
\label{sec:PS1}

The \PS\ system is a high-etendue wide-field imaging system, designed
for dedicated survey observations. The system is installed on the peak
of Haleakala on the island of Maui in the Hawaiian island chain.  We
provide below a terse summary of the \PS\ survey instrumentation.  A
more complete description of the \PS\ system, both hardware and
software, is provided by \cite{PS1_system}.

The \PS\ optical design \citep{PS1_optics} uses a 1.8~meter diameter
$f$/4.4 primary mirror, and a 0.9~m secondary.  The telescope delivers
images with low distortion over a field diameter of 3.3 degrees.  An
individual CCD cell has 800x800 pixels, with 10~$\mu$m pixels that
subtend 0.258~arcsec. 64 of these CCD cells are grouped into an 8x8
array. The focal plane consists of 60 of these independent arrays, for
a total of 1.4~Gigapixel. The detectors are back-illuminated CCDs
manufactured by Lincoln Laboratory, which are read out using a
StarGrasp CCD controller in 7 seconds for a full unbinned
image. Initial performance assessments are presented in
\cite{PS1_GPCB}.

The \PS\ observations are obtained through a set of 5 broadband
filters, which we have designated as \gps, \rps, \ips, \zps, and \yps\
(\grizy). These filters are similar to the ones used in SDSS, with the
most significant difference in the $g$ band.  We use the instrumental
response functions determined by \citet{Tonry_photcal}.

In addition to covering the entire sky at $\delta>-30\deg$ in 5 bands
(3$\pi$ survey), the \PS\ survey has obtained deeper, higher cadence
images in the \grizy\ bands of the MDF fields listed in Table
\ref{tab:fields}.  In this paper, we exclusively use SN~Ia detected in
the MDFs.  
\begin{deluxetable}{lccccccc}
\tablecaption{Pan-STARRS1 MDFs \label{tab:fields}}
\tablehead{
\colhead{Field} &
\colhead{RA} &
\colhead{Dec} &
%\colhead{First} &
%\colhead{Last} &
\colhead{$N_{\rm SN}$} &
%\colhead{FWHM \griz}\\
%~ & J2000 & J2000 &  ~ & arcsec
\multicolumn{4}{c}{FWHM in arcsec}\\
~ & J2000 & J2000 &  ~ & \gps & \rps  & \ips& \zps
}
\startdata
MD01  & 035.875  &  $-$04.250 & 15 & 1.25 &  1.15 & 1.05 & 1.03 \\
MD02  & 053.100  &  $-$27.800 & 16 & 1.31 &  1.20 & 1.11 & 1.06 \\
MD03  & 130.592  &  $+$44.317 & 20 & 1.18 &  1.09 & 1.06 & 1.03 \\
MD04  & 150.000  &  $+$02.200 & 22 & 1.17 &  1.09 & 1.07 & 1.03 \\
MD05  & 161.917  &  $+$58.083 & 13 & 1.24 &  1.17 & 1.06 & 0.99 \\
MD06  & 185.000  &  $+$47.117 & 15 & 1.25 &  1.18 & 1.14 & 1.05 \\
MD07  & 213.704  &  $+$53.083 & 15 & 1.23 &  1.13 & 1.14 & 1.08 \\
MD08  & 242.787  &  $+$54.950 & 24 & 1.27 &  1.14 & 1.07 & 1.09 \\
MD09  & 334.188  &  $+$00.283 & \phantom{1}9 & 1.26 & 1.15 & 1.02 & 1.02 \\
MD10  & 352.312  &  $-$00.433 & \phantom{1}7 & 1.26 & 1.18 & 1.01 & 1.03 
\enddata
\end{deluxetable}

The MD field exposure times in the 5 filters
are listed in Table~\ref{tab:MDFcadence}.  Observations of 3-5 MD
fields are taken each night and the filters are cycled through in the
pattern: \gps\ and \rps\ in the same night (dark time), followed by
\ips\ and \zps\ on the subsequent second and third night,
respectively. Around full moon only \yps\ data are taken.  Any one
epoch consists of 8 dithered exposures of 8$\times$113 seconds for
\gps\ and \rps\ or 8$\times$240 seconds for the other three, giving
nightly stacked images of 904 and 1920 seconds duration.
For the current analysis, we do
not use the \yps\ band.  However in the future, the \yps\ band data
may prove useful for cosmological analysis of SN Ia because
measurements in this band are less affected by dust in the near
infrared.  Due to weather and the occasional technical downtime, the
effective cadence varies significantly from season to season. On
average, our cadence is 6 detections in 10 days, with a 5 day gap
during bright time when the MDFs are exclusively observed in \yps.

\begin{deluxetable}{cccc}
\tablecaption{MDF cadence\label{tab:MDFcadence}}
\tablehead{
\colhead{Night} &
\colhead{Filter} &
\colhead{Exposure Time} &
\colhead{$5\sigma$ Depth} \\
\colhead{} &
\colhead{} &
\colhead{(seconds)} &
\colhead{(AB mag)} 
}
\startdata
1 & \gps, \rps & 8$\times$113 each & 23.1, 23.3 \\
2 & \ips & 8$\times$240  & 23.2 \\
3 & \zps & 8$\times$240  & 22.8 \\
repeats \\
FM$\pm$3 & \yps & 8$\times$240  & 21.9 
\enddata
\end{deluxetable}

\section{Transient Alert System}
\label{sec:alerts}

The depth of the MDF survey is ideal for detecting SN~Ia up to a
redshift of $z \approx 0.7$, and the cadence provides 
well-sampled, multi-band light curves. In this section, we describe
how the MDF data are reduced and the objects of interest are
identified. 

The \PS\ IPP system \citep{PS1_IPP} performs flat-fielding on each
individual image, using white-light flat-field images of a dome
screen, in combination with an illumination correction obtained by
rastering sources across the field of view.  After determining an
initial astrometric solution \citep{PS1_astrometry}, the flat-fielded
images were then warped onto the tangent plane of the sky, using a
flux conserving algorithm.

For the MDFs, there are several (typically eight) dithered images per
filter in a given night.  This allows for the removal of defects like
cosmic rays and satellite streaks before they are combined into a
nightly stacked image using a variance-weighted scheme.  The nightly
MDF stacked images 
are processed through a
frame-subtraction analysis using the \photpipe\ pipeline that members
of our team developed for the SuperMACHO and ESSENCE surveys
\citep{Rest05,Garg07,Miknaitis07}.  This robust and well-tested system
determines the appropriate spatially varying convolution
kernel\footnote{http://www.astro.washington.edu/users/becker/hotpants.html}
needed to match an image with a template image, and then performs a
subtraction of the two images.  
We then
detect significant flux excursions in the difference images using a
modified version of {\it DoPHOT} \citep{Schechter93}.  

The SNe~Ia discussed in this work were discovered during the first
$\sim$1.5~yr of the survey.  At the beginning of the survey, we had
not yet obtained sufficient observations to construct deep template
images for the SN search.  Instead, we typically obtained two epochs
of observations in each filter with double the usual exposure time to
use as templates.  We subtracted these templates from later
observations to search for transient objects.  The use of two
templates reduced false detections from imperfections in a single
template.  After this initial period, we generated deep reference
images from a subset of images with good image quality and low sky
background.  These deep reference images were subsequently used as
templates for the SN search.  The template images used for the search
with \photpipe\ are different from the deeper ones used for the final
light curves created by \transphot, as discussed in Section
\ref{sec:template}.

Once photometry is performed on the difference images, we apply a set
of conditions before flagging an excursion in flux at a given
position as a potential transient source.  Typical conditions are:
\begin{itemize}
\item Positive detections with a signal-to-noise ratio, SNR, $\ge$4 in
  at least 3 \griz\ images within a time window of 10 days.
\item Detections in subtractions using at least two distinct templates.
\item No previous alert at that position.
\end{itemize}

The SNR cut significantly reduces the number of false
positives and artifacts.  Similarly, requiring detections in
subtractions versus two templates reduces spurious detections related
to artifacts in a particular template.  Our requirement of no previous
alert at a particular position prevents variable sources from being
continually flagged.  The parameters related to the search were
adjusted several times throughout the survey.  There was a constant
evolution of the parameters to optimize source detection as we gained
experience with the data.  But there were also cases where we would
modify the parameters at a given time for testing purposes or for
immediate goals such as increasing the number of young detected
objects just prior to a spectroscopic follow-up observing run.

An important aspect of our reduction scheme is to perform PSF fitting
photometry on all difference images at the location of any transient
candidate, which we call ``forced photometry.''  This is particularly
helpful for discriminating between SN and AGN, since many AGN show
low-level, long timescale variability in previous epochs that may not be
above the detection limit in a single image, but clearly detectable
over many epochs.  It also can provide additional data when a SN is
faint (particularly when it is rising) or if the image quality is poor
(for instance when there are clouds).

Each transient candidate that passes our stated conditions is sorted
into one of five categories: transient, possible transient, variable,
asteroid, and artifact.  Transients are identified as sources with no
variability in previous epochs, a ``smooth'' light curve, and either
previous ``in-season'' non-detections or a clear offset from a host
galaxy.  Both because of possible confusion with AGN and subtraction
artifacts, transients that are visually offset from a galaxy nucleus
are easier to identify than those that are coincident with a nucleus.
In cases where the identification is ambiguous either due to low SNR
or possible variability in previous epochs, the object is classified
as a possible transient.

We attempt to err on the side of inclusion to maximize the number of
transients discovered.  We refrain from classifying objects as
``non-transient'' until there is significant evidence that they are not
transients (such as past variability, clear indication of the source
being an artifact, or spectroscopy from an exterior catalog). We
re-evaluate and possibly re-classify each candidate object as we
acquire more points on its light curve.

\begin{figure}[h!]
\includegraphics[width=250pt]{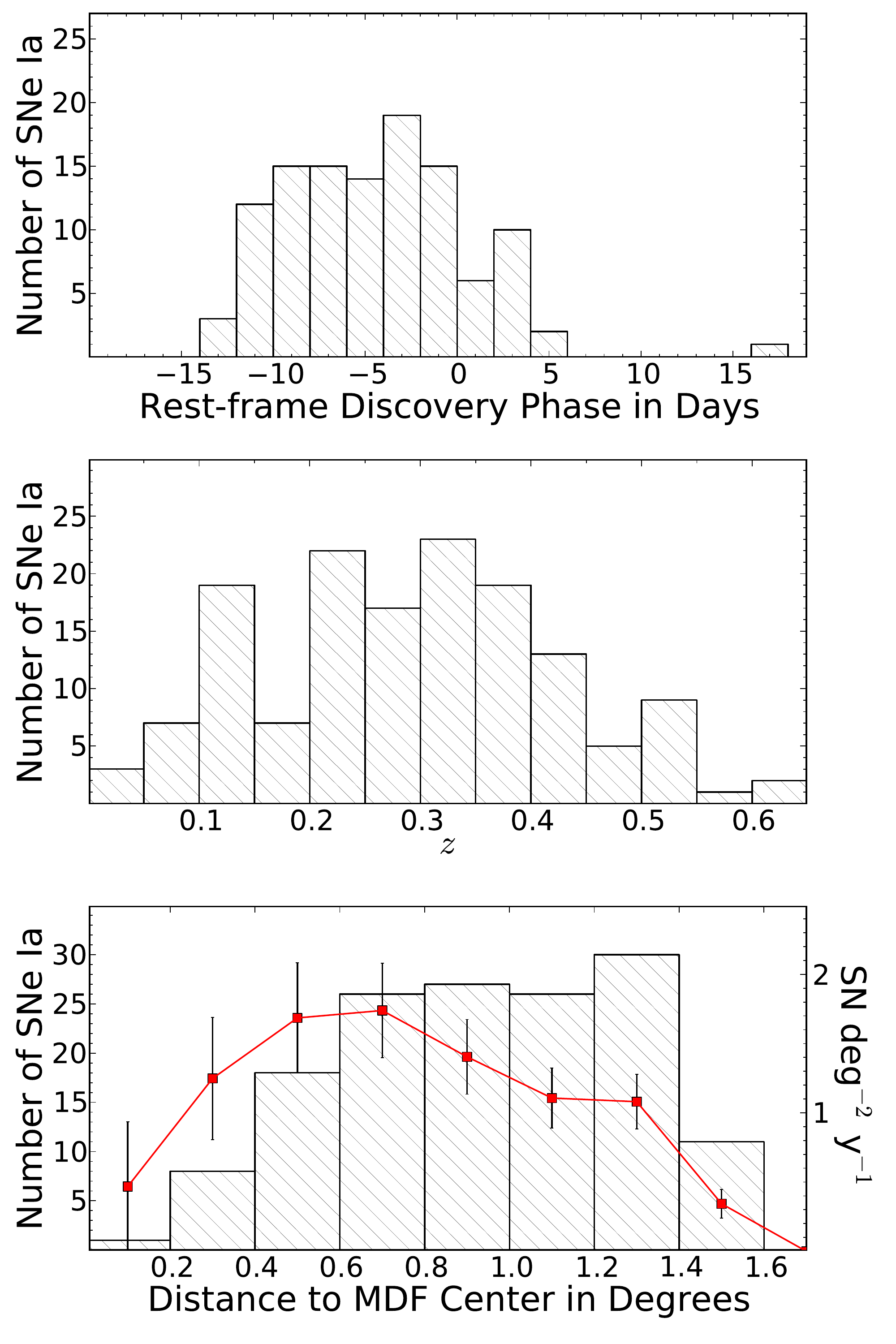}% 
\caption[]{PS1 MDF survey characteristics for all spectroscopically
  confirmed PS1 SN~Ia.
{\it Upper panel:} Histogram of the number of SN~Ia with respect
to the phase of discovery in the rest-frame.
{\it Middle panel:} Histogram of the redshift distribution. 
{\it Lower panel:} Histogram of SN~Ia as a function of distance to the
MDF center.  The observed SN rate per constant area is shown with the red
symbols and line.  The rate is constant within the uncertainties to a
distance of \about 1.3 degrees from center.
\label{fig:surveyhistos}}
\end{figure}

As shown by the upper panel of Figure~\ref{fig:surveyhistos}, we
discovered almost all of our spectroscopically confirmed SNe~Ia at
phases between $-12$ and $+4$~days.

A parallel, independent search for transients is done at Queen's
University Belfast based on difference images which are generated in
Hawaii through the IPP system and this is described in
\cite{McCrum13}. The correlation between the discoveries in \photpipe\
and the IPP based ``Transient Science Server'' is excellent, with
virtually all high significance transients detected in both
systems. Some of the SN in this sample were originally selected for
spectroscopic classification from the IPP-based catalogs, but all were
also detected through \photpipe\ and all the photometry discussed in
this paper is exclusively from the \photpipe\ system.

\section{Spectroscopic Prioritization \& Follow-up Spectroscopy}
\label{sec:specfollowup}

Several spectroscopic programs rely on targets generated from the
PS1/\photpipe\ data stream.  The primary source of classification
spectra is a multipurpose CfA program at the MMT to observe PS1
transients (PI Berger). Other spectra were obtained through Gemini
programs to observe exotic transients and superluminous SN (PIs
Berger, Smartt, and Chornock), a Gemini program to observe transients
with faint hosts (PI Tonry), a Magellan program to obtain high-SNR
spectra of SN Ia (PI Foley), and Magellan and William Herschel
programs to observe unusual transients and superluminous SN (PIs
Berger, Chornock, and Smartt). Ongoing programs that will create a
larger sample of well-observed SN Ia from \PS\ include a program on
Gemini (Foley, PI) and at Keck using NASA time (Kirshner, PI.) to
select targets for further study in the rest frame IR with HST.
The target selection for each program
varies significantly, but all begin with identifying potential targets
from the PS1/\photpipe\ data stream.  We only have sufficient
spectroscopic resources to obtain spectra of $\sim$10\% of the PS1
transients, so potential targets must then pass several selection
criteria, which again depend on the program.  The most important
criteria are position and brightness.  Magellan and Gemini-South can
only point to the five equatorial and southern MDFs.  Observations at
Magellan and the MMT are generally limited to targets with $\rps <
22$~mag, the WHT is limited to $\rps < 21.5$~mag, while fainter
targets can be observed at Gemini.  Generally, an effort was made to
observe transients close to their peak brightness.  Several programs
specifically target transients with very faint host galaxies, while
others attempt to observe transients with host galaxies that have a
particular photo-$z$.  Although many transients are targeted
specifically as potential SN~Ia, a significant number of SN~Ia in our
sample were originally targeted as possible non-SN~Ia transients.

Spectroscopic observations of PS1 targets were obtained with a wide
variety of instruments: the Blue Channel Spectrograph
\citep{Schmidt89} and Hectospec \citep{Fabricant05} on the 6.5-m MMT,
the Gemini Multi-Object Spectrographs (GMOS; \citealt{Hook04}) on both
Gemini North and South, the Low Dispersion Survey Spectrograph-3
(LDSS3\footnote{http://www.lco.cl/telescopes-information/magellan/instruments-1/ldss-3-1})
and the Magellan Echellette (MagE; \citealt{Marshall08}) on the 6.5-m
Magellan Clay telescope, and the Inamori-Magellan Areal Camera and
Spectrograph (IMACS; \citealt{Dressler11}) on the 6.5-m Magellan Baade
telescope, and the ISIS spectrograph on the WHT.  Nod-and-shuffle
techniques \citep{Glazebrook01} were used with some GMOS (North and
South) observations to improve sky subtraction in the red portion of
the spectrum.

Standard CCD processing and spectrum extraction were accomplished with
IRAF\footnote{IRAF: the Image Reduction and Analysis Facility is
  distributed by the National Optical Astronomy Observatory, which is
  operated by the Association of Universities for Research in
  Astronomy, Inc. (AURA) under cooperative agreement with the National
  Science Foundation (NSF).}.  The data were extracted using the
optimal algorithm of \citet{Horne86}.  Low-order polynomial fits to
calibration-lamp spectra were used to establish the wavelength scale.
Small adjustments derived from night-sky lines in the object frames
were applied.  For the MagE spectra, the sky was subtracted from the
images using the method described by \citet{Kelson03}.  We employed
our own IDL routines to flux calibrate the data and remove telluric
lines using the well-exposed continua of the spectrophotometric
standards \citep{Wade88, Foley03, Foley09:08ha, Silverman12:bsnip}.

With fully calibrated spectra, we classify the PS1 objects
by physical origin.  SNe, having broad (\about 10,000~km~s$^{-1}$) spectral
features, are very distinct from AGNs, galaxies, stars, and other
astrophysical objects.  However, it can occasionally be difficult to
distinguish among SN types.  High-redshift SN spectra typically have low
signal-to-noise ratio (S/N) and considerable host-galaxy contamination,
while spectra of the highest redshift SNe will lack the \ion{Si}{2}
$\lambda$6355 feature.  We have implemented the SNID algorithm
\citep{Blondin07} to aid in SN classification.

For SNID, as well as any SN spectral fitting routine, the output is
dependent on the input parameters such as wavelength range.  Because
of the different approaches and the various input parameters, it is
possible for different fitters to suggest different classifications.
However, humans ultimately classify each object, and reasonable inputs
to the fitters should yield similar results. Unless there is a
confident classification, the SN is discarded.

\begin{figure}
\includegraphics[width=380pt]{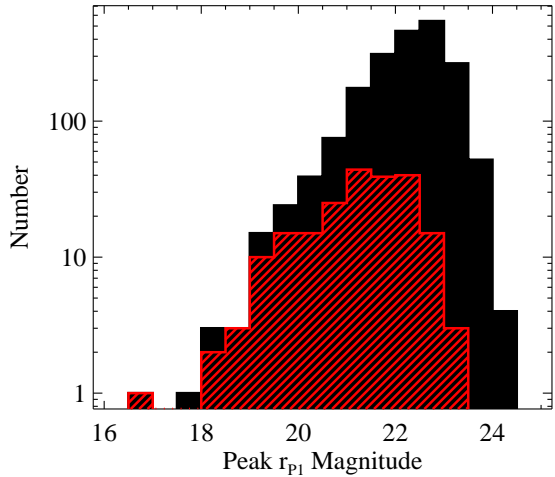}% 
\caption[]{Histogram of \rps\ peak magnitudes of spectroscopically
  confirmed SN~Ia (red) and events classified as highly likely SN~Ia
  (black) based on light-curve fitting with PSNID \citep{Sako11}.
\label{fig:specfollowup}}
\end{figure}

In Table~\ref{tab:spec}, we present a full list of our observations
(date of observation, telescope/instrument, and exposure times) of all
spectroscopically confirmed SNe Ia detected in the \PS\ photometric
data spanning from 2009 September to the end of 2011 May.  We also
include information about the nature of each object (redshift, phase,
and light-curve shape) in Table~\ref{tab:SN}.  The middle panel of
Figure~\ref{fig:surveyhistos} shows the redshift distribution of the
spectroscopically confirmed SN~Ia.  Since this period includes the
ramp-up of \PS\ operations, our detection efficiency increased over
time due to improved reductions, deeper templates, and longer history
to identify and differentiate between the transients and variables.
In addition, access to telescopes, weather, and \PS\ downtime
influenced our spectroscopic follow-up efficiency.  In particular, the
annual monsoon rainstorms at the MMT in July along with the August
telescope shutdown resulted in poor follow up for the summer fields
(primarily MDFs 9 and 10).  Our spectroscopic follow-up rate for SN~Ia
is \about 1.5~deg$^{-2}$~yr$^{-1}$ and constant within the
uncertainties out to a radius of \about 1.3 degrees in each field (see
lower panel of Figure~\ref{fig:surveyhistos}). In
Figure~\ref{fig:specfollowup}, we show the peak magnitudes of
spectroscopically confirmed SN Ia as well as events having a high
likelihood of being SNe~Ia based on light-curve only classification
\citep{Sako11}. Unsurprisingly, spectroscopic follow-up favored
brighter transients. As Figure~\ref{fig:specfollowup} indicates, our
5-sigma detection limit
for identifying transients
provides good light curves for SN Ia with $m < 24$, while our
spectroscopic sample consists principally of objects with $m < 22$.
The Malmquist bias introduced by this selection is discussed in
\S\ref{sec:results}.

\section{Light Curves with the Transient Photometry Pipeline}
\label{sec:transphot}

The goal of the transient alert system is to produce light curves of
potential transients in a quick and robust way to distribute them for
validation and follow-up spectroscopy. However, for measuring
distance, the main focus is on minimizing the random and systematic
uncertainties in the light curves.  Therefore, we have set up a
transient photometry pipeline (\transphot), which is closely related
to \photpipe\ and uses most of its features, but differs in some
significant ways, described below.
The overall process can be summarized as such: Templates from stacking
multiple images at a given sky position are constructed to allow for
accurate WCS registration and image subtraction to measure photometry.
From the subtracted images of supernovae, a SN centroid is determined
and forced photometry is performed at that position on all images,
using the PSF determined from neighboring stars for that epoch.
Afterwards, the errors of the SN measurements and the baseline flux is
adjusted so that the reduced chi-squared of the measurements of the
supernova that do not have supernova light will be unity.

\subsection{Templates}
\label{sec:template}

The templates are created through a custom calibration and stacking
process. The 8 dithers from a single night are reduced and combined
into a ``nightly stack'' by IPP with a variance-weighted scheme (see
\S\ref{sec:alerts}). 

The ``deep stack'' is constructed by combining nightly stacks
(typically \about 30), weighted by the product of the inverse variance
and the inverse area of the PSF.  This prescription is nearly optimal
for point-source detection and photometry \citep{Tonry_WD}. The
typical seeing values determined by {\it DoPHOT} for deep stacks that
use all available epochs are shown in Table~\ref{tab:fields}
\citep{Tonry_WD}.  The typical 5$\sigma$ detection limit in these deep
stacks is \about 25.2,25.4,25.4,25.0 for \griz, respectively.  To
create the deep templates for a given SN, we only use epochs with no
SN flux in the images. This ensures the spatial consistency of the
PSF, and also minimizes errors introduced by imperfect image
subtraction kernels.  In general, if there is no SN flux in the
template, only the extended flux contribution from the galaxy is
subtracted from the image. Such spatially extended sources are less
susceptible to imperfections in the kernel than point sources,
minimizing the difference image residuals at the position of the
SN. The main disadvantage of templates that only use epochs that
contain no SN flux is the template's lower SNR than
templates that use all epochs.  However, since even in these
templates the SNR is still significantly higher than the SNR in the
single epochs (nightly stacks), the SNR in the difference image is
only marginally degraded.

\subsection{SN Centroids}
\label{sec:centroids}

Determining the correct position of the SNe is important for two
reasons.  (1) For many epochs, the flux of the SN is below the
detection limit. (2) Even if the flux of the SN is above the detection
limit, its measurement is biased toward higher fluxes due to Poisson
fluctuations in the sky background (see \S\ref{sec:forcedphot} for a
detailed discussion).

In order to achieve good centroiding, an accurate world coordinate
system (WCS) solution for each image is needed.  Additionally, the
uncertainties in the position vary significantly due to their strong
dependence on the SNR of the flux and the seeing of the image.
Therefore it is important to have an accurate understanding of the
uncertainties in order to calculate an unbiased weighted average
centroid. In Appendix~\ref{sec:wcs}, we determine the accuracy of our
WCS solution. The astrometric uncertainty $\sigma_a$ of a given
detection has a floor mostly due to pixelization ($\sigma_{a1}$), and
in addition a random error $\sigma_{a2}$, which is scaled by the
square of the ratio of the FWHM of the image and the SNR of the detection:
\begin{equation}
  \sigma_a^2 = \sigma_{a1}^2 + \sigma_{a2}^2\left( \frac{\rm FWHM}{\rm
      SNR}\right)^2.
\end{equation}

As described in Appendix~\ref{sec:wcs}, for \PS\ we conservatively use
$\sigma_{a1} = 40$~mas and $\sigma_{a2} = 1.5$ to calculate the
astrometric uncertainty of a single detection. The detections are
grouped into transient objects, for which the 3$\sigma$ clipped and
weighted average centroids are calculated. 
The accuracy of the centroid has a small effect on the accuracy of the
photometry (see Figure~\ref{fig:comparison_forced_regdiff}), and we
discuss how we correct for this effect as described in
\S\ref{sec:forcedphot} and Appendix~\ref{sec:centroiderrordmcorr}.

\subsection{Light Curves}
\label{sec:lcs}

The modeling and fitting of the PSF is one of the main potential
sources of systematic biases in the photometry. For the alert system,
we use a customized version of {\it DoPHOT}, which is quick, robust,
and produces adequate photometry for alerts.  However, this {\it
DoPHOT} version uses an analytic PSF that does not capture the
non-Gaussian PSF tails in the \PS\ images, which was especially
important during the first year while the optical system was being
perfected.  This can introduce biases at the level of a few percent
between faint and bright stars.  For more precise photometry, we
employ {\it DAOPHOT} \citep{Stetson87} as implemented in IDL, which
fits an empirical correction in addition to the Gaussian model.

\subsubsection{Linearity}
\label{sec:linearity}

\begin{figure}[h!]
\includegraphics[width=250pt]{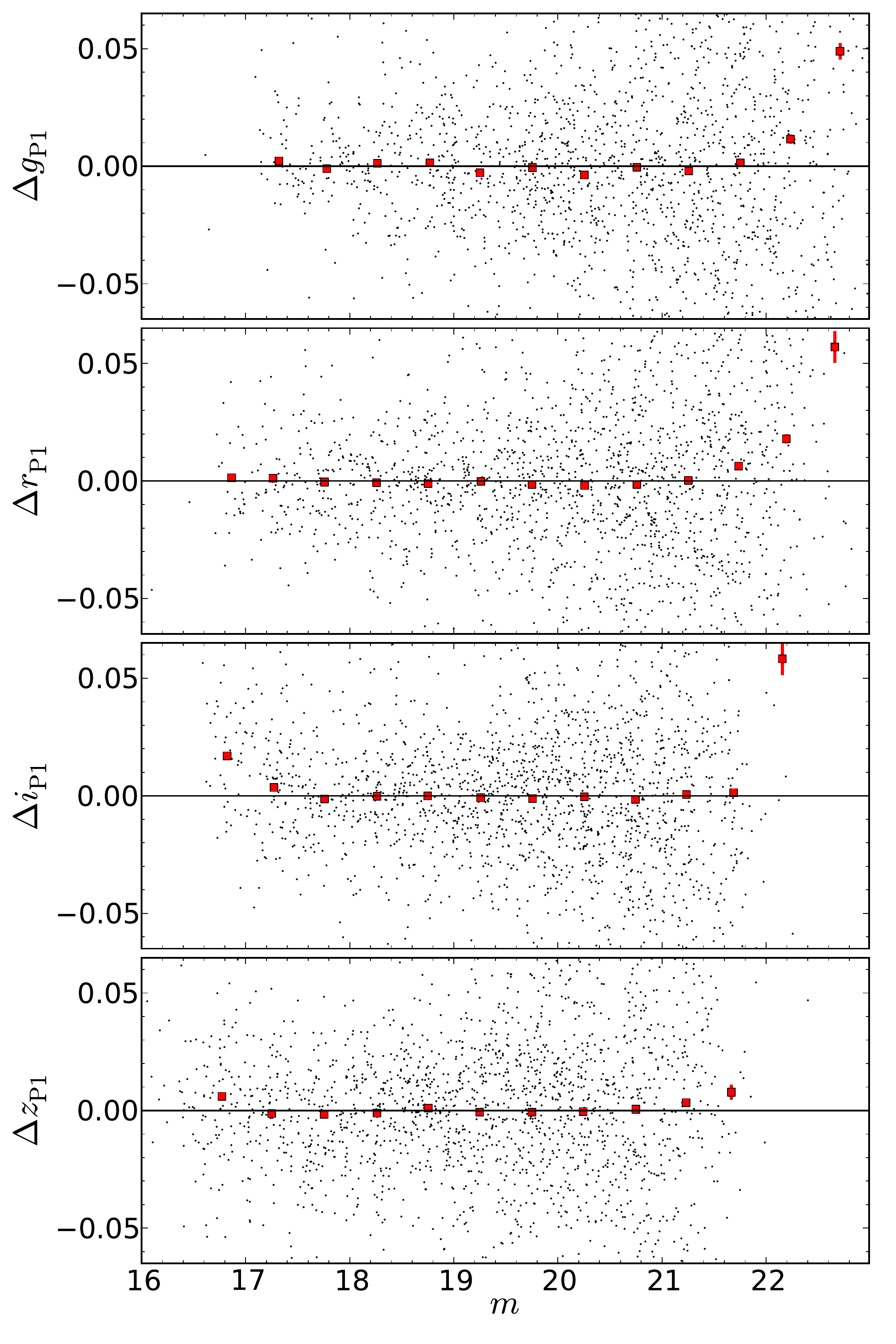}
\caption[]{Comparison between deep and nightly stack photometry for
  stars in the MDFs. The red circles indicate the weighted average of
  the magnitude difference $\Delta m = m_{\mathrm{deep}} -
  m_{\mathrm{nightly}}$ from 100 randomly selected images for each of
  \griz\ from top to bottom, respectively. They are consistent with
  zero within the errors at all magnitudes except at the very
  brightest and faintest ends. In particular \ips\ and \zps\ show significant deviation from
  zero for $m < 17$~mag.  Malmquist bias is likely the cause of the
  deviation at faint magnitudes. The black dots show a small subset of
  $\Delta m$ values for illustration.
\label{fig:dMversusM}}
\end{figure}

Figure~\ref{fig:dMversusM} shows the magnitude difference between the
deep and nightly stacks, $\Delta m = m_{\mathrm{deep}} - m_{\mathrm{nightly}}$,
versus the deep stack magnitudes (black dots) for the various \PS\
filters.  For magnitudes fainter than 17.5, the agreement is
excellent, and the average residual is on the order of 1~mmag and in
general within the uncertainties (with the exception of the faintest
magnitudes, where the expected Malmquist bias can be seen).  A small
systematic bias can be seen for the very brightest magnitude bins. In
particular, for the \ips\ filter, the average $\Delta m = 0.017 \pm
0.0013$~mag for the bin with $i_{\rm p1, deep} \approx 17$~mag is
significantly above zero. This is most likely the result of imperfect
stacking algorithms of stars close to or at the saturation limit, and
we therefore exclude all stars brighter than $m = 17.5$~mag for
determining the zero point of an image or the difference image
kernel. No detections from our SN light curves fall into this
magnitude range, therefore we set the systematic bias due to
non-linearity of the photometry to 1~mmag (see Table~\ref{tab:systematics_cal}).

\begin{figure}[h!]
\includegraphics[width=250pt]{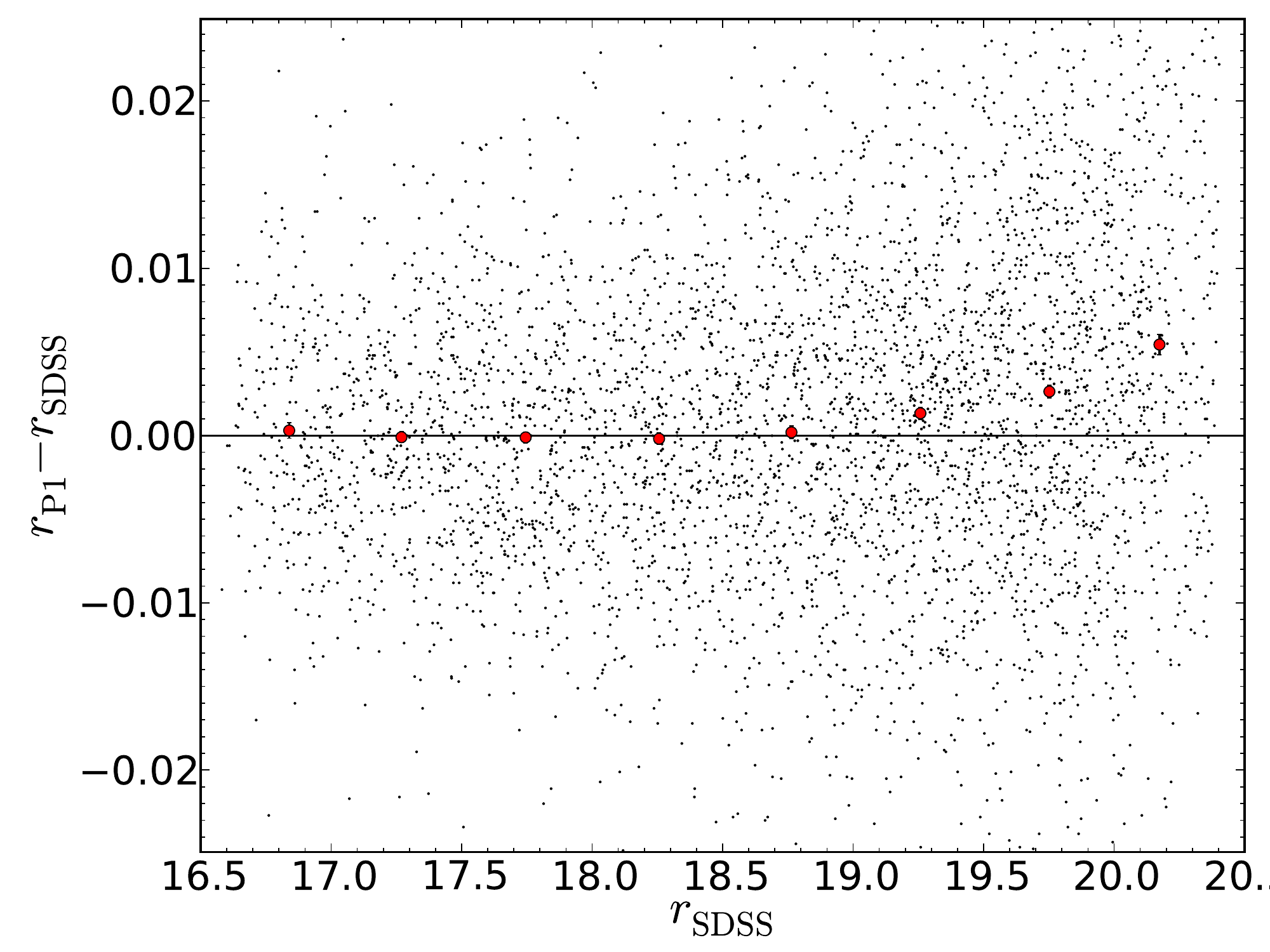}
\caption[]{
Comparison between \rps\ and $r_{\mathrm{SDSS}}$, where
$r_{\mathrm{SDSS}}$ is the SDSS $r$ band magnitude converted into \PS\
natural system magnitudes using Equation~6 and Table~6 from
\Tonryphot.
The red circles indicate the
average of the magnitude difference.
\label{fig:linearity}}
\end{figure}

We also tested the linearity of the PS1 detectors by comparing it to
an outside catalog.  We convert $r_{\mathrm{SDSS}}$ from stripe 82
into \PS\ natural system magnitudes, and calculate the difference,
shown in Figure~\ref{fig:linearity}. For $r_{SDSS}<19$, the average
differences are smaller than 1~mmag, which is an upper constraint on
the linearity over this magnitude range. For larger magnitudes we are
susceptible to a Malmquist bias in the SDSS magnitudes.

\subsubsection{Forced Photometry}
\label{sec:forcedphot}

For each SN, we calculate the weighted average position and then
perform forced photometry in every difference image at this position.
Because we convolve the template, the PSF shape in each difference
image is the same as the one in the original image and is kept
unchanged.  The only free parameter is the peak value of the PSF.

\begin{figure}[h!]
\includegraphics[width=250pt]{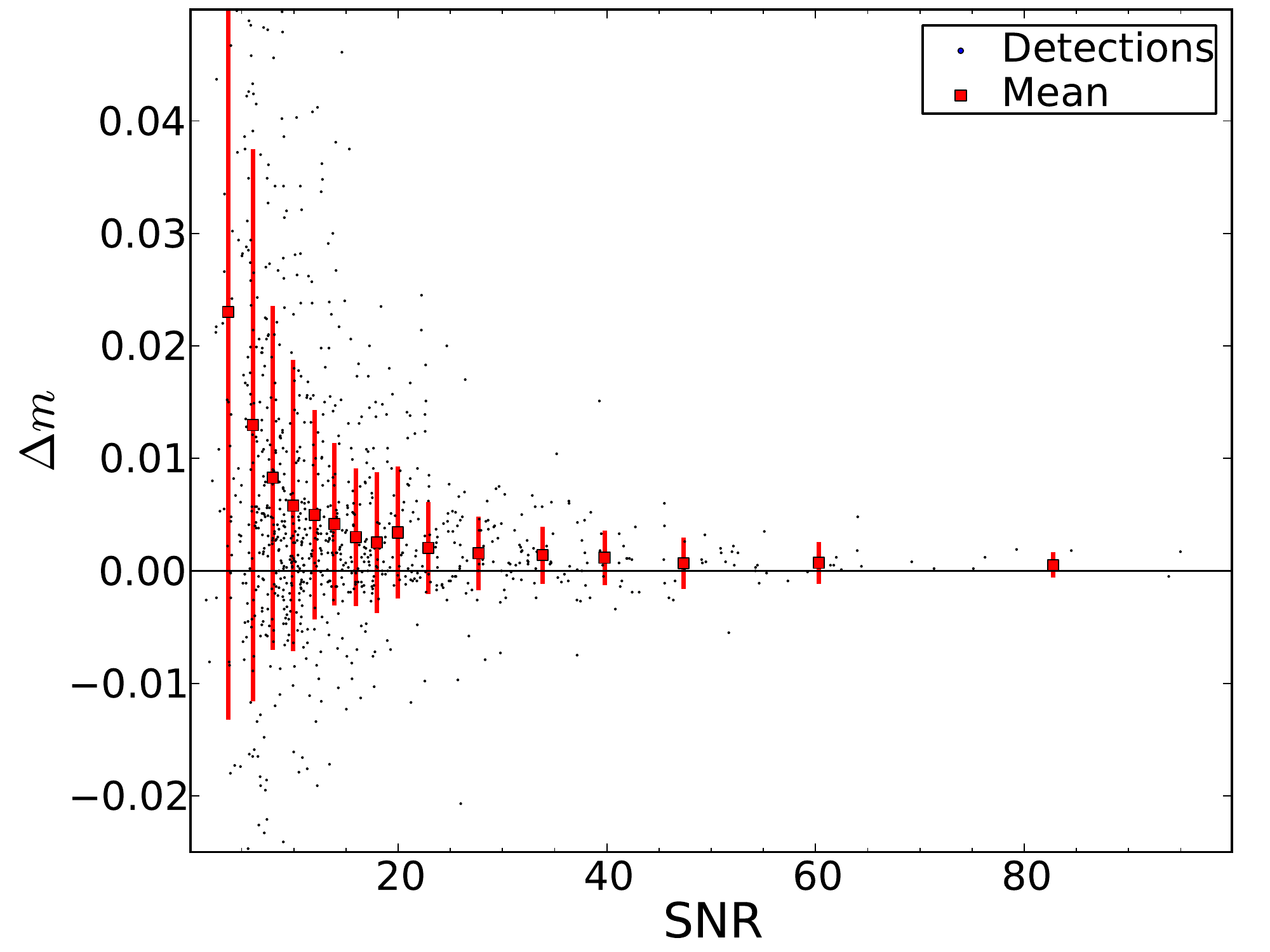}% 
\caption[]{Comparison of forced and regular photometry for
  PS1 SN~Ia as a function of SNR. The red circles indicate the
  weighted average of the magnitude difference $\Delta m =
  m_{\mathrm{forced}} - m_{\mathrm{regular}}$ of SN detections for
  different SNR bins. The error bars indicate the standard
  deviation. The black dots show a small subset of $\Delta m$ values
  for illustration. At high SNR, the average magnitude difference is
  an insignificant 0.5~mmag different from zero, indicating that there
  are no issues with the astrometric calibration, centroiding, or
  differences between forced and unforced PSF fitting.
\label{fig:comparison_forced_regdiff}}
\end{figure}

The success of this photometry method strongly depends on the accuracy
of the WCS solution.  If the WCS accuracy is poor, the forced position
is off the true peak and the flux is underestimated.  This effect is
strongest for high-SNR measurements.  In
Figure~\ref{fig:comparison_forced_regdiff}, we show the magnitude
difference $\Delta m$ between the forced photometry and photometry
where the centroid of the SN is determined in each individual image
versus the SNR of the forced photometry.  At high SNR, the average
magnitude difference is an insignificant 0.5~mmag different from zero.
Though there is some scatter in the magnitude differences, the scatter
is small compared to the Poisson noise. This makes sense, because the
same data are used with both methods.  The important result is that,
for high SNR, the scatter is small, symmetric, and shows no
significant bias.  If there are any problems with the astrometric
calibration, centroiding, or differences between forced and unforced
PSF fitting, systematic differences would occur. 
In Appendix~\ref{sec:centroiderrordmcorr}, we
quantify the photometric bias introduced by centroiding uncertainties,
and calculate the expectation value of this bias for each SN depending
on its centroid uncertainty (see Figure~\ref{fig:dmcorrcentroid}).
Since the uncertainty in the centroid is larger at higher
redshift, there is a redshift dependence of this bias of \about 2~mmag
from low to high redshift.  We correct the light curves for this
photometric bias and estimate that the residual photometric bias is
less than 1~mmag (see Table~\ref{tab:systematics_cal}).
Another advantage of forced photometry is that it does not suffer
from a photometric Malmquist bias at low SNR.  This effect can be seen
in Figure~\ref{fig:comparison_forced_regdiff}.  At ${\rm SNR} = 10$
and 3, the regular photometry is on average brighter by 0.5\% and
2.5\%, respectively.

There is additional uncertainty since we perform forced photometry on
the SNe, but ``regular'' photometry (i.e. free x,y positions) on our
field stars from which we determine the zeropoint. We have chosen
reference stars for our photometric calibration that have a
SNR$\ge$20. For such stars, the difference between
regular and forced photometry is smaller than 2~mmag (see
Figure~\ref{fig:comparison_forced_regdiff}). This has been added to
our systematic error table (see Table~\ref{tab:systematics_cal}).  A
technique to mitigate this effect is to do forced photometry for both
the field stars and SNe \citep[e.g.,][]{Astier06}.

\subsubsection{Empirical Adjustment of Uncertainties}
\label{sec:d2s}

In our pipeline, the variance terms are propagated through all of the
image processing. However, this underestimates the true uncertainties
in the measurements since the resampling process as well as
kernel-matching one image to the other introduces covariance between
the pixels.  In order to empirically determine by how much the
uncertainties are underestimated, we measure the flux $f_{r}$ and its
uncertainty $\sigma_{r}$ at random positions in a given difference
image in exactly the same way as we measure the SN flux. We calculate
the weighted mean of the flux measurements $\bar{f_{r}}$, excluding
outliers by a 3-sigma cut. This $\bar{f_{r}}$ is an empirical estimate
of the systematic bias in the photometry at any position for that particular
difference image, and we correct all photometry of that image with
$\bar{f_{r}}$. This value is typically on the order of $\sim$3 ADU.
We then estimate the factor $s_r$ by which the
uncertainties are underestimated using the reduced chi-square
$s_r=\sqrt{\chi_r^2}$, as described in more detail in
Appendix~\ref{sec:empadjust}.  Figure~\ref{fig:diffimstats} shows the
histograms of $s_{r}$ for all difference images and filters used for
the SN light curves.  
\begin{figure}[h!]
\includegraphics[width=250pt]{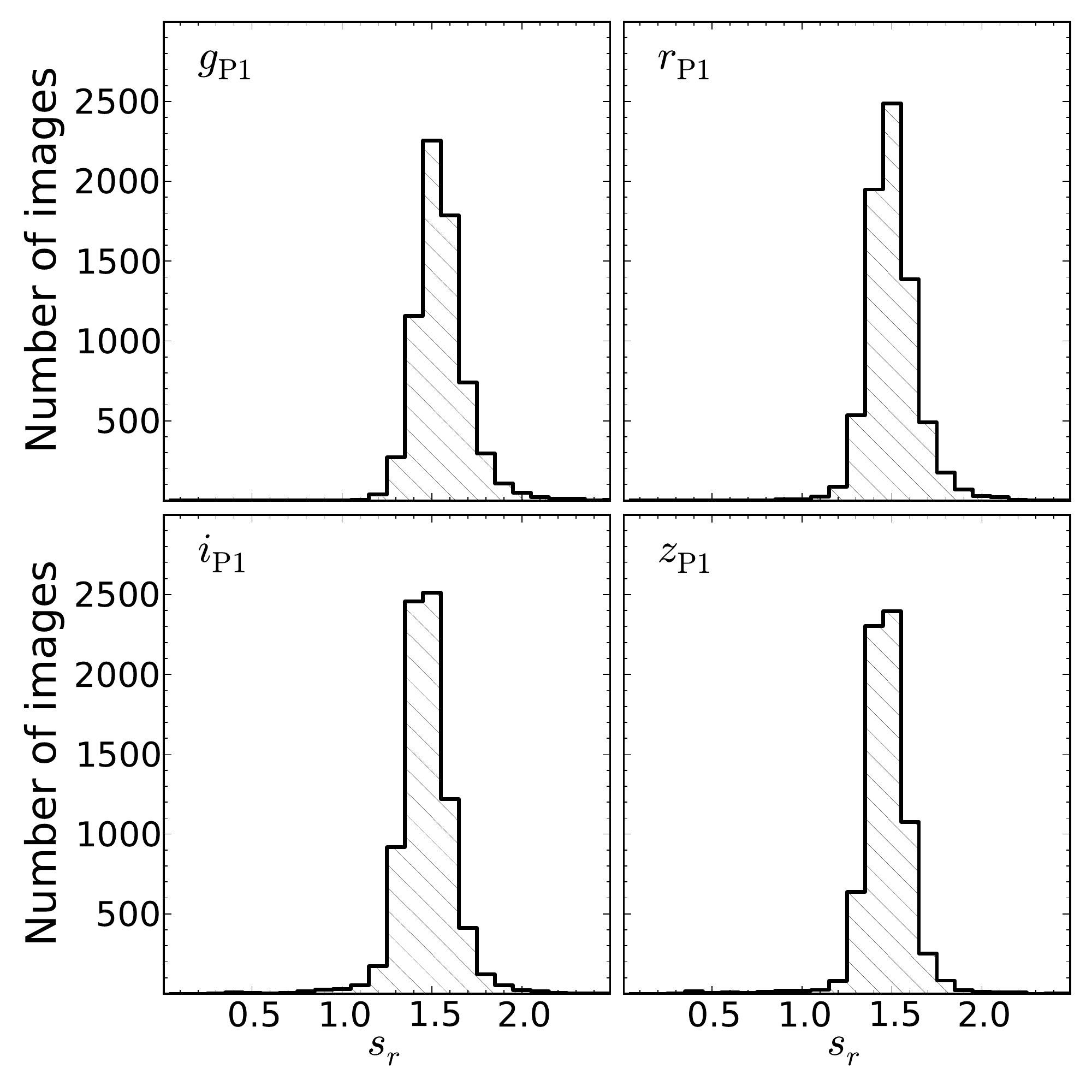}% 
\caption[]{
Histogram of the empirical multiplicative factor $s_{r}$ which is used to correct
the flux uncertainties of a given difference image.
\label{fig:diffimstats}}
\end{figure}
The histograms for \griz\ are nearly identical
with a peak at $s\sim1.5$, with the bluer filters having slightly
higher values since on average their larger PSF leads to larger
convolution kernels and thus to more covariance. We empirically
correct the difference image uncertainties by multiplying them with
$s_{r}$. Note that this is a multiplicative correction since the ratio
of covariance to variance is independent of the flux level in a given
pixel.  In \S\ref{sec:fluxoffset} we show that the baseline
(pre-SN) flux uncertainties are in general very good representations
of the true uncertainties, since their reduced chi-square distribution
has a peak at 1.0.
A similar approach of empirically adjusting the uncertainties has been
applied to the SNLS data \citep{Astier06}. An alternative approach is
to apply ``scene modeling'' \citep{Holtzman08}, which fits a scene
that includes the SN and a model of its environment to the data in the
original, unwarped image, thus avoiding the complication of correlated
pixels.

\subsubsection{Baseline Flux Adjustment}
\label{sec:fluxoffset}

We construct the templates on the assumption that there is no SN flux
present. However, stacking images is not a perfect process and there
may be some small biases in the template construction (e.g., SN flux
included in some images, or artifacts from different PSF sizes).
Since we always subtract the same template from each image, this
single realization of noise creates the same systematic offset in
every difference image measurement of a given SN, and directly affects
the peak magnitude of the SN. We therefore correct all the light
curves by the average baseline flux calculated from the fluxes where
there is known to be no SN flux, as described below.  This correction
is equivalent to creating a deep stack with all available images, but
in addition, it corrects for any imperfection in the template stack
creation and photometry.

We note that even after the baseline flux adjustement, there is still
some small Poisson uncertainty left since a finite number of
measurements were used. However, this uncertainty is very small
compared to the uncertainty in a single epoch, and it does not bias
the light curve fits in a significant way.

We calculate the weighted average of all forced photometry difference
image measurements at epochs that are $<-30$ days or $>200$ days from
maximum for each SN, and subtract this baseline flux from all SN light
curve measurements.  We find that the calculated offsets are on
average zero. However, the standard deviation of the distribution of
these offsets is 24~ADUs for a zeropoint of 30.0. The significance of
correcting with these offsets depends on the magnitude of the SNe. For
example, 24 ADUs correspond to 3 and 41~mmag to SNe of magnitude 19
and 23, respectively. Thus, this
correction is non-negligible for an individual SN, in particularly for
faint ones.  However, since the baseline flux correction distribution
is symmetric and centered around zero, we find that any systematic
bias possibly introduced to the peak magnitude can only be very small,
and we assign an upper limit of 1~mmag to it (see
Table~\ref{tab:systematics_cal}).
\begin{figure}[h!]
\includegraphics[width=250pt]{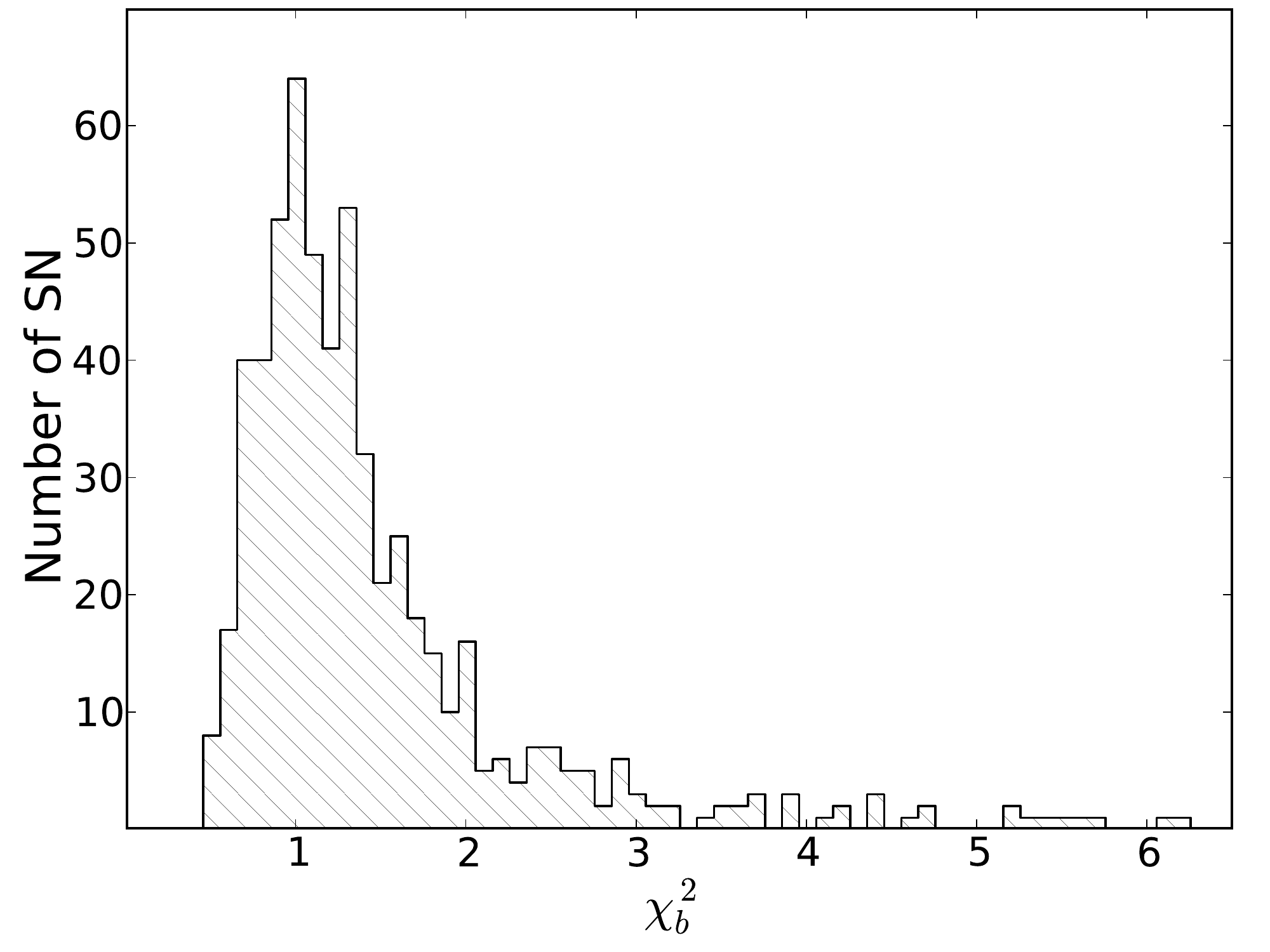}% 
\caption[]{Histogram of the baseline flux reduced chi-square
$\chi^2_{b}$. Note that the peak of the distribution is very close to
unity, indicating that the difference image uncertainties are a good
representation of the true errors. The tail toward large
$\chi^2_{\mathrm{norm}}$ is most likely due to difference image artifacts due
to bright host galaxies in close proximity to the SN.
\label{fig:fluxoffsetX2}}
\end{figure}

The normalized $\chi^{2}_{b}$ from the baseline flux calculation is
shown in Figure~\ref{fig:fluxoffsetX2}.  The peak of the distribution
is very close to one, indicating that our measured uncertainties are
generally good estimates of the true uncertainties.  There is a tail
toward large $\chi^{2}_{b}$, which we have determined is due to
difference image artifacts introduced by bright extended sources near
the position of the SN.  This effect is particularly strong at the
center of bright sources, and the outliers in the $\chi^{2}_{b}$ are
exclusively SNe~Ia close to the nucleus of a bright galaxy.

Since these artifacts have a net flux of zero, they do not introduce a
systematic bias into the photometry, but the large background
variation near these artifacts artificially increases $\chi^{2}_{b}$.
This effect reduces the quality of the light curve fits for some
supernovae, which do not pass our quality cuts for further analysis.
Future analyses that seek the largest useful sample should address
this problem.
\begin{deluxetable}{ccccc}[h!]
\tabletypesize{\scriptsize}
\tablecaption{Light curve of PS1-10hu
\label{tab:lcexample}}
\tablehead{
\colhead{Name} & \colhead{MJD} & \colhead{Filter} &  \colhead{Flux $f$} & \colhead{Magnitude $m$}
}
\startdata
 PS1-10hu & 55230.60229 &   g &           \phantom{1}\phantom{1}20.0\,$\pm$\,21.5 &        \nodata\\
 PS1-10hu & 55233.53140 &   g &           \phantom{1}\phantom{1}-5.8\,$\pm$\,24.8 &        \nodata\\
 PS1-10hu & 55236.59489 &   g &           \phantom{1}\phantom{1}-0.9\,$\pm$\,15.4 &        \nodata\\
 PS1-10hu & 55248.52526 &   g & \phantom{1}\phantom{1}-1.4\,$\pm$\,\phantom{1}7.8 &        \nodata\\
 PS1-10hu & 55272.56808 &   g &                               1550.3\,$\pm$\,25.6 & $19.52\pm0.02$\\
 PS1-10hu & 55275.51514 &   g &                               1387.7\,$\pm$\,34.2 & $19.64\pm0.02$\\
 PS1-10hu & 55326.41381 &   g &                     \phantom{1}119.6\,$\pm$\,20.2 & $22.31\pm0.17$\\
\hline
 PS1-10hu & 55230.61622 &   r & \phantom{1}\phantom{1}\phantom{1}7.2\,$\pm$\,20.6 &        \nodata\\
 PS1-10hu & 55233.54702 &   r &                     \phantom{1}-32.2\,$\pm$\,28.7 &        \nodata\\
 PS1-10hu & 55236.60945 &   r &           \phantom{1}\phantom{1}18.3\,$\pm$\,13.7 &        \nodata\\
 PS1-10hu & 55248.53908 &   r & \phantom{1}\phantom{1}-3.9\,$\pm$\,\phantom{1}8.4 &        \nodata\\
 PS1-10hu & 55251.52352 &   r &           \phantom{1}\phantom{1}34.9\,$\pm$\,32.9 &        \nodata\\
 PS1-10hu & 55266.55484 &   r &                               1487.1\,$\pm$\,36.2 & $19.57\pm0.02$\\
 PS1-10hu & 55272.58426 &   r &                               1652.1\,$\pm$\,25.6 & $19.45\pm0.02$\\
 PS1-10hu & 55275.52937 &   r &                               1614.2\,$\pm$\,98.5 & $19.48\pm0.07$\\
 PS1-10hu & 55305.35085 &   r &                     \phantom{1}471.3\,$\pm$\,20.1 & $20.82\pm0.04$\\
 PS1-10hu & 55326.42714 &   r &                     \phantom{1}221.3\,$\pm$\,31.0 & $21.64\pm0.14$\\
 PS1-10hu & 55332.32633 &   r &                     \phantom{1}202.3\,$\pm$\,10.8 & $21.73\pm0.05$\\
\hline
 PS1-10hu & 55231.58320 &   i & \phantom{1}\phantom{1}\phantom{1}9.8\,$\pm$\,13.7 &        \nodata\\
 PS1-10hu & 55234.55779 &   i &                     \phantom{1}-16.7\,$\pm$\,13.4 &        \nodata\\
 PS1-10hu & 55237.57200 &   i &                     \phantom{1}-15.3\,$\pm$\,12.1 &        \nodata\\
 PS1-10hu & 55240.53452 &   i &           \phantom{1}\phantom{1}20.3\,$\pm$\,11.1 &        \nodata\\
 PS1-10hu & 55243.53133 &   i &           \phantom{1}\phantom{1}-5.7\,$\pm$\,23.1 &        \nodata\\
 PS1-10hu & 55246.46399 &   i &           \phantom{1}\phantom{1}18.1\,$\pm$\,15.7 &        \nodata\\
 PS1-10hu & 55249.55923 &   i &           \phantom{1}\phantom{1}-9.7\,$\pm$\,11.0 &        \nodata\\
 PS1-10hu & 55252.55308 &   i &                     \phantom{1}178.6\,$\pm$\,32.6 & $21.87\pm0.18$\\
 PS1-10hu & 55288.44303 &   i &                     \phantom{1}771.3\,$\pm$\,25.1 & $20.28\pm0.03$\\
 PS1-10hu & 55297.34091 &   i &                     \phantom{1}709.2\,$\pm$\,20.4 & $20.37\pm0.03$\\
 PS1-10hu & 55303.48521 &   i &                     \phantom{1}585.3\,$\pm$\,17.4 & $20.58\pm0.03$\\
 PS1-10hu & 55327.42882 &   i &                     \phantom{1}185.2\,$\pm$\,28.8 & $21.83\pm0.16$\\
 PS1-10hu & 55330.30423 &   i &                     \phantom{1}185.8\,$\pm$\,12.9 & $21.83\pm0.07$\\
 PS1-10hu & 55339.28203 &   i &                     \phantom{1}283.2\,$\pm$\,56.7 & $21.37\pm0.20$\\
 PS1-10hu & 55348.34014 &   i &                     \phantom{1}126.0\,$\pm$\,19.1 & $22.25\pm0.15$\\
\hline
 PS1-10hu & 55235.58945 &   z &           \phantom{1}\phantom{1}16.3\,$\pm$\,21.6 &        \nodata\\
 PS1-10hu & 55238.56725 &   z & \phantom{1}\phantom{1}\phantom{1}0.2\,$\pm$\,16.4 &        \nodata\\
 PS1-10hu & 55241.49175 &   z &           \phantom{1}\phantom{1}10.3\,$\pm$\,25.7 &        \nodata\\
 PS1-10hu & 55247.53664 &   z &                     \phantom{1}-24.9\,$\pm$\,16.5 &        \nodata\\
 PS1-10hu & 55250.57895 &   z & \phantom{1}\phantom{1}\phantom{1}0.7\,$\pm$\,16.0 &        \nodata\\
 PS1-10hu & 55268.58466 &   z &                     \phantom{1}865.0\,$\pm$\,33.8 & $20.16\pm0.04$\\
 PS1-10hu & 55280.35431 &   z &                     \phantom{1}548.8\,$\pm$\,14.8 & $20.65\pm0.03$\\
 PS1-10hu & 55298.39193 &   z &                     \phantom{1}572.0\,$\pm$\,23.1 & $20.61\pm0.04$\\
 PS1-10hu & 55304.31987 &   z &                     \phantom{1}548.5\,$\pm$\,18.1 & $20.65\pm0.03$\\
 PS1-10hu & 55322.38723 &   z &                     \phantom{1}254.3\,$\pm$\,18.1 & $21.49\pm0.07$\\
 PS1-10hu & 55325.36617 &   z &                     \phantom{1}171.4\,$\pm$\,16.9 & $21.92\pm0.10$\\
 PS1-10hu & 55334.33428 &   z &                     \phantom{1}186.9\,$\pm$\,11.0 & $21.82\pm0.06$\\
 PS1-10hu & 55340.28431 &   z &                     \phantom{1}203.3\,$\pm$\,36.0 & $21.73\pm0.18$\\
\enddata
\tablecomments{An abbreviated example PS1 light curve - PS1-10hu.  The
  light curves of all SNe in the cosmological sample are available in
  machine-readable format in the electronic edition.  Only photometry
  within 40 days before and 100 days after maximum at MJD=55270 is
  presented here.  All fluxes $f$ are with respect to a zeropoint of
  27.5, and the magnitudes are accordingly calculated as
  m=-2.5log10(f)+27.5.  We only show magnitudes $m$ for fluxes with at
  least $3\sigma$ significance.
}
\end{deluxetable}

\begin{figure}[h!]
\includegraphics[width=250pt]{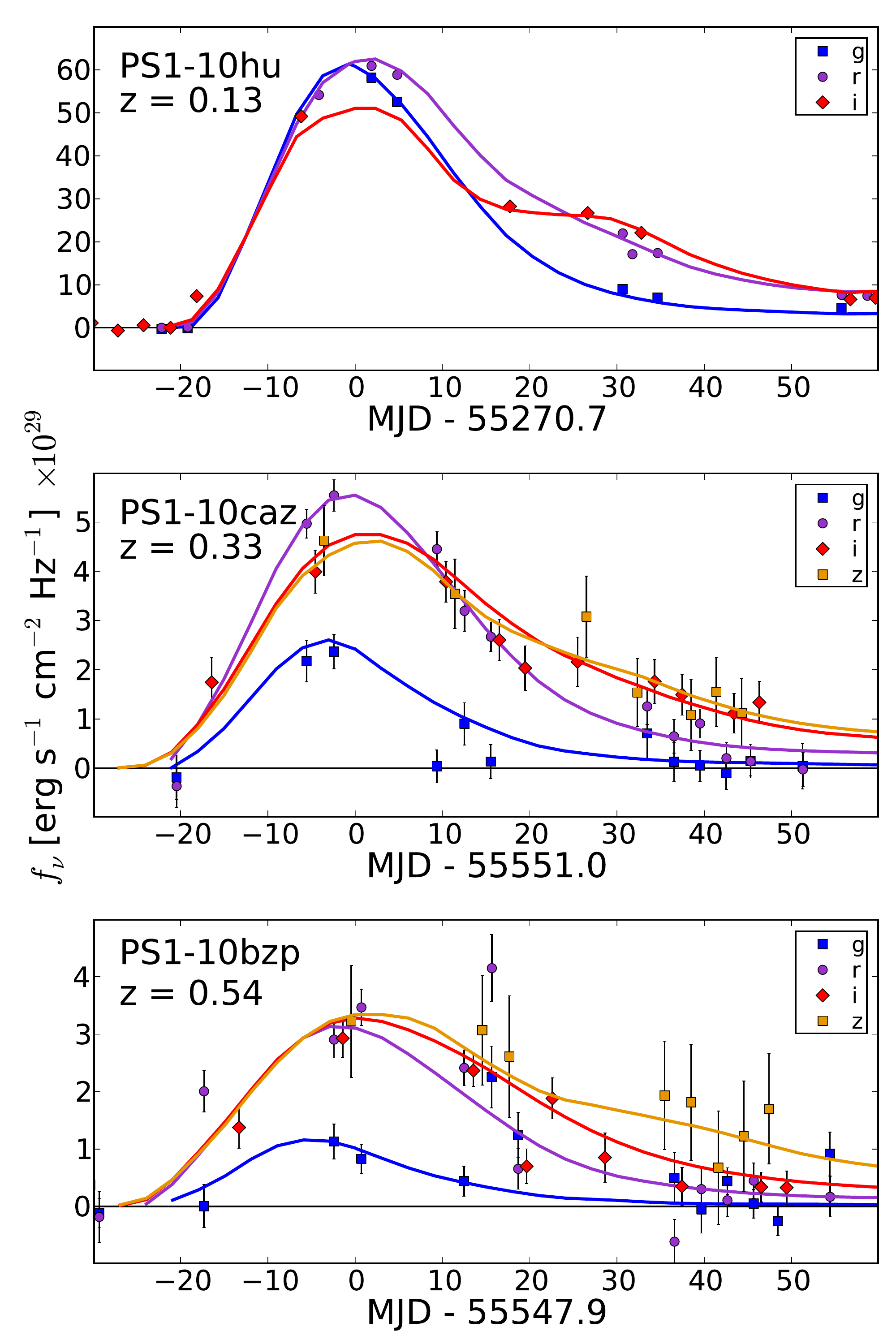}% 
\caption[]{Example light curves for three \PS\ SN~Ia at different redshifts.
MJD offset are applied so that zero is the time of maximum in rest-frame B.
The SALT2 fits are shown with the solid lines.
\label{fig:lcs}}
\end{figure}

In Figure~\ref{fig:lcs} we show representative \PS\ SNe~Ia light
curves: PS1-10hu, PS1-10caz, and PS1-10bzp with redshifts of 0.13,
0.33, and 0.64 from top to bottom, respectively. The solid lines are
their respective light curve fits as described in \S\ref{sec:lcfit}.
The light curves are available in machine-readable format in the 
electronic edition. As an abbreviated example, we show the light curve of SN~PS1-10hu in
Table~\ref{tab:lcexample}.

\section{Photometric Calibration}
\label{sec:photcal}

As \cite{Sullivan11} has pointed out, our ability to constrain the
properties of dark energy from SN~Ia observations is limited by
systematic errors from the photometric calibration. The unique area
and depth of our SN survey creates the opportunity to observe
supernovae over a wide redshift range with the same telescope and
filters.  This improves the potential for minimizing calibration
errors.

\cite{Tonry_photcal} measures the
system passbands, including both the instrumental sensitivity and
atmospheric transmission functions.  The PS1 photometric system is
fundamentally based on HST Calspec spectrophotometric observations.
For 7 Calspec standards, \cite{Tonry_photcal} determines slight
adjustments to the bandpasses so that the photometry of the standards
best agrees with the spectrophotometry of the standards with HST
spectra.  The fields with Calspec standards are passed to
\cite{Schlafly_uebercal}, which performs the ubercal calibration that
ties the observations of stars in these fields with stars across the
entire 3pi survey.  The zeropoints set by \cite{Tonry_photcal} are
mostly preserved in this stage, though may be slightly adjusted so
that there is optimal relative calibration across the sky.  The Medium
Deep fields are calibrated in this process.  \Scolnicsys\ then checks
this process by analyzing the photometry of all the Calspec
standards observed in the full 3pi survey and comparing the photometry
to the synthetic photometry of the positions.  \Scolnicsys\ finds the
AB offsets so that the photometry and spectrophotometry agree.  The
offsets found here are applied to all of the Medium Deep field
photometry and SN light curves.

The total systematic uncertainty in the PS1 calibration may be broken
down into four parts: the instrumental response function, the
propagation of zeropoints across the medium deep fields, the adherence
of the PS1 zeropoints to the AB system, and spatial/temporal variation
in the photometric calibration.  The uncertainty of the instrumental
response function is given in \cite{Tonry_photcal} (\Tonryphot\
hereafter).  This analysis also details the full systematic
uncertainties in the \PS\ implementation of the AB system, though we
have redone a significant amount of the error accounting.  The
uncertainty in the propagation of zeropoints across the sky must be
included as there are 10 Medium Deep Fields with large separations in
distance. We use the results of \cite{Schlafly_uebercal} to correct
for this.  The adherence of the PS1 calibration to a true AB system is
analyzed in detail by the companion paper \Scolnicsys.  Finally, the spatial
and temporal variation of the instrumental response function is
analyzed in this paper.

Our entire systematic error budget for calibration is summarized in
Table~\ref{tab:systematics_cal}.  We find that after accounting for
the full systematic error budget, the overall systematic uncertainty
for each filter is $\sim0.012$ mag without including the uncertainty
in the {\it HST} Calspec definition of the AB system.  We briefly review the
uncertainties that are discussed in external work, and analyze the
remaining uncertainties below.
\begin{deluxetable}{lll}
\tablecaption{Photometric Calibration Error Budget
\label{tab:systematics_cal}}
\tablehead{
\colhead{Source} &
\colhead{Uncertainty} & \colhead{Section} \\
~ & [Millimag]
}
\startdata
Photometric Non-linearity      & 1        & \ref{sec:linearity}   \\
Centroid Accuracy              & $\la$1   & App. \ref{sec:centroiderrordmcorr} \\
Forced photometry              & 2        & \ref{sec:forcedphot}   \\
Baseline Flux Correction       & $\la$1   & \ref{sec:fluxoffset}   \\
Instr. Response Function       & 3        & \ref{sec:abscalallsky}   \\
Flux determination             & 5        & \ref{sec:abscalallsky}   \\
Net offset wrt Calspec         & 8,5,6,3  & 3.2 in \Scolnicsys   \\
Zeropoint propagation          & 3        & \ref{sec:abscalallsky}   \\
Spatial variation              & 5,6,4,6  & \ref{sec:instrresponse}   \\
Temporal variation             & 3        & \ref{sec:instrresponse}   \\
\hline
Total Internal PS1                & $\sim$12&   \\
\hline
SED conformity to AB           & 5-15 & 3.2 in \Scolnicsys
\enddata
\tablecomments{The dominant systematic uncertainties in defining the
\PS\ photometric system. If four numbers are given, they refer to
\griz, respectively. We assign a range of uncertainty to the SED
conformity to the AB system, since the color dependence is not easily
enumerated and is discussed in more detail in \Scolnicsys. The
bandpass uncertainties is included into the instrumental response
function and also discussed in \Scolnicsys.}
\end{deluxetable}

\subsection{Absolute Calibration and All-Sky Photometry}
\label{sec:abscalallsky}

The system's response function naturally divides into the instrumental
response function (mirrors, optics, filters, and detectors) and the
atmospheric extinction. In principle it is possible to determine these
two components independently without the reliance on standard stars:
The atmospheric extinction can be determined with measurements and
modeling, while the instrumental response function can be traced with
a high-accuracy NIST-calibrated photodiode in combination with a
tunable laser system \citep{Stubbs06,Stubbs10}.  In practice, however,
the calibration accuracy is improved by including observations of
spectrophotometric standards, which provide the overall normalization
of the photometric system and can verify the derived system and
atmospheric response functions. \Tonryphot\ describes in detail how
the \PS\ photometric system has been determined with this recipe.

The Calspec standards, and thus also the \PS\ photometry, are
fundamentally based on models of hydrogen white dwarf atmospheres and
the absolute flux for Vega in the {\it V} band \citep{Bohlin07}.
Substantial work has been invested in improving the {\it HST}
spectrophotometric standards
\citep[e.g.][]{Colina94,Bohlin96,Bohlin01}, and they have the distinct
advantage that the Calspec standard star observations are not marred
by the atmosphere.  However, even in space, complications like charge
transfer efficiencies, non-linearity of detectors, and secular
degradation of the optics and detectors contribute to systematic
measurement biases \citep{Bohlin07}.  Combining all this, we adopt a
systematic error of 5-15~mmag from non-conformity of the {\it HST}
Calspec standards to the AB system, depending on wavelength
\citep{Bohlin04}. This wavelength dependence is further discussed in
\Scolnicsys.

\begin{figure}[t]
\includegraphics[width=250pt]{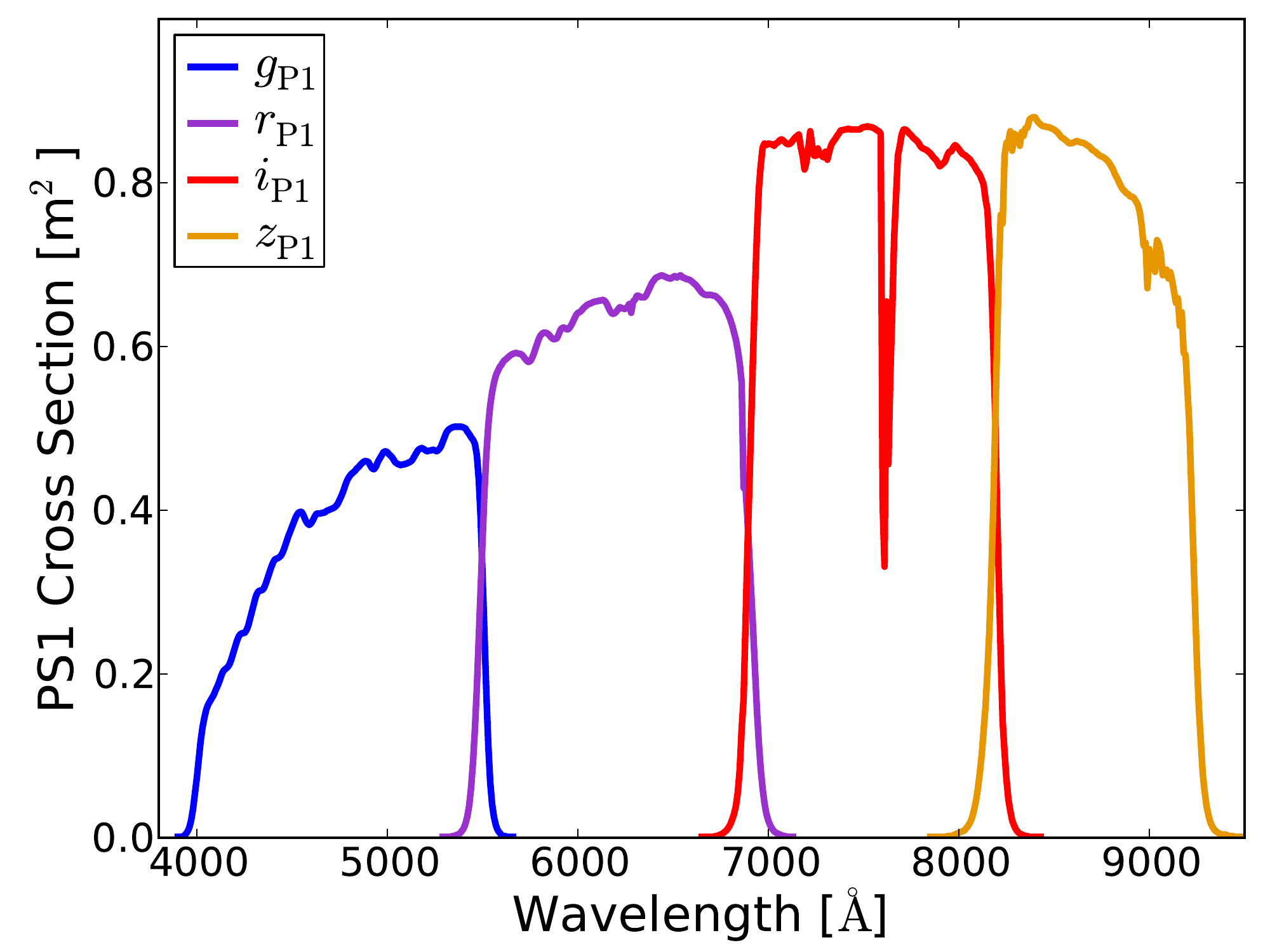} 
\caption[]{System response function for the four \PS\ filters \griz
at an airmass of 1.2, assuming representative precipitable water
vapour and aerosol exponent parameters of 0.65~cm and 0.7,
respectively (\Tonryphot).
\label{fig:ps1trans}}
\end{figure}

The \PS\ survey observed 7 standard stars (1740346, KF01T5, KF06T2,
KF08T3, LDS749B, P177D, and WD1657-343) on MJD 55744 (UT 02 July 2011)
in photometric conditions at a wide range of airmasses in all filters.
The stars were placed on the same CCD detector/amplifier, which we chose
to be away from the center of the field, where spatial variation in the
PSF adds additional uncertainty to photometric measurements as shown
in \S\ref{sec:instrresponse}.  In
theory, a common normalization to all filters should be sufficient to
match the observations to the synthetic magnitudes. However, the
synthetic colors of the standard stars deviate from the observed
colors.  \Tonryphot\ attempts to correct for these differences by
correcting the instrument response function.  For the filters \griz,
they determine the corrections to be 0.012, 0.019, 0.009, and
-0.009~mag, respectively.  The correction is small because the
agreement between the in-situ NIST-based response function and the
manufacturer's benchmark measurements of the filters and the CCD is
very good.  The system response function (instrument response function
and atmosphere) of the \PS\ photometric system for \griz\ is shown in
Figure~\ref{fig:ps1trans} at an airmass of 1.2.

The photometric normalization in the limited set of fields with the
Calspec standards is then propagated across the sky, encompassing all
of the Medium Deep survey fields, using the same \"{u}bercalibration
method \citep{Padmanabhan08} that has been successfully used for the
SDSS all-sky survey.  The details of this \"{u}bercalibration are described in
\cite{Schlafly_uebercal}. Comparing the \PS\ and SDSS photometry,
they find that both surveys show similar photometric calibration
errors.  They estimate that the relative precision in \gps, \rps, and
\ips\ is $<10$~mmag, and \about10~mmag in \zps.  Since we have 10 MDFs
distributed over the full sky, we estimate that the uncertainty
introduced by zeropoint variation is $10/\sqrt{10}$~mmag.
\Scolnicsys\ repeat the process outlined in \Tonryphot\ of adjusting
the filter transmissions based on agreement between observed and
synthetic photometry.  They increase the number of Calspec standards
used to 10.  \Scolnicsys\ determines that systematic uncertainties in
the photometry of the standard stars are no larger than 5~mmag.  They
also find that small adjustements $\Delta
\griz$=$[-0.008,-0.0095,-0.004,-0.007]$ should be added to the
zeropoints defined by \Tonryphot\ and \cite{Schlafly_uebercal}.  This
improves the measurement of the offset for each filter, and reduces
the systematic uncertainty due to the adjustments from 10~mmag to
those given in Table~\ref{tab:systematics_cal} (``Net offset wrt Calspec'').

To test the calibration, \Tonryphot\ finds an excellent
agreement between the stellar locus constructed from observed \PS\
photometry from a field with very low Galactic extinctions with its 
synthetic colors, with systematic differences of \about 1\%. They also
transform SDSS stripe 82 magnitudes into the \PS\ photometric system,
compare them to the observed \PS\ magnitudes, and find differences at
the 1\%-2\% level. \Tonryphot\ also notes that there are
differences at the same level between the different SDSS data
releases. A more detailed exploration of the source of these
differences is given in \Scolnicsys.

\subsection{Spatial and Time Variation of the \PS\ Instrumental Response Function}
\label{sec:instrresponse}

The system response function can vary with wavelength, position on the
focal plane, and time.  This section quantifies the systematic biases
introduced by errors in the determination of the instrument throughput.

\begin{figure}[t]
\includegraphics[width=250pt]{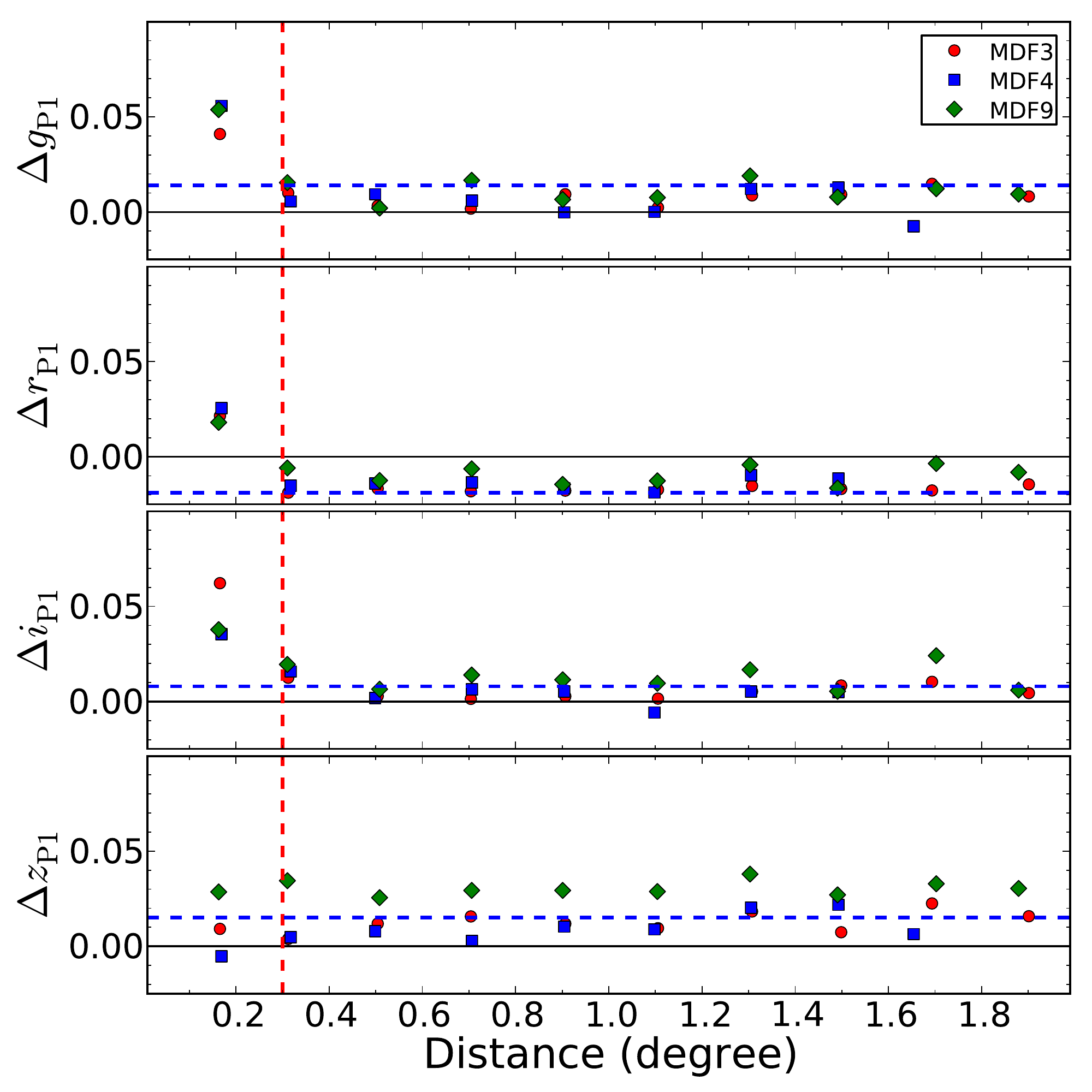} 
\caption[]{The median differences between the PS1 photometry and the
synthetic PS1 photometry derived from SDSS using Equation~6 and
Table~6 from \Tonryphot\ using stars in the color range
$0.5<(g-r)_{SDSS}<0.8$ for $gri_{\mathrm{P1}}$ and
$0.8<(r-i)_{SDSS}<1.1$ for $z_{\mathrm{P1}}$. The median is calculated
in bins of distance between the stars and the center of the MDFs.  We
exclude all SNe that are within 0.3~deg of the field center (red
dashed lines), since in the photometric calibration in the field
center is unreliable due to strong spatial variation in the PSF. The
blue lines indicate the known offsets between SDSS DR8 and the \PS\
calibration by \Tonryphot, as shown in Table~1 of
\Scolnicsys.
\label{fig:dmspatial}}
\end{figure}

For \PS, the spatial variation of the instrument response function is
removed by flat-fielding and the on-sky illumination correction, which
is constructed using dithered images of a dense stellar field
\citep{PS1_IPP}. However, \cite{Stubbs10} have suggested that after
these corrections there may still be a residual spatial variation in
the photometry that is band-dependent.  To test this, we compared PS1
photometry of stars in 3 MDFs against SDSS photometry of the same
stars - converted to the PS1 system of \Tonryphot\ (see \Scolnicsys\
for further details).  Figure~\ref{fig:dmspatial} shows the results of
this test, plotting the median difference between the \PS\ and SDSS
photometry.  The most striking disagreement is in the first bin, which
lies at the center of the focal plane where the optics of the
telescope generate strong spatial variation of the PSF.  Because the
central region is so difficult to handle, we discard from our
cosmological analysis the 3 SN Ia that fall in the central region
(PS1-11ams, PS1-10f, and PS1-11yu) because we cannot be certain they
meet our threshold for photometric fidelity.

Across the rest of the focal plane, the PS1 photometry in each filter
exhibits a steady constant offset relative to SDSS, ranging from 0.5\%
in \rps\ to 3\% in \zps.  These overall differences between the
synthetic and observed PS1 magnitudes are known and described in more
detail in \Tonryphot\ and \Scolnicsys.  They can be taken as
conservative upper limits on the systematic biases in the PS1
photometry.  After removing this offset, the remaining difference
between SDSS and \PS\ photometry is less than 1\% in each filter - and
that is further reduced to $\sim$5~mmag for all filters when
systematic shifts between the MDFs are accounted for\footnote{We
attribute the MDF-to-MDF variation to large scale spatial variations
in the SDSS photometric system
\citep{Schlafly11,Schlafly_uebercal}}. 
We therefore use 5~mmag as the
limit for spatial consistency in our error budget
(Table~\ref{tab:systematics_cal}).
The differences between PS1 and SDSS shown here are most likely due to
flat-fielding issues, as the expected differences due to variations in
the filter passbands across the focal plane are $<3$~mmag for stars
with colors around $0.4<g-i<1.4$.  The photometry of SNe has a
stronger dependence on these spatial variations in the filter
passbands due to the broad absorption features in their spectra. As
described in \Scolnicsys, we fit each SN light curve using the
expected filter function at the position of the SN on the focal plane.
Accounting for the variation of the filter functions may change the
distances from their nominal values by \about 3~mmag.  These
differences are shown in \Scolnicsys\ (Fig. 2) and are included in the
distances presented in our Table~\ref{tab:lcfitparams}.

\begin{figure}[t]
\includegraphics[width=250pt]{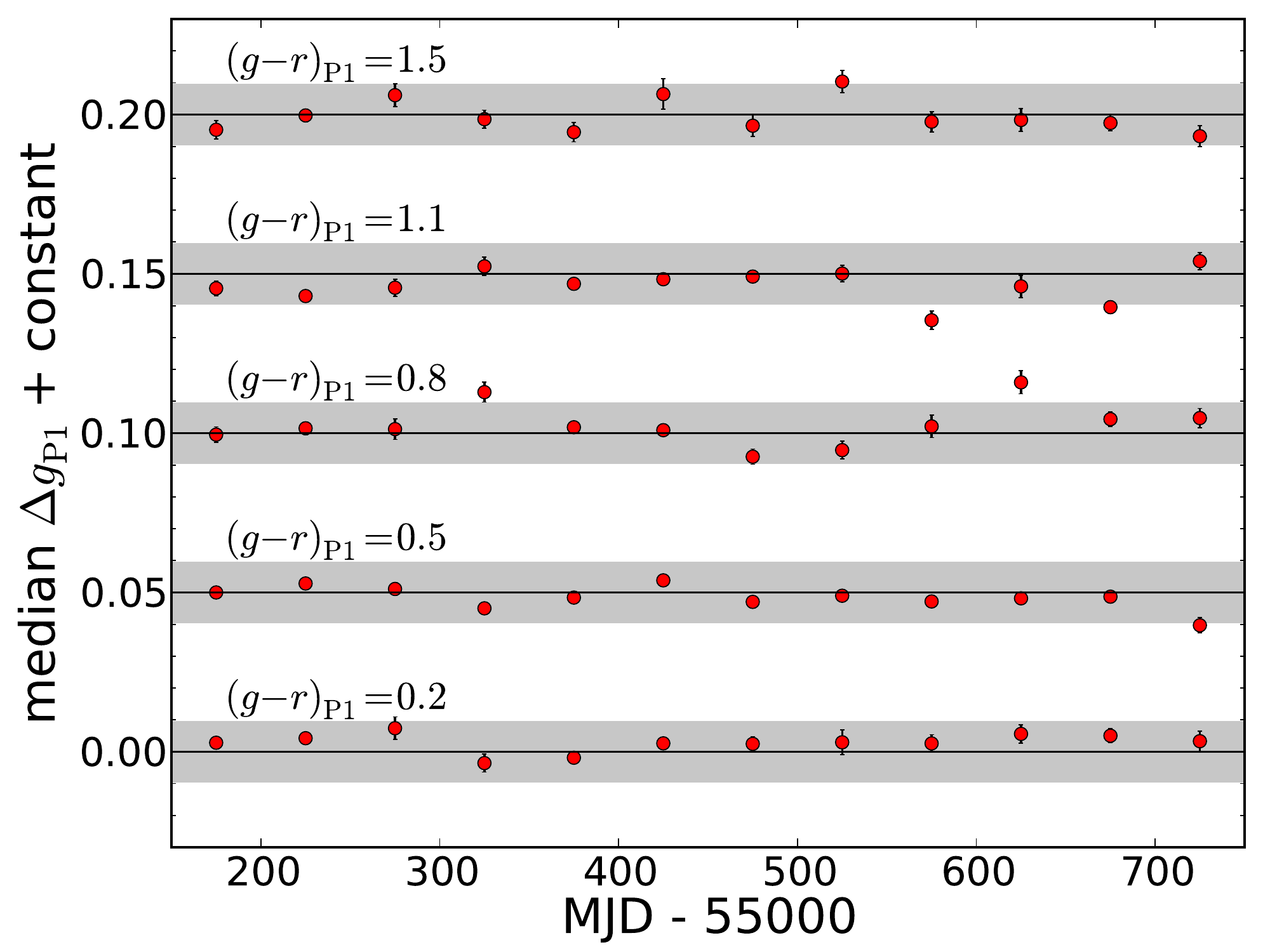} 
\caption[]{Median difference between the average magnitude of stars
and their associated detections at different epochs for color bins of
$(g-r)_{P1}$ in [0.2,0.5,0.8,1.1,1.5]. The size of the color bins is
0.1 magnitude, and only stars with $\rps<19$ were used.  For clarity,
offsets in steps of 0.05 are added to $\Delta\gps$, and the gray
shaded bars indicate a plus-or-minus 1\% level deviation.  No long term variation is
apparent above $\pm$3~mmag, which we assign as the systematic
uncertainty.
\label{fig:phot_time}}
\end{figure}

Over time, the optical components may degrade or
change. Color-dependent changes could introduce a systematic bias into
the photometry.  In Figure~\ref{fig:phot_time}, we
show the median variation $\Delta\gps$ in the average magnitude of
stars at various epochs for different colors $(\gps-\rps)$. Similar
results are seen in each band. Typically, the median is not different
than zero to within 3~mmag, and the standard deviation in a given bin
is on the order of 5~mmag. No long-term trends can be discerned, and
we set the upper limit on any long-term changes of the optical system
to 3~mmag.

\section{Light Curve Fits and Sample Cuts}
\label{sec:lcfit}

\subsection{SALT2 Light Curve Fits}

Many light-curve fitters have been developed over the last decade
(e.g., MLCS2K2, \citealt{Jha07}; SALT, \citealt{Guy07}; SiFTO,
\cite{Conley08_SIFTO}; and BayeSN, \citealt{Mandel09, Mandel11}).
Each method makes corrections for the light-curve shape and observed
color of the SN.  However, there are two fundamentally different ways
to apply the color correction. 
The first empirically determines the correlation between observed
color and uncorrected distance residuals.  The second uses the
assumption that the observed color is the combination of intrinsic SN
color, photometric errors, and reddening due to dust. The former is
less model dependent, while the latter is more physically motivated.

Ultimately, these choices have small, but noticeable consequences for
cosmological inferences.  \citet{Kessler09_SDSS} made an in-depth
analysis of the different light curve fitters, and found that in
general they agree reasonably well given the same assumptions.
Currently the most widely used light curve fitter is SALT2
\citep{Guy07}, and we use this fitter in our analysis.  SALT2
explicitly matches the observations in the filters of any given survey
to integrals of the warped model spectra for those passbands.  It
treats the color of SNe~Ia entirely empirically, and is used to find
an overall relation between luminosity and color. \Scolnicsys\
explores whether the linear models used by SALT2 to describe the color
and stretch-luminosity relation are adequate to fit the data.

We use the most up-to-date published version of SALT2 \cite{Guy10_lc}
implemented in SNANA\footnote{SNANA\_v10\_23}
\citep{Kessler09_SNANA}. We transform our lightcurve
fit parameter into distances using the Tripp formula \citep{Tripp98}
\begin{equation}
  \mu_{B} = m_{B} - M + \alpha x_{1} - \beta c, \label{eqn:salt2}
\end{equation}
where $\mu_{B}$ is the distance modulus, $m_B$ is the peak $B$-band
brightness, $x_1$ is a light-curve shape parameter, and $c$ is a color
parameter.  The parameters $\alpha$, $\beta$, and $M$ are nuisance
parameters.  $\alpha$ is determined by the relation between luminosity
and stretch while $\beta$ is determined by the relation between
luminosity and color.  $M$ is the absolute $B$-band magnitude of a
fiducial SN~Ia with $x_1 = 0$ and $c = 0$.  Motivated by
\cite{Schlafly11}, we make one modification to SALT2 by replacing the
``CCM'' \citep{CCM89} Milky Way Galaxy (MWG) reddening law with that
from \cite{Fitzpatrick99}. \cite{Schlafly11} also finds that the MWG
extinction values from \cite{SFD98} are overestimated by 6-14\%. We
therefore correct our extinction values accordingly. These changes are
explained further in Section 7 in \Scolnicsys.

\begin{figure}
\centering
\epsscale{1.15} 
\plotone{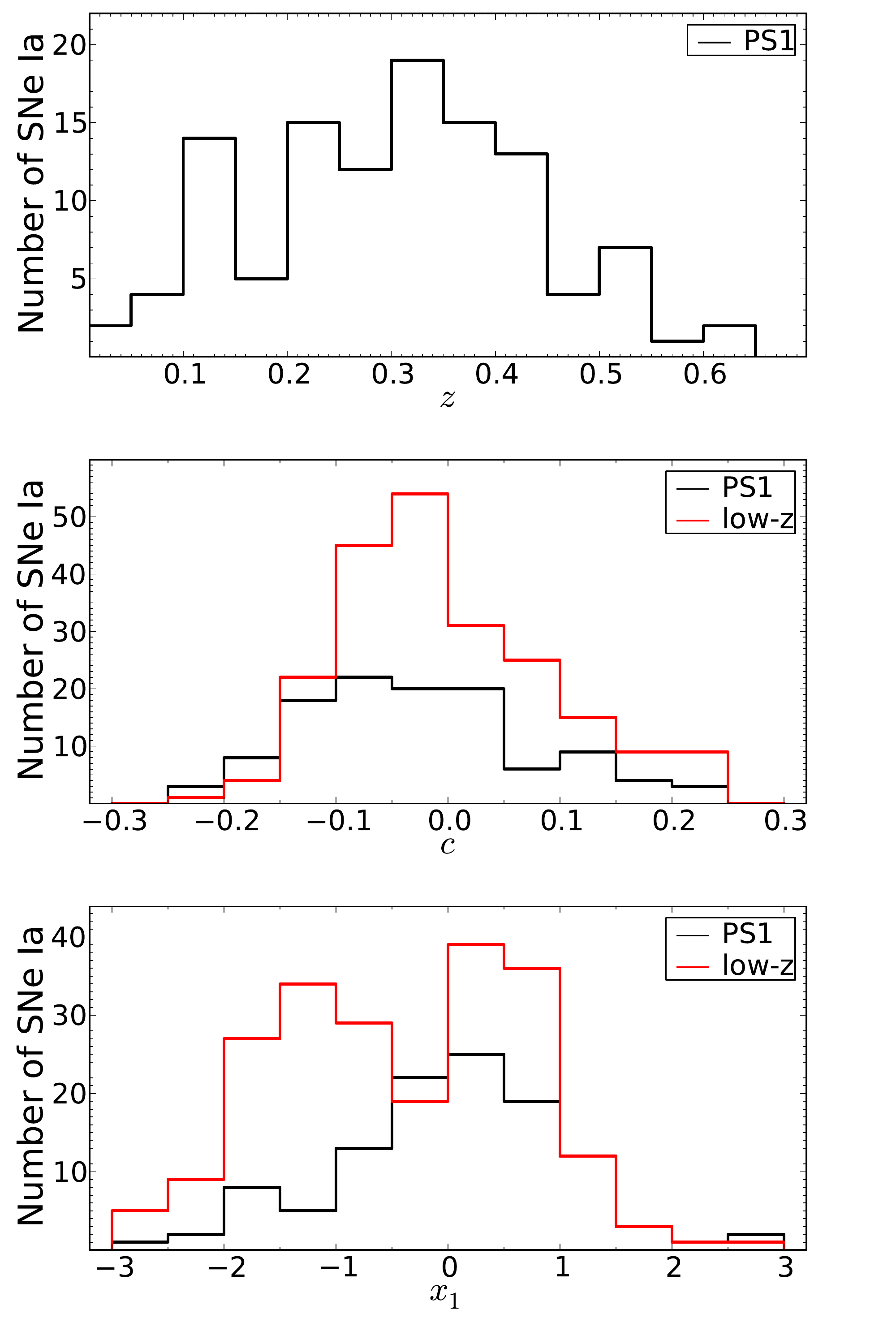}
\caption{Histograms of redshift, color, and stretch for the \PS\
(black solid) and low-redshift (red dashed) samples.  The first bin of
the low-redshift sample has \SNIAlowztot\ light
curves from \SNIAlowzused\ SNe~Ia.  For the samples presented both
color and stretch cuts were applied (see \S\ref{sec:cuts}), but not
the minimum redshift cut.}
\label{fig:ps1char}
\end{figure}

We present in Table~\ref{tab:lowz}~and~\ref{tab:lcfitparams} the SALT2
parameters for the entire set of cosmologically useful SNe~Ia from the
PS1 plus low-z sample (PS1+lz). The cuts that are used to remove
SNe~Ia from the cosmological sample are described in the second part
of this section.  Distributions of the SALT2 fit parameter are shown
in Figure~\ref{fig:ps1char}.

Discrepancies in the $x_{1}$ and $c$ distributions between the \PS\
and low-z samples are likely due to selection effects.  Most
low-redshift SNe in our sample were discovered in surveys that target
specific nearby galaxies \citep{Li11}.  These surveys are capable of detecting
SNe~Ia with more extinction than the
untargeted sample from \PS.  We use stringent cuts on the light curve
properties, as described below, that pass only about 1/2 of the low-z
sample but 4/5 of the PS1 sample.

\subsection{Sample Cuts}
\label{sec:cuts}

The \PS\ survey spectroscopically confirmed \SNIAPSall\ SNe~Ia during
the first 1.5 years in the MDFs.  This was only a very small fraction
of the \about 1700 transients with SN~Ia-like light curves.  While it
would be ideal to use all spectroscopically confirmed \PS\ SNe~Ia to
constrain cosmological parameters, we use sharp quality cuts to improve the
analysis.  We require that every SNe~Ia has adequate light-curve
coverage to properly measure a distance and that it has a light-curve
fit, redshift, and Milky Way extinction that limits systematic bias
in the distance.  Applying these cuts, the cosmological analysis of
\S\ref{sec:results} employs \SNIAPSused\ SN Ia from the initial
sample of \SNIAPSall\ spectroscopically-confirmed \PS\ objects.

For our low-z SN~Ia, we use the following SNe~Ia samples: we denote
JRK07 as the compilation of SNe~Ia collected by Cal\'an/Tololo
\citep[][29~SNe]{Hamuy96:lc}, CfA1 \citep[][22~SNe]{Riess99}, CfA2
\citep[][44~SNe]{Jha06}, and other sources \citep[][28~SNe]{Jha07}.
We also use the SN~Ia compiled more recently by CfA3
\citep[][185~SNe]{Hicken09}, CSP \citep[][85~SNe]{Contreras10}, and
CfA4 \citep[][94~SNe]{Hicken12}.  After applying our cuts, the low-z
sample is trimmed from \SNIAlowzall\ to \SNIAlowztot\
light curves from \SNIAlowzused\ SNe~Ia.  Table~\ref{tab:fitoverview}
shows the effect of each cut on the different samples.

Below, we detail the cuts.  We apply these criteria in three steps.
Initial criteria do not require any light-curve fitting. Light curves
that do not meet these standards are eliminated. First-pass criteria
are determined from polynomial fits to the light curves.  Light curves
that pass these tests are subjected to our final criteria, which use
the output of the full SALT2 light curve fit.
\\\\
\indent{\bf Initial:}
\begin{enumerate}
\item Unambiguous spectroscopic classification as a SN~Ia.
\item Not a Iax- or 91bg-like SN~Ia.
\item Outside the central 0.3~deg of the MDF (\PS\ only).
\item $z > 0.01$.
\item Galactic reddening along the line of sight of $E\left(B -
    V\right)_{\rm MWG} < 0.5$~mag. 
\item Measurements in 2 or more filters with ${\rm SNR} \geq 5$.
\end{enumerate}

{\bf First-pass:}
\begin{enumerate}
  \setcounter{enumi}{6}
  \item At least 1 measurement with $-10 < t < +5$~days.
  \item At least 1 measurement with $+5 < t < +20$~days.
  \item 5 or more measurements with $-10 < t < +35$~days.  
  \item 2 or more filters with a measurement between $-8 < t < +10$~days.
\end{enumerate}

{\bf Final:}
\begin{enumerate}
  \setcounter{enumi}{10}
\item Light curve fit converges.
\item $\Pfit > 0.001$, where $\Pfit$ is the SALT2 light-curve fit
  probability based on the $\chi^{2}$ per degree of freedom.
\item $-0.25 < c < 0.25$ or $-3.0 < x_{1} < 3.0$. 
\item Outlier rejection (Chauvenet's criterion from \cite{Taylor97}, 4$\sigma$)
\end{enumerate}

\subsubsection{Initial Cuts}

We require spectroscopic confirmation of all SN~Ia used in our
cosmological analysis.  We follow a method similar to that presented
by \citet{Foley09:year4} to determine the classification of each
potential SN~Ia.  Briefly, we use {\it SNID} \citep{Blondin07} to
match a SN spectrum with a library of high-SNR spectra.  {\it SNID}
provides a quantitative assessment that a particular SN is of a given
class at a given redshift.  Some redshifts come from host galaxy
emission lines.  Each supernova is ultimately classified by
spectroscopists: co-authors RC and RJF.  Supernovae with ambiguous
classifications are eliminated. Although the classification is
subjective, almost every SN in the \PS\ cosmology sample has a
spectrum of high quality and it is unlikely that any non-SN Ia have
leaked into this analysis.  Similarly, there should be no catastrophic
failures in the determination of the redshifts.

We specifically exclude SNe of the ``Iax'' subclass
\citep{Foley13a:Iax} from all samples.  Although this mostly affects
the low-redshift samples, SN~2009ku, a SN~Iax discovered by \PS\
\citep{Narayan11}, is removed from our final sample.  We also exclude
PS1-11yj, a SN~1991bg-like object at $z = 0.107$.  We decided not to
include this peculiar SN because it has not yet been shown whether
this subclass of SN~Ia can be well represented by the training sample
and has been excluded in some past surveys
\citep[e.g.,][]{Hicken09b}. The 3 SNe~Ia (PS1-10f, PS1-11yu,
PS1-11ams) that are within 0.3~deg of the field center are excluded,
since the absolute photometric calibration is uncertain due to strong
PSF variation (see \S\ref{sec:instrresponse}).

We exclude all SNe with $z < 0.01$ to avoid objects affected by
departures from Hubble's Law due to bulk flows or a regional Hubble
bubble.  We exclude SN with a Milky Way reddening $E\left(B -
V\right)_{\mathrm{MWG}} > 0.5$~mag to avoid introducing any substantial error
due to the extinction correction.  These cuts affect only the
low-redshift sample, and \Scolnicsys\ provides a detailed explanation
of the effects of these cuts.

\subsubsection{First-pass cuts}

We require high-quality light curves which when fit result in accurate
distance measurements.  One requirement is that the SNR is
sufficiently large in a minimum number of observations.  A large
subset of requirements can be considered ``coverage'' cuts, where each
light curve must have observations in certain phase ranges with
certain filters.  Finally, we require that the resulting light-curve
fit parameters fall within ranges known to have low distance biases
from simulations.  We outline the cuts below.

For our sample, we require that there be measurements in at least two
bands with ${\rm SNR} > 5$. This requirement does not affect the
low-redshift or \PS\ spectroscopic sample; however, it does reflect
our detection limits, and is a necessary requirement for simulating
the survey.

The light-curve cuts applied to the PS1 sample are taken from
\citet{Guy10_lc}. Most requirements are expressed in terms of the
rest-frame phase $t = (t_{\rm obs} - t_{\rm max})/(1 + z)$.  While
\citet{Kessler09_SDSS} required at least one measurement with $t <
0$~days, \cite{Guy10_lc} found that a more flexible requirement of
needing one measurement in the range of $-8 < t < 5$~days provided a
similar constraint. Using simulations of the \PS\ and low-z samples
from \Scolnicsys, we find that SNe~Ia that pass the \citet{Guy10_lc}
cut, but not the \citet{Kessler09_SDSS} cut only introduce a very small
bias of 0.2\% into the distance modulus.  This bias increases to 0.4\%
for redshifts larger than 0.5. However, there are no high-redshift
SNe~Ia in our \PS\ sample that pass the \citet{Guy10_lc} but not the
\citet{Kessler09_SDSS} cut.

\subsubsection{Final Cuts}

We require that the light curve fit converges and has a SALT2
light-curve fit probability $\Pfit > 0.001$ based on the $\chi^{2}$
per degree of freedom.  We remove all SNe~Ia in our sample with $|c| >
0.25$ or $|x_{1}| > 3$.  Colors or stretch values that deviate far
from zero are not well represented in the training sample and such
objects could bias the measurement of $\beta$.  
Where we have light curves of the same SNe from two different surveys,
we take the average distance of the two so that the supernova is only
treated as a single independent data point.

Finally, when fitting cosmological parameters, we apply Chauvenet's
criterion \citep{Taylor97} to reject outliers, removing SNe for which
we could expect less than half of an event in our full sample
(assuming a Gaussian distribution of intrinsic luminosities).  For the
PS1+lz sample, this is 4$\sigma$.  This criterion does not depend
significantly on our choice of cosmological parameters, and thus the
same SNe are excluded for all cosmological choices. Chauvenet's
criterion removes no SN~Ia from the \PS\ sample and two (SN~2004gc and
SN~2008cm) from the low-z sample.

We make all data and tools used for this analysis publicly available
at http://ps1sc.org/transients/
%We make all data used in
%this analysis publicly available, including light curve
%fit parameters\footnote{http:wachowski.pha.jhu.edu/\textrm{\about}dscolnic/PS1\_Public}.

\section{Cosmological Results}
\label{sec:results}

\subsection{Luminosity Distance Measurements}

The ultimate goal of this analysis is to put constraints on the
cosmological parameters $\OM$, $\OL$, and $w$.  We first transform the
SALT2 fit parameters into distances using the SALT2mu program
\citep{Marriner11}, which finds the $\alpha$ and $\beta$ parameters
that minimize the distance modulus residuals for a given cosmology.
While there is uncertainty in whether these parameters evolve with
redshift \citep{Kessler09_SDSS,Conley11}, here we assume that $\alpha$, $\beta$
and $M$ are all constant with redshift.  In order for the
$\chi^2_{\nu}$ of the distance residuals to be unity, an intrinsic
dispersion $\sigma_{\mathrm{int}}$ is added in quadrature to the error
of each SN distance (which includes the distance error from redshift
uncertainty)\footnote{$\sigma^2=\sigma^2_{\mathrm{N}}+\sigma^2_{\mathrm{int}}+\sigma^2_{\mu-z}$,
where $\sigma^2_{\mathrm{N}}$ is the photometric error of the SN
distance, $\sigma^2_{\mathrm{int}}$ is the intrinsic scatter, and
$\sigma^2_{\mu-z}$ is the distance error due to redshift uncertainty}.
Both the intrinsic dispersion of the sample and the photometric errors
of each SN distance include a dependence on the nuisance parameters
and covariances between the fit parameters.  The SALT2mu procedure
propagates these errors and determines the values for $\alpha$ and
$\beta$.
\begin{deluxetable}{lll}
\large
\tablecaption{Effects of Choices for Intrinsic Scatter
\label{tab:marriner}}
\tablehead{
\colhead{Intrinsic Scatter} &
\colhead{$\alpha$} &
\colhead{$\beta$} 
}
\startdata
$\sigma_{\mathrm{int},m_{B}}  = 0.115$ & $0.147 \pm 0.010$ & $3.13 \pm 0.12$\\
$\sigma_{\mathrm{int},c\phantom{x..}} = 0.025$    & $0.141 \pm 0.010$ & $3.71 \pm 0.15$
\enddata
\tablecomments{
 Intrinsic scatter $\sigma_{\mathrm{int},m_{B}}$ and $\sigma_{\mathrm{int},c}$
 in the  PS1+lz sample, and how $\alpha$ and $\beta$ vary for each
 method. The magnitudes of each scatter given above is such that the
 total reduced $\chi^{2}$ of the sample is \about 1.0.}
\end{deluxetable}

There is ongoing debate about the source of the intrinsic scatter seen
in SNe Ia distances.  As shown in \cite{Kessler13} and
\cite{Scolnic13_color}, the determined values of $\alpha$ and $\beta$
depend on assumptions about the source of the intrinsic scatter.  For
the PS1+lz sample, $\alpha$ and $\beta$ are given in
Table~\ref{tab:marriner} after 
attributing the remaining intrinsic distance
scatter to either luminosity variation or color variation. This is done after
the SALT2 light curve fit includes a small amount of color variation in its model.
We find that the intrinsic dispersion of the sample is
0.115~mag if we attribute intrinsic scatter to luminosity variation
and 0.025~mag if we attribute intrinsic scatter to color variation.
There is a large difference in the values of $\beta$ found for these
two different assumptions: $\beta=3.13\pm0.12$ and $\beta=3.71\pm0.15$
for the luminosity and color variation respectively.  The value of
$\beta$ found for the color variation case is within $3\sigma$ of a
MW-like reddening law.  Interestingly, the low-z sample by itself
pulls $\beta$ to a higher value ($\beta \sim 3.9$) than the PS1 sample
($\beta \sim 3.0$) when we attribute scatter to color variation.  This
may be due to the different selection effects in the low-z and PS1
samples, or the incompleteness of the SALT2 training sample for blue
colors (\Scolnicsys).  Most likely related, the total intrinsic
scatter of the PS1 distances is half as large
($\sigma_{\textrm{int}}=0.07$) as that for the low-z sample
($\sigma_{\textrm{int}}=0.123$).  The scatter seen for the low-z
sample is larger than that found in past studies
\citep[e.g.,][]{Kessler09_SDSS} as the assumed peculiar velocity
errors in this analysis are smaller. We fix different values of
$\sigma_{\textrm{int}}$ for our high and individual low-z
subsamples. These values are given in \Scolnicsys.

\begin{figure}
\centering
%\epsscale{1.15}  % 1.15 for emulateapj
\includegraphics[width=250pt]{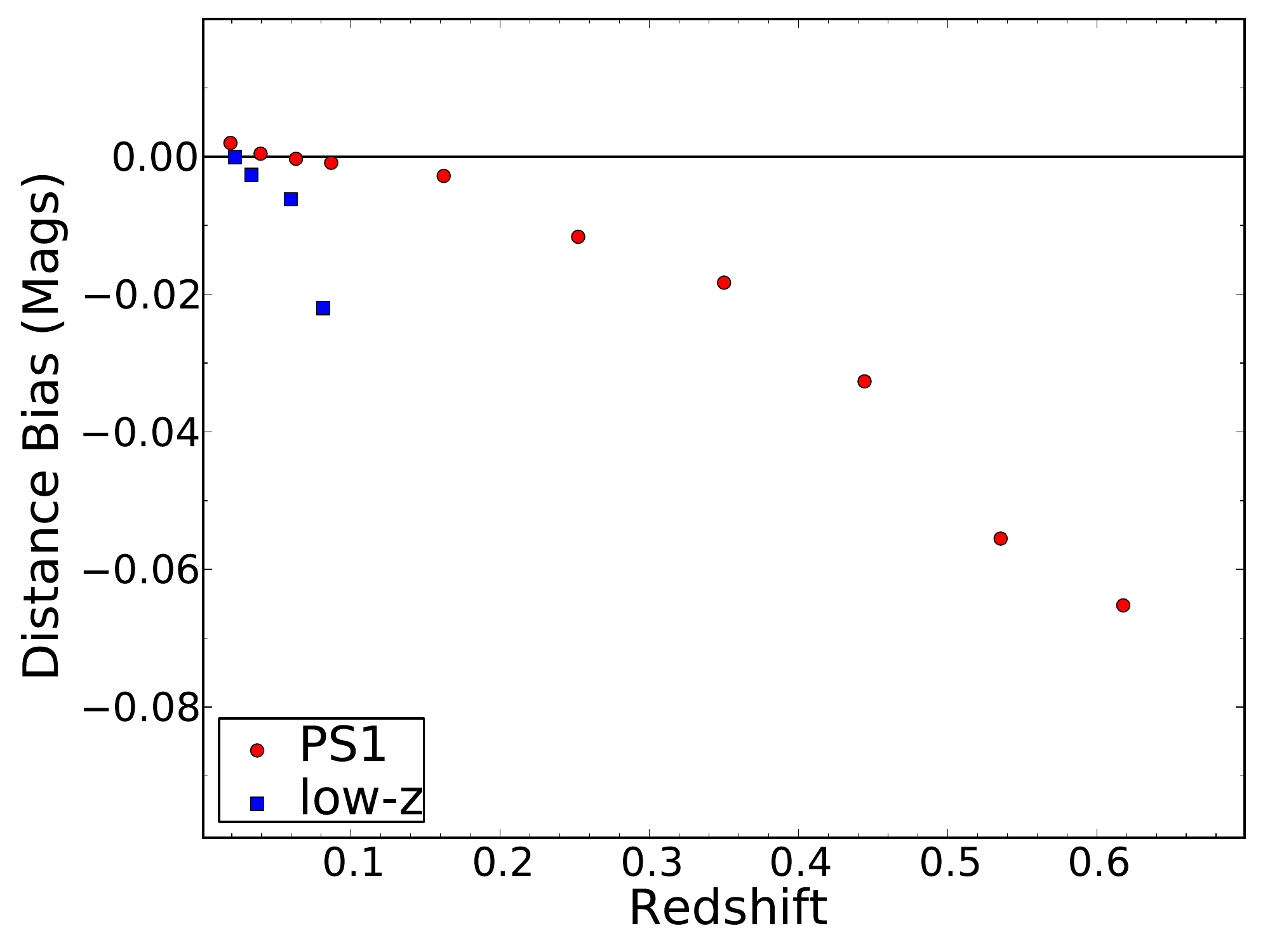}% 
\caption{
Distance corrections for the low-z and \PS\ 
samples as a function of redshift.
}
\label{fig:malm}
\end{figure}

The source of intrinsic scatter is explored in the companion paper
\Scolnicsys.  To summarize briefly, we create two simulations of the
PS1+lz samples in which one has $\beta=3.1$ and scatter is dominated
by luminosity variation, and another in which $\beta$ is consistent
with a MW-like value ($\beta=4.1$) and scatter is dominated by color
variation.  We then find the biases in distance over the entire
redshift range from subtracting our recovered distances from the
simulated distances.  For both simulations, we assume the scatter
comes from luminosity variation alone.  Since we simulate the full
sample, the discrepancies between recovered and simulated distances
will also be composed of the Malmquist bias at the upper limits of our
redshift ranges.  The differences in distance modulus correction from
the two simulations can be as high as 4\%.  To correct the distances,
we take the average correction from these two simulations at each
redshift.  We do not choose one model or the other as there is
insufficient empirical evidence to favor either model (\Scolnicsys).
Understanding and correcting for the intrinsic variation of SN~Ia is
one of the largest systematic uncertainties in our analysis.  In
Figure~\ref{fig:malm}, we show the bias in distances for both the
low-z and PS1 sample. For any given redshift, we then interpolate the
bias from this average correction vector. This bias is subtracted from
all \PS\ distance moduli.

\begin{figure*}
\includegraphics[width=500pt]{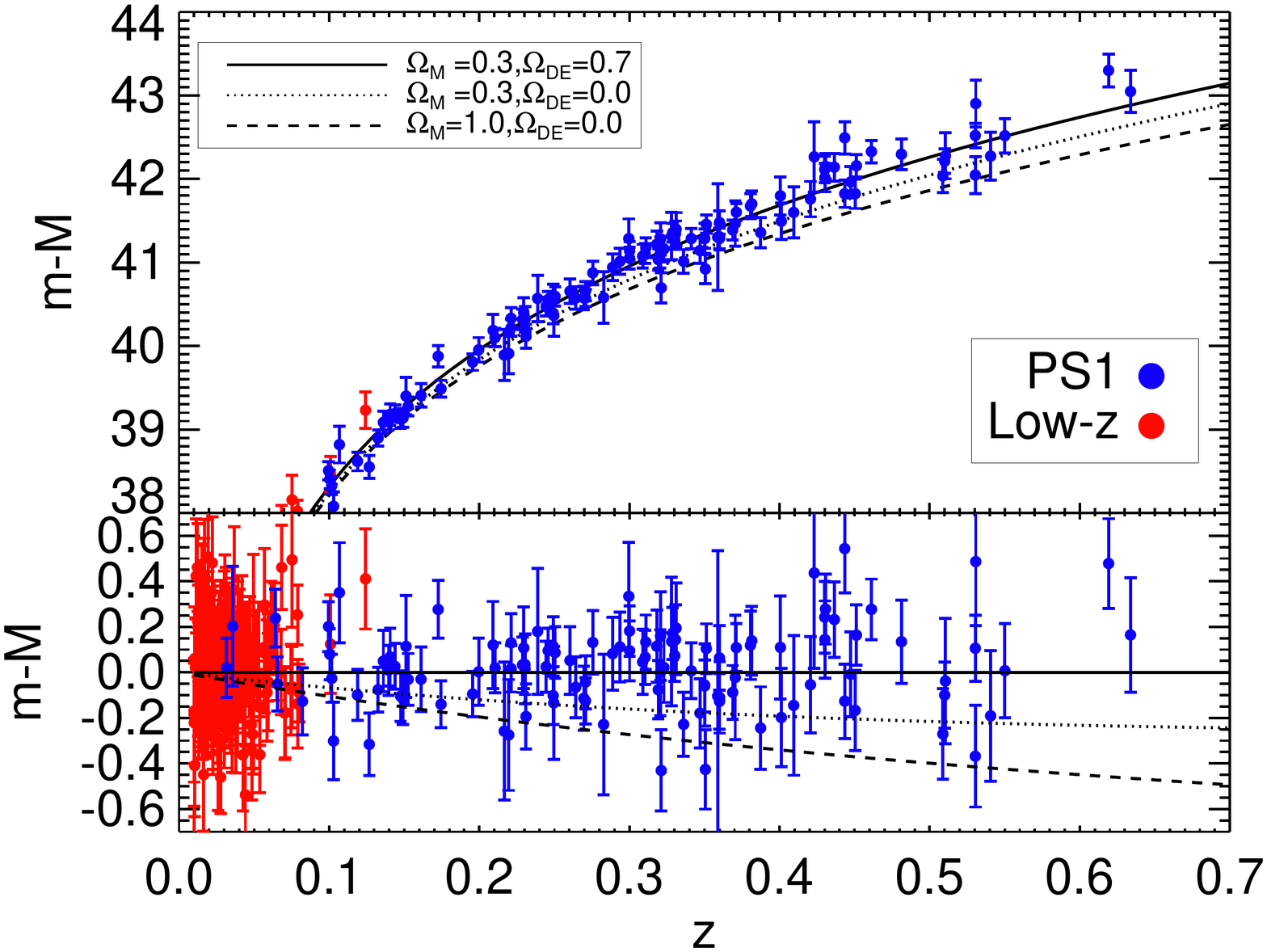} 
\includegraphics[width=500pt]{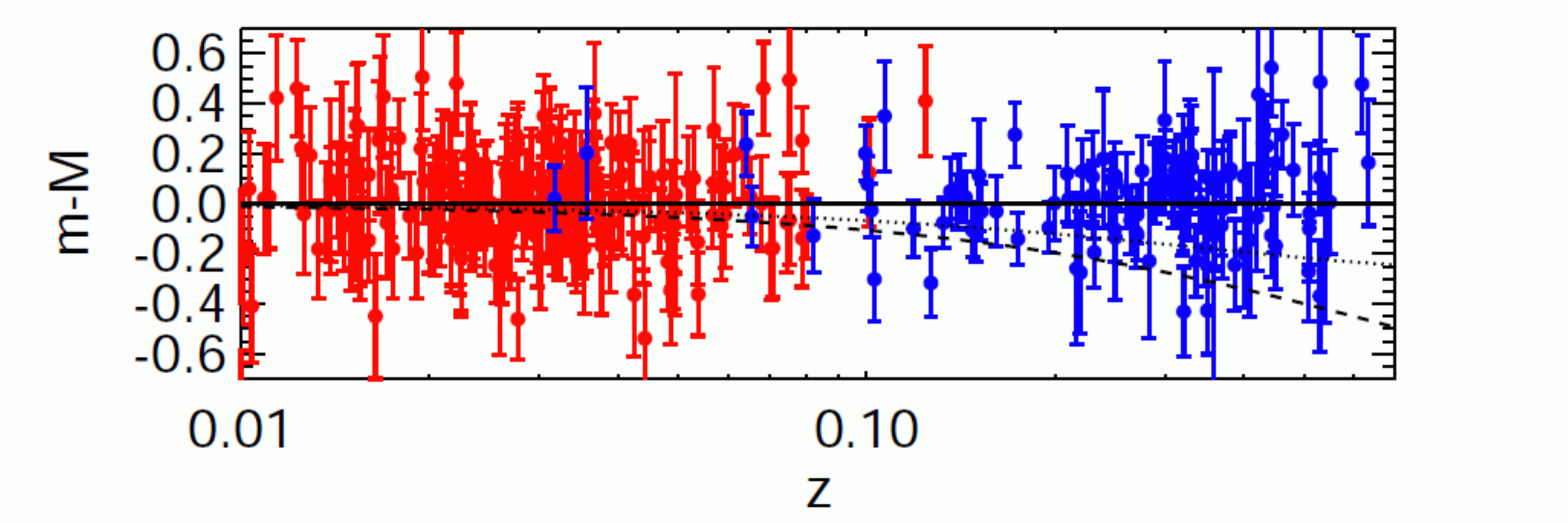} 
\caption[]{Hubble diagram for the combined PS1 and low-redshift
  sample. The bottom panel shows the difference modulus residuals 
  versus the logarithmic redshift in order to visualize the 
  the low-z SN~Ia residuals.
  \label{fig:hubble}}
\end{figure*}

The PS1+lz Hubble diagram with the corrected distances is shown in
Figure~\ref{fig:hubble}.  Three model universes are given: the
$\Lambda$CDM universe, a universe with $\OM=1$, and a universe
with $\OM=0.3$.  The distance modulus
$\mu(H_0,\OM,\OL,w,z)$ is found from the luminosity
distance $\DL$ such that $\mu=5 \times \log (\DL)+25$.  The luminosity
distance is commonly expressed as
\begin{eqnarray}
 & & \DL(z;w,\OM,\OL,H_0) =  \nonumber \\
 & &(1+z)|\Ok|^{-1/2}~\SINSINHFUN
   \left[
       {c|\Ok|^{1/2}} \int_{0}^{z} {dz' \over H(z')}
   \right] ~,
  \label{eq:dl}
\end{eqnarray}
where the curvature density $\Ok \equiv 1 -\OM-\OL$, 
and the function 
$\SINSINHFUN(x)=\sin(x)$ for $\Ok <0$, 
$\SINSINHFUN(x)=\sinh(x)$ for $\Ok >0$, and 
$\SINSINHFUN(x)=x$ for a flat Universe with $\Ok=0$.
Finally, the function $H(z)$ is defined as:
\begin{eqnarray}
 H(z) &=&   H_0 \left[ \OM(1+z)^3 + \OL(1+z)^{3(1+w)} + \Ok(1+z)^2 \right]^{1/2}\nonumber \\
\end{eqnarray}

Cosmological constraints can be found from the SN sample with only
statistical errors by measuring the $\chi^2$ value for a grid of
$\OL$ versus $\OM$ values with $w=-1$ and $w$ versus
$\OL$ for a flat universe
($\OL+\OM=1$).  The
\PS\ sample does not probe high enough redshifts to provide much
constraining power on evolving equations of state, and we thus assume
a constant $w$ and a flat universe when we determine the constraints
for $w$ and $\OM$.

\subsection{Systematic Uncertainties}

We briefly summarize the systematic uncertainty analysis detailed by
\Scolnicsys.  They derive a covariance matrix for the systematic
uncertainties that may be included when determining the \PS\
cosmological results. The most important systematic uncertainties are
due to calibration uncertainties, assumptions in the light curve
fitting and selection effects.

Uncertainties in the PS1 photometric system due to calibration are
given in Table~\ref{tab:systematics_cal}.  When combining the PS1
sample with the low-z sample, we include calibration uncertainties
from each of the low-z samples.  These uncertainties are composed of
uncertainties in bandpasses and zeropoints, as well as the uncertainty
in the Landolt standards.  The calibration uncertainties make up $>50\%$
of the total systematic uncertainty of the sample.  The other largest
uncertainties are due to incomplete understanding of the intrinsic
color of SNe~Ia, selection effects and Milky Way extinction
corrections.  Further possible systematic uncertainties are due to
dependencies of fitted distances on host galaxy properties and
coherent flow effects.  While we correct the redshifts of the low-z
supernova for coherent flow effects \citep{Neill07}, we currently do
not correct for the dependence on host galaxy properties.  This
adjustment is not applied as the difference in Hubble residuals for
low and high mass host galaxies is only $0.037\pm0.032$
(\Scolnicsys). For the use of future studies, we present the host
galaxy masses $M_{\mathrm{host}}$ in Table~\ref{tab:lcfitparams}. The
derivation of the masses is described in \Scolnicsys.

Following \cite{Conley11}, the uncertainties are propagated through
a systematic error matrix. We define a total uncertainty matrix
$\mathbf{C}$ such that $\mathbf{C} =\mathbf{D}_{\mathrm{stat}} +
\mathbf{C}_{\mathrm{sys}} \label{eqn:cdef}$.  The statistical matrix
$\mathbf{D}_{\mathrm{stat}}$ has only a diagonal component that
includes errors from the fit parameters and intrinsic scatter.  While
statistical covariances may arise from errors in the SALT2 template,
\cite{Conley11} finds these to be negligible.  The systematic error
matrix is determined by varying a given uncertainty parameter, and then
finding the difference between the original distance and a newly
determined distance.  While the statistical covariance matrix
includes components of the error budget that can in principle be
reduced by adding more SNe~Ia to the sample, the systematic
uncertainties can be reduced with improved analysis or external data.
Given a vector of distance residuals of the SN sample $\Delta
\vec{\mathbf{\mu}} = \vec{\mu_B} -
\vec{\mu(H_0,\OM,\OL,w,z})$ then $\chi^2$ may be
expressed as
\begin{equation}
 \chi^2 = \Delta \vec{\mu}^T \cdot \mathbf{C}^{-1} \cdot  \Delta \vec{\mu} .
  \label{eqn:chieqn}
\end{equation}

\begin{figure*}
\includegraphics[width=260pt]{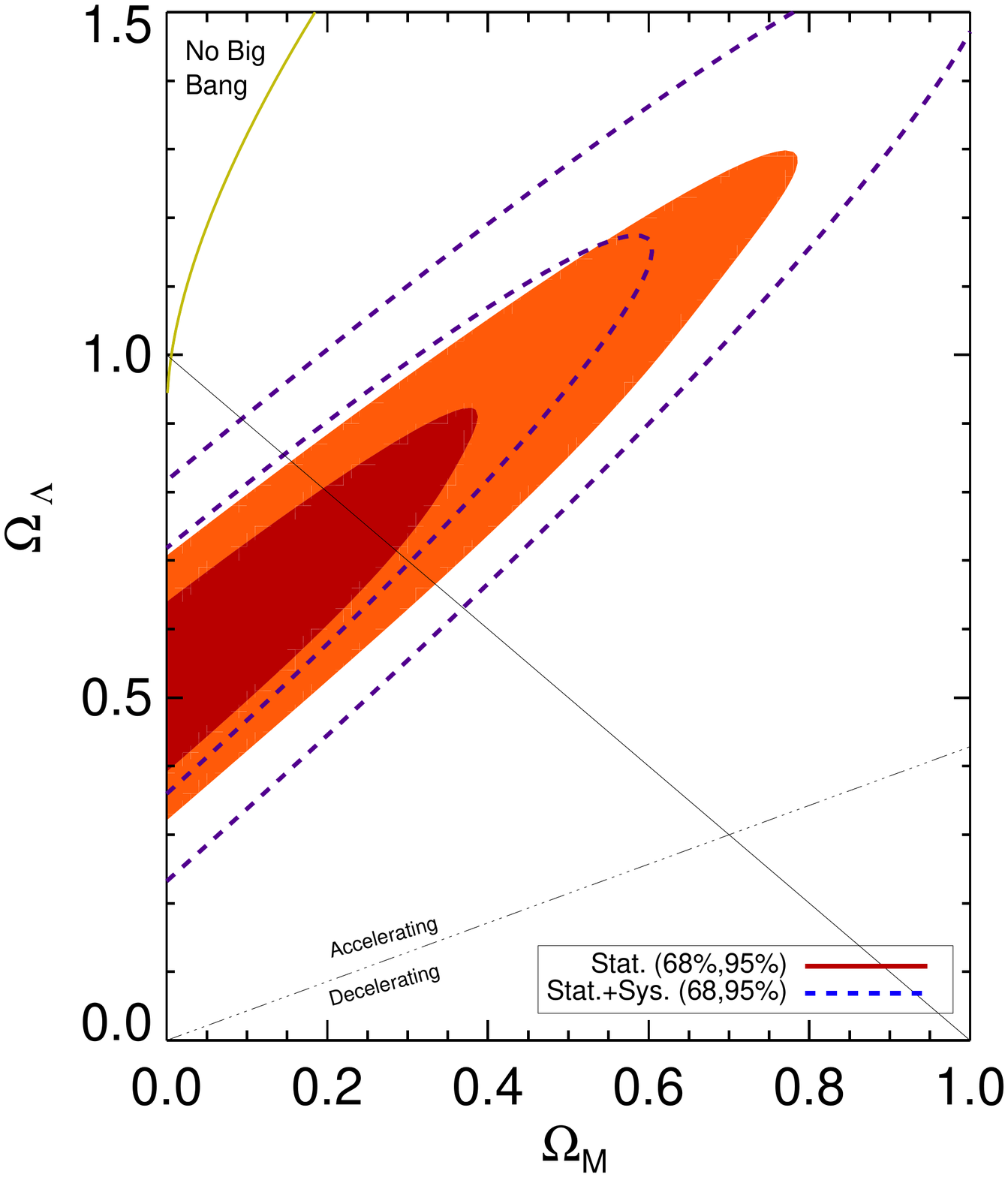} 
\includegraphics[width=260pt]{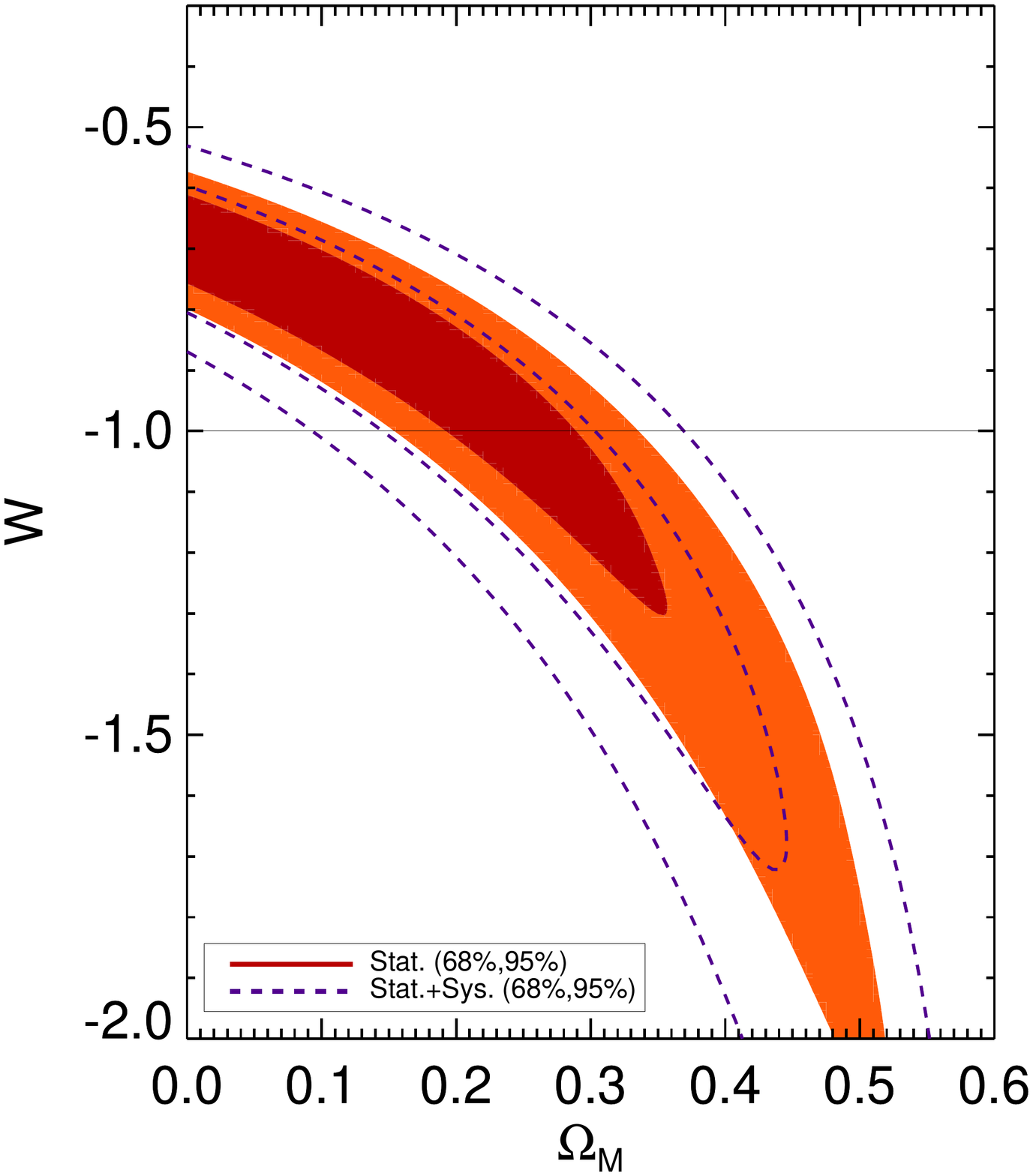}
\caption[]{The 1$\sigma$ and 2$\sigma$ cosmological constraints using
the PS1-lz SN sample only.  The statistical constraints as well as
when statistical and systematic uncertainties are combined are shown
with the solid red and dashed blue lines, respectively.
{\it Left:} Constraints on $\OM$ and $\OL$
assuming a cosmological constant ($w = -1$).
{\it Right:} Constraints on $\OM$ and $w$ assuming a constant
dark energy equation of state and flatness.
  \label{fig:OmOlw_snonly}}
\end{figure*}

These constraints are shown 
Figure~\ref{fig:OmOlw_snonly} and compared to the constraints from the
statistical only sample.  As the systematic uncertainties weight the
errors of SNe differently, the best fit values of the recovered
parameters will be shifted.  

\subsection{External Constraints}

To better determine cosmological parameters, we include constraints
from baryon acoustic oscillations \citep[BAO;][]{Blake11}, the CMB
\citep{Planck13}, and the Hubble constant \citep{Riess11}.  In order to
focus on the constraints from the PS1 sample and simplify the
analysis, we do not include additional high-redshift SN~Ia samples
(e.g., SNLS and {\it HST}).  A combination of all SN data will be left
for a future study.

We follow \cite{Planck13} to include constraints from other surveys,
as they have made the most precise measurements on the CMB, and gather
data from all of the various BAO surveys to determine this constraint.  The
likelihood of cosmological parameters is found from the Markov chains
given as an extension of \cite{Planck13}. We note that there are still
unresolved calibration discrepencies between {\it Planck} and {\it WMAP}
\citep{Hinshaw12}, and the constraints from these two surveys are
compared in \Scolnicsys.

For {\it Planck}, we use their Markov chains for determining
cosmological parameters.  When we wish to explore a flat $w$CDM model,
we use their $+w$ model.  When we want to examine non-flat
$\Lambda$CDM models, we use their $+k$ model.  For all CMB
constraints, we include data from the {\it Planck} temperature power
spectrum data, {\it Planck} temperature data, {\it Planck} lensing,
and {\it WMAP} polarization at low multipoles.  For the BAO
constraint, we take data from a multitude of surveys, as shown in
Table~\ref{tab:BAOvec}.
\begin{deluxetable}{llll}
\tabletypesize{\scriptsize}
\tablecaption{BAO data vector
\label{tab:BAOvec}}
\tablehead{
\colhead{Parameter}&\colhead{Value}&\colhead{Survey}&\colhead{Source}\\
}
\startdata
$D_{\rm V}(0.106)$ & $457 \pm 27 \,{\rm Mpc}$ & 6dF & \cite{Beutler11} \\
$r_{\rm s}/D_{\rm V}(0.20)$ & $0.1905 \pm 0.0061$ & SDSS & \cite{Percival10}\\
%$r_{\rm s}/D_{\rm V}(0.35)$ & $0.1097 \pm 0.0036$ & SDSS & \cite{Percival10}\\
$A(0.44)$ & $0.474 \pm 0.034$ & WiggleZ& \cite{Blake11}\\
$A(0.60)$ & $0.442 \pm 0.020$ & WiggleZ& \cite{Blake11}\\
$A(0.73)$ & $0.424 \pm 0.021$ & WiggleZ& \cite{Blake11}\\
$D_{\rm V}(0.35)/r_{\rm s}$ & $8.88 \pm 0.17$ & SDSS(R)& Padmanabhan (2012)\\
$D_{\rm V}(0.57)/r_{\rm s}$ & $13.67 \pm 0.22$ & BOSS& \cite{Anderson12}
\enddata
\tablecomments{BAO data vector from the different surveys used for the
cosmological analysis. 
Markov chains for these constraints are taken from the Planck data release \citep{Planck13}.
\phantom{\cite{Padmanabhan12}}}
\end{deluxetable}

For the Hubble constraint, we follow \cite{Riess11}
which uses {\it HST} observations of Cepheid variables in the host galaxies
of eight SNe Ia to calibrate the supernova magnitude-redshift
relation. Their best estimate of the Hubble constant is $H_0= 73.8
\pm 2.4~{\rm km}\,{\rm s}^{-1}\,{\rm Mpc}^{-1}$.

%Armin - can you rewrite with definitions for full constraints.
\subsection{Cosmological Parameter Constraints}

\begin{figure*}
\includegraphics[width=250pt]{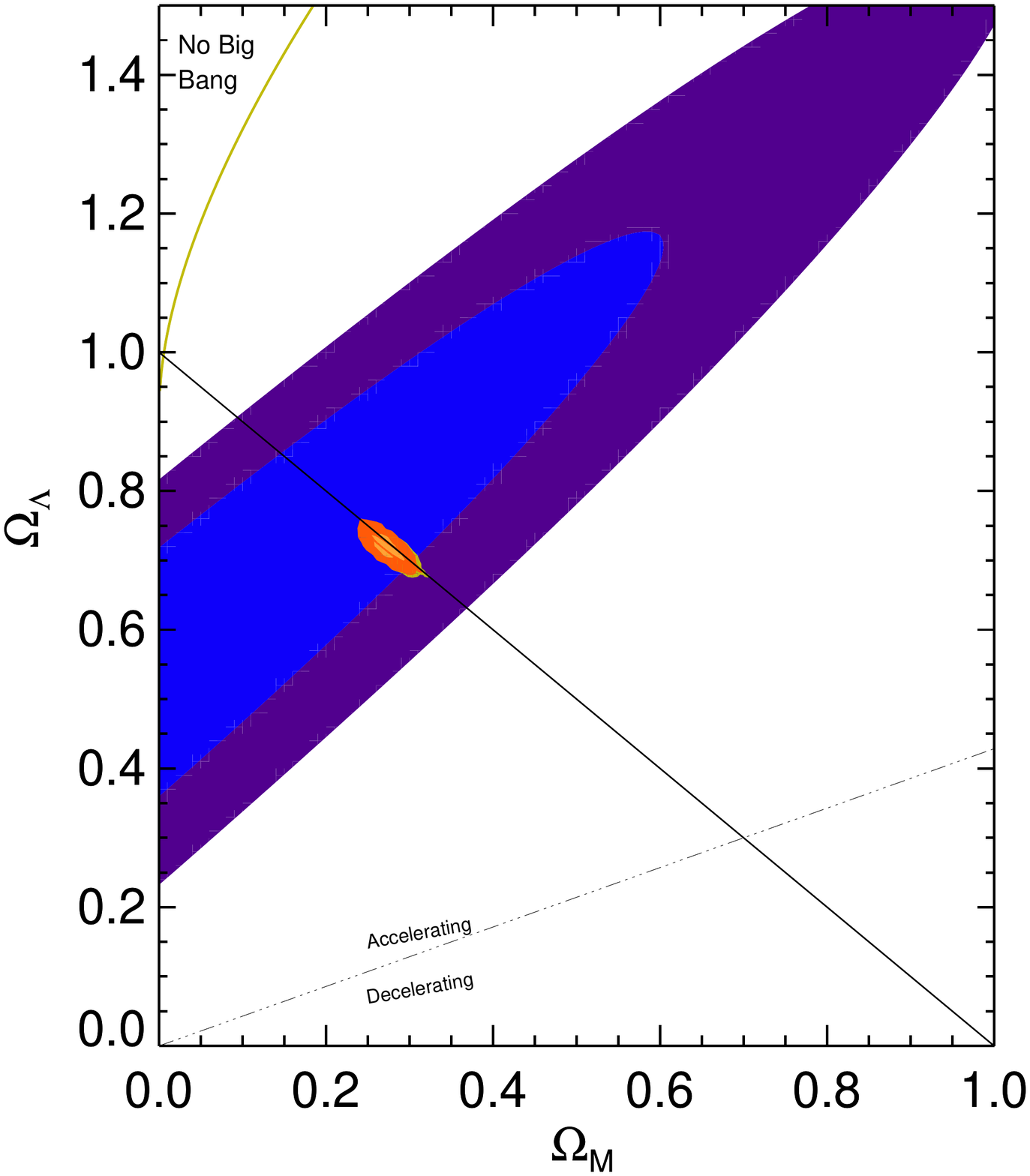}
\includegraphics[width=250pt]{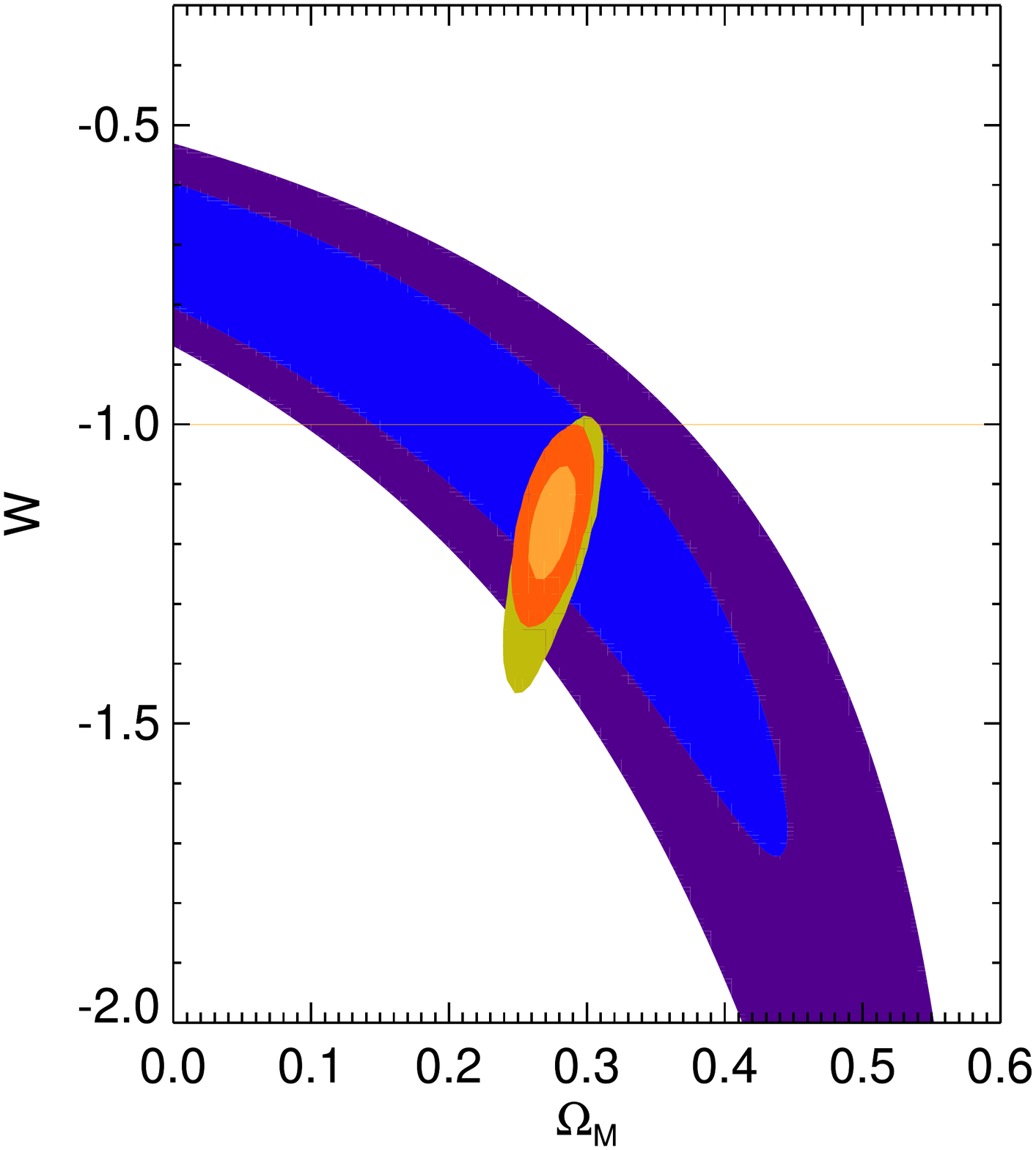}
\caption[]{The 68\% and 95\% cosmological constraints using
PS1-lz, {\it Planck}, BAO and $H_0$ measurements. Here the statistical and
systematic uncertainties are propagated.
{\it Left:} Constraints on $\OM$ and $\OL$ assuming a cosmological constant ($w = -1$).
{\it Right:} Constraints on $\OM$ and $w$ assuming a constant
dark energy equation of state and flatness.
\label{fig:OmOlw_all}}
\end{figure*}

Using the set of distances and redshifts for the PS1+lz sample, we are
able to constrain possible cosmological models.  First, we assume a
$\Lambda$CDM model ($w = -1$) and measure constraints on $\OM$ and
$\OL$. Using only the PS1+lz sample and assuming flatness, we find
$\OM = \snOMsysf$ and $\OL = \snOLsysf$ including systematic
uncertainties.  We present confidence contours for these parameters in
the left panel of Figure~\ref{fig:OmOlw_snonly}.  The evidence for
dark energy when not assuming a flat universe, from the SN only
sample, is \accel\% when including all systematic uncertainties.  We
also combine the SN constraints with BAO, CMB, and $H_0$ constraints
to find values of $\OM = \cOMsys$ and $\OL = \cOLsys$ (See left panel
of Figure~\ref{fig:OmOlw_all}). We summarize our constraints for $\OM$
and $\OL$ under these different assumptions in Table~\ref{tab:OMOL}.
%
%\begin{deluxetable*}{l|cc|cc|cc|cc}
%\tabletypesize{\scriptsize}
%\tablecaption{Cosmological Parameter Constraints
%\label{tab:OMOL}}
%\tablehead{
% & \multicolumn{2}{c|}{Stat Only} & \multicolumn{2}{c|}{Stat+Sys} & \multicolumn{2}{c|}{Stat Only} & \multicolumn{2}{c}{Stat+Sys}\\
% & $\OM$ & $\OL$  & $\OM$ & $\OL$ & $\OM$ & $w$  & $\OM$ & $w$
%}
%\startdata
%PS1-lz only       &\snOMstat&\snOLstat&\snOMsys&\snOLsys & \fomSNstat &  \wSNstat & \fomSNsys & \wSNsys \\
%PS1-lz+PL+BAO+$H_0$ &\cOMstat&\cOLstat&\cOMsys&\cOLsys & \fomCstat &  \wCstat & \fomCsys & \wCsys 
%\enddata
%\tablecomments{{\it Left side:} Constraints on $\OM$ and $\OL$
%assuming a $\Lambda$CDM model ($w = -1$) with statistical only and
%systematic + statistical uncertainties. {\it Right side:} Constraints
%on $\OM$ and $w$ assuming a flat universe with statistical only and
%systematic + statistical uncertainties.}
%\end{deluxetable*}
%
\begin{deluxetable*}{l|cc|cc}
\tabletypesize{\scriptsize}
\tablecaption{
%Cosmological Parameter Constraints
Constraints on $\OM$ and $w$
\label{tab:OMOL}}
\tablehead{
 & \multicolumn{2}{c|}{Stat Only} & \multicolumn{2}{c|}{Stat+Sys} \\
 & $\OM$ & $\OL$  & $\OM$ & $\OL$ 
}
\startdata
PS1-lz only (flat Universe)      &\snOMstatf&\snOLstatf&\snOMsysf&\snOLsysf  \\
PS1-lz+PL+BAO+$H_0$ &\cOMstat&\cOLstat&\cOMsys&\cOLsys  
\enddata
\tablecomments{Constraints on $\OM$ and $\OL$
assuming a $\Lambda$CDM model ($w = -1$) with statistical only and
systematic + statistical uncertainties. For the PS1-lz only sample,
the constraints on $\OM$ are not significant, we therefore present the constraints
assuming flatness.
}
\end{deluxetable*}
\begin{deluxetable*}{l|cc|cc}
\tabletypesize{\scriptsize}
\tablecaption{
%Cosmological Parameter Constraints
Constraints on $\OM$ and $w$
\label{tab:OMw}}
\tablehead{
 & \multicolumn{2}{c|}{Stat Only} & \multicolumn{2}{c|}{Stat+Sys}\\
 & $\OM$ & $w$  & $\OM$ & $w$
}
\startdata
PS1-lz only         & \fomSNstat &  \wSNstat & \fomSNsys & \wSNsys \\
PS1-lz+PL+BAO+$H_0$ & \fomCstat &  \wCstat & \fomCsys & \wCsys 
\enddata
\tablecomments{Constraints
on $\OM$ and $w$ assuming a flat universe with statistical only and
systematic + statistical uncertainties.}
\end{deluxetable*}

Relaxing the assumption of a cosmological constant, we can attempt to
measure $w$, the dark energy equation-of-state parameter.  For these
$w$CDM models, we assume a flat Universe ($\Omega_{k} = 0$).  With
only the PS1+lz SN sample, we measure $w = \wSNstasys$ (see right
panel of Figure~\ref{fig:OmOlw_snonly}).  We also combine the SN
constraints with CMB, BAO and $H_0$ constraints to determine
$\OM=\fomCsys$ and $w=\wCsys$ (see right panel of
Figure~\ref{fig:OmOlw_all}). 
The first number represents the mean, and
the uncertainties represent the distance between the mean and the
1$\sigma$ limits.  This formalism is similar to what \cite{Planck13}
uses in its Markov-Chain release.

In Table~\ref{tab:OMw}~and~\ref{tab:cosmoresults}, we
compare how the different cosmological probes impact the constraints
on $\OM$ and $w$.  As shown in
Figure~\ref{fig:OmOlw_all}, the constraints of PL+BAO+$H_0$ is within
the 2$\sigma$ of our SN-only constraints. Using the {\it Planck}
measurements alone leads to a very low value of $w=-1.485$, albeit still
within 2$\sigma$ of -1 due to its uncertainties of $0.253$.
\cite{Planck13} claim that there is significant tension between their
measurements and measurements of $H_0$ and/or SNe (SNLS3), but not
with the BAO measurements. We show in Table~\ref{tab:cosmoresults}
that BAO, $H_0$ and SN measurements all have a similar pull toward a
cosmological constant when combined with {\it Planck}.  When combining
our PS1+lz measurements with {\it Planck}, BAO, and $H_0$, we find
$w=\wCsys$ inconsistent with -1 at the 2.3$\sigma$ level. We note that
this is very similar to the 2.6$\sigma$ inconsistency found when
substituting SNLS3 for PS1+lz \citep[see
Table~18.27\footnote{http:\/\/www.sciops.esa.int/SYS/WIKI/uploads/
Planck\_Public\_PLA/9/9b/Grid\_limit68.pdf} in][]{Planck13}.
It is important to point out that a tension with a cosmological
constant is greatest when $H_0$ and Planck constraints only are used.
\Scolnicsys\ analyzes the tension between SN and the CMB using {\it WMAP}
\citep{Hinshaw12} instead of {\it Planck}, and finds the tension to be
smaller. 

\begin{deluxetable*}{l|cc|cc|cc}
\tabletypesize{\scriptsize}
\tablecaption{Cosmological Parameter Constraints Using different Cosmological Probe Combinations
\label{tab:cosmoresults}}
\tablehead{
& \multicolumn{2}{c|}{Without PS1-lz} & \multicolumn{2}{c|}{With PS1-lz (Stat. only)}& \multicolumn{2}{c}{With PS1-lz (Sys. \& Stat.)}\\
Sample & $ \OM $ & $w$  & $\OM$ & $w$ & $\OM$ & $w$
}
\startdata
PL & $ 0.218^{+0.023}_{-0.079}$ &$ -1.485^{+0.253}_{-0.426}$  & $ 0.289^{+0.015}_{-0.019}$ & $-1.102^{+0.058}_{-0.061}$ & $0.281^{+0.018}_{-0.022}$ & $-1.136^{+0.077}_{-0.079}$ \\
PL+BAO & $ 0.287^{+0.021}_{-0.020}$ &$ -1.133^{+0.138}_{-0.104}$  & $ 0.291^{+0.010}_{-0.012}$ & $-1.102^{+0.055}_{-0.058}$  & $ 0.288^{+0.014}_{-0.014}$ & $-1.124^{+0.077}_{-0.066}$ \\
PL+$H_0$ & $ 0.258^{+0.016}_{-0.021}$ &$ -1.240^{+0.095}_{-0.093}$ & $ 0.277^{+0.011}_{-0.015}$ & $-1.131^{+0.044}_{-0.052}$ & $ 0.269^{+0.016}_{-0.015}$ & $-1.174^{+0.064}_{-0.059}$\\
PL+BAO+$H_0$ & $ 0.275^{+0.014}_{-0.014}$ &$ -1.205^{+0.102}_{-0.087}$ & $ 0.284^{+0.010}_{-0.010}$ & $-1.131^{+0.049}_{-0.049}$ & $ 0.280^{+0.013}_{-0.012}$ & $-1.166^{+0.072}_{-0.069}$ \\
\enddata
\tablecomments{
Comparison of the $\OM$ and $w$ constraints using different
variations of external constraints {\it Planck} \citep{Planck13}, BAO, and $H_0$ \citep{Riess11}.
}
\end{deluxetable*}

\section{Discussion}

Using the first 1.5 years of the \PS\ MDF survey, we have discovered
thousands of transients, \SNIAPSall\ of which we have
spectroscopically confirmed to be SNe~Ia.  Combining novel calibration
techniques with a well-tested photometric data reduction pipeline, we
have obtained precise photometry for these SNe. We estimate that the
photometric uncertainty is 1.2\%, excluding the uncertainty in the
{\it HST} Calspec definition of the AB system.  Using the SALT2
light-curve fitter, we have measured distances to a carefully selected
sample of \SNIAPSused\ SNe~Ia.  After correcting for biases related to
detection and spectroscopic follow-up efficiency, we used these SNe to
constrain cosmological parameters.

\subsection{Comparison to Previous Work}

For our cosmological analysis, we used very recent constraints coming
from BAO and CMB experiments.  Previous SN analyses did not have
access to those data, and thus a direct comparison is more difficult.
Similarly, our low-redshift sample is larger than previous
compilations, again, complicating any comparison.  Nonetheless, we
report previous results in an attempt to place the current \PS\
analysis in context.

The ESSENCE survey, using 60 high-redshift SNe~Ia, 45 low-redshift
SNe~Ia, and the initial SDSS BAO results \citep{Eisenstein05}, found
$w = -1.05$ with statistical and systematic uncertainties on the order
of 13\% \citep{Wood-Vasey07}.  The SDSS-II SN survey, using 103
high-redshift SNe~Ia, 33 low-redshift SNe~Ia, WMAP5, and the initial
SDSS BAO results, found $w = -0.92 \pm 0.13$ (stat)
\citep{Kessler09_SDSS}.  From combinations with SNLS and {\it HST} SNe
(resulting in 288 total SNe~Ia), their constraints decreased to $w =
-0.96 \pm 0.06 \mathrm{(stat)} \pm 0.12 \mathrm{(sys)}$.  The 3-year
SNLS analysis used a combined sample of 242 high-redshift SNe~Ia from
SNLS, 93 SDSS, 14 {\it HST}, and 132 low-redshift SNe~Ia (a total of
472 SNe), and when combined with the initial SDSS BAO results, WMAP5,
and the $H_0$ constraint, they find $w = -1.061$ with both statistical
and systematic uncertainties on the order of 5\% \citep{Conley11,
Sullivan11}.  The \PS\ cosmological constraints are derived using the
most recent external constraints from {\it Planck}, the Wiggle-Z BAO
results, and the $H_0$ constraint.  Combining these external
constraints with \SNIAPSused\ high-redshift \PS\ SNe~Ia and
\SNIAlowztot\ light curves from \SNIAlowzused\ low-redshift SNe~Ia, we find $w=\wCsys$, inconsistent
with -1 at the 2.3$\sigma$ level. It is currently still unclear
whether the tension with flat $\Lambda$CDM is a feature of the model
or a combination of chance and systematic errors.

All the other SN~Ia surveys of the past decade constrain $w$ to the
cosmological constant value of -1 within 1$\sigma$.  An outside
observer might expect 1/3 of the results to lie outside this bound,
and a confirmation bias could be in effect \citep{Croft11}.
Superficially these may appear to be independent tests, providing
mounting evidence for a $w=-1$ universe.  However, there is
significant overlap in the low-redshift SN data used in all analyses,
and we should thus expect some degree of correlation in the results.
Furthermore, most SN surveys also share the same calibration
sources.

\subsection{Future Improvements to Photometric Calibration}

Our systematic uncertainties are dominated by the photometric
calibration.  Although our calibration effort is sufficient for our
current analysis and comparable to other SN surveys, there are
additional areas in which significant improvements can be made.  Here,
we identify areas for future improvement.

Since photometric calibration is currently the largest component of
the systematic error budget, it should be the primary focus for future
efforts.  In order to reduce the systematic uncertainty of the
calibration to the level of other systematic uncertainties,
photometric accuracy must be significantly better than 1\%.  For the
current \PS\ photometric system, all indications are that we have an
accuracy of 1.2\% before accounting for the {\it HST} Calspec
uncertainty in the AB magnitude system.  Over the coming years, we
expect that the NIST-based tunable laser system will continue to
improve, resulting in decreased ghosting and a more uniform
illumination of the flat-field screen.  The \PS\ photometry is also
still improving through changes in reduction software and the
acquisition of new data (which will continue until early in 2014 for
the MDF survey).

Recently, the SNLS and SDSS teams have undertaken an effort to
precisely calibrate their photometric systems onto roughly a single
system.  They have reached a 0.5\% consistency between their
photometric systems, using methods similar to those employed in the
\PS\ analysis.  Specifically, they used observations of the {\it HST}
Calspec standard stars in both systems to provide consistency.  An
attempt of this analysis is done by \Scolnicsys, which finds
discrepencies of up to 2\% between the \PS\ and SDSS AB systems.  The
large overlap between the SDSS photometric footprint and the \PS\
3$\pi$ survey has already been used in an \"{u}bercalibration of the
\PS\ system \citep{Schlafly_uebercal}. This type of analysis can
provide further improvement. In addition, the ongoing {\it HST}
program GO-12967 measures fluxes of DA white dwarf stars, one of which
resides within a \PS\ MDF.  This will allow an in situ direct
comparision of white dwarf colors with models that incorporate
$\log{(g)}$ and temperature from ground-based spectroscopy.

We believe that we can also improve our estimation of flux
uncertainties.  Specifically, we know that there is a tail to the
reduced $\chi^{2}$ distribution shown in Figure~\ref{fig:fluxoffsetX2}
to large $\chi^{2}$.  We have shown that these outliers are SNe close
to bright objects (usually a host galaxy nucleus), resulting in an
underestimate of the uncertainty.  We have performed preliminary tests
that indicate that modeling the uncertainties with an additional term
proportional to the distance to the nearest bright static object will
remove this tail and provide a better estimate of the uncertainty.
Only a small fraction of SNe are significantly affected by this bias,
so it does not affect our current results, but should become more
important as we reduce other systematic uncertainties.

\subsection{Improvements to Cosmological Constraints}

The data presented here comprises the first 1.5~years of \PS.  As is
common with large projects, the data quality and cadence at the
beginning of the survey were not as good as those achieved later in
the survey.  We expect that typical light curve coverage and SNR for
the full \PS\ sample will be better than in the current sample.  At
the end of the survey, the total exposure time for the MDF survey will
exceed the exposure time of the current sample by nearly a factor of
three.  In addition, our spectroscopic follow-up efficiency has
increased; we expect a final spectroscopic sample will consist of
$\gtrsim$ 400 SN~Ia. Thus we can expect that our current statistical
uncertainties will improve by a factor of \about 2, comparable to the
best statistical uncertainties currently reported.

Along with the large spectroscopic sample, we will observe several
times as many SNe~Ia without spectra.  We can generate a large
photometrically classified sample of SNe~Ia with relatively small
contamination from other SN classes \citep{Sako11}.  The SDSS-II
supernova survey produced such a sample consisting of 752
photometrically classified SNe~Ia and produced cosmological
constraints with this sample \citep{Campbell13}.  Combining distance
estimates of these SNe (and no other data sets, i.e., no low-redshift
SNe) with constraints on $H_0$, data from {\it WMAP}, and large scale
structure measurements, they find constraints on $w$ with statistical
uncertainties of 10\%.

The advantage of this method is that the sample size is not limited by
the availability of spectroscopic follow-up resources, which has not
scaled with the number of SNe discovered.  \PS\ has discovered
$\gtrsim$4500 transients with SN-like light curves. This number will
likely increase to $\gtrsim$6000 by the end of the \PS\ survey. In
magnitude-limited surveys like \PS, 79\% of all SNe are SN~Ia
\citep{Li11}, thus with a classification efficiency of 70\%, we can
expect \about 3300 photometrically classified SNe~Ia from the \PS\
survey.

The photometric classification of a transient is significantly
improved with a host-galaxy spectroscopic redshift, which can be
obtained after the SN has faded.  Follow-up observations at the end of
the survey with a large FOV multi-object spectrograph such as
Hectospec on the MMT should provide redshifts for the vast majority of
candidate SNe~Ia \citep{Jones14}.

With this large set of new SNe~Ia, more work can be done to identify a
``third parameter'' that correlates with luminosity independent of
light curve shape and color.  This would be particularly important if
there is some redshift-dependent evolution of the SN~Ia properties.
Candidates are, for example, ejecta velocity \citep{Wang09,Foley11a,
Foley11b}, other spectral features \citep{Foley08:uv,Bailey09,
Blondin11,Silverman12:spec,Foley13b:uv}, and host galaxy properties
\citep{Kelly10, Lampeitl10, Sullivan10, Hayden12}.

We are also conducting a large (100-orbit) {\it HST} program, RAISIN
(GO-13046; PI Kirshner) to obtain rest-frame NIR photometry of \PS\
SNe~Ia ($ 0.2 < z < 0.4$).  Although calibration is currently the
largest systematic uncertainty in our analysis, the largest
astrophysical systematic uncertainty comes from the treatment of
intrinsic SN colors and dust reddening, which directly affects
distance measurements.  When combining the {\it HST} and \PS\
photometry, the large wavelength range (from UV to NIR) provides
additional constraints on the reddening law, which should
significantly improve our understanding of systematics related to dust
extinction, reddening, and SN~Ia colors.  SN Ia are more nearly
standard candles when observed in restframe near-infrared bands and
extinction is significantly lower at these wavelengths
\citep{Krisciunas04, Wood-Vasey08, Mandel09, Folatelli10,Barone-Nugent12,Kattner12}.  The
systematic errors that result from lightcurve shape corrections and
inferences about extinction from color are distinctly smaller when
using these bands \citep{Mandel11}.

\section{Conclusion}

We have presented the light curves and analysis of \SNIAPSall\
spectroscopically confirmed SNe~Ia from the first 1.5 years of the
\PS\ Medium Deep Survey. We have described the SN discovery and
spectroscopic follow-up of the survey.  We analyze the relative and
absolute photometric calibration in the \PS\ natural system using
on-site measurements of the instrument response function,
spectro-photometric standard star observations, and
\"{u}bercalibration, with an emphasis on any systematic biases
introduced.  We find that the systematic uncertainties in the
photometric calibration are currently 1.2\% without accounting for the
uncertainty in the {\it HST} Calspec definition of the AB system.

From the sample of \SNIAPSall\ SNe~Ia in the redshift range $0.03 < z
<0.65$, \SNIAPSused\ passed the various quality cuts in our
cosmological analysis.  The spectral information and SALT2 light curve
fit parameters of each SN are presented so that this sample can be
used in joint analyses with other supernova samples.  When combining
the \PS\ sample with low-z SNe~Ia (\SNIAPSused\ PS1 SNe~Ia +
\SNIAlowztot\ light curves from \SNIAlowzused\ low-z SNe~Ia), we find $w=\wSNsys$ assuming a flat
Universe. A Universe devoid of Dark Energy is rejected at the at
\accel\% level with the SN sample alone including all identified
systematics. When combined with external constraints (BAO, CMB, and
$H_0$), our cosmological analysis yields $\OM=\fomCsys$ and $w=\wCsys$
including all identified systematics, consistent with a cosmological
constant. This is in agreement with the results from previous SN~Ia
surveys like SNLS and SDSS.  We show the pull on the recovered
cosmological parameters from the SN measurements.  Compared to when
CMB, BAO and $H_0$ measurements are used to constrain $w$, including
the SN measurements reduces the total uncertainty in $w$ by
$\sim40\%$.

Further analysis of the systematic errors of this sample is presented
in the companion paper by \Scolnicsys.  In the future analysis of the
full 4-year \PS\ sample, we can reduce the dominant systematic
uncertainty in absolute photometric calibration with more \PS\
observations, additional Calspec spectrophotometric standards, and
improved reduction and analysis.

This paper and the companion paper by \Scolnicsys\ represent the first
in a series of cosmological analysis using the PS1 sample.  It is
necessary to accurately measure SN photometry and quantify systematic
uncertainties with the spectroscopically confirmed sample before doing
so for the much larger photometric sample.  The photometric sample is
currently $>5\times$ bigger than any published spectroscopic sample of
SN~Ia and will represent a significant step forward for constraining
the nature of dark energy.

\begin{deluxetable*}{lccc}
\tabletypesize{\scriptsize}
\tablecaption{Pan-STARRS1 Spectroscopic Sample
\label{tab:spec}}
\tablehead{
\colhead{PS1 name} & \colhead{Spectra MJD's} & \colhead{Telescopes} & \colhead{$z_{hel}$}
}
\startdata
 PS1-0909006 & 55098                   & Gem-N                    & 0.284(0.001)\\
 PS1-0909010 & 55105                   & Gem-N                    & 0.270(0.010)\\
 PS1-0910012 & 55116,55118,55121,55142 & Gem-S,NOT,Magellan,Gem-S & 0.079(0.001)\\
 PS1-0910016 & 55125                   & Gem-N                    & 0.230(0.010)\\
 PS1-0910017 & 55125                   & Gem-N                    & 0.320(0.010)\\
 PS1-0910018 & 55125                   & Gem-N                    & 0.265(0.001)\\
 PS1-0910020 & 55124                   & Gem-S                    & 0.242(0.001)\\
 PS1-0910021 & 55125                   & Gem-S                    & 0.256(0.001)\\
\enddata
\tablecomments{
Spectroscopic information for all spectroscopically classified \PS\
SN~Ia from 2009 September to 2011 May. 
}
\end{deluxetable*}
\pagebreak

\begin{deluxetable*}{lccccc}
\tabletypesize{\scriptsize}
\tablecaption{Properties of the Pan-STARRS1 Spectroscopic Sample
\label{tab:SN}}
\tablehead{
\colhead{PS1 name} & \colhead{R.A. (J2000)} & \colhead{Dec. (J2000)} & \colhead{MDF} & \colhead{$t_{disc}$}  & \colhead{Cut}
}
\startdata
 PS1-0909006 & 22:15:48.000 & \phantom{-}01:11:05.39 & 09       &        55083 & \nodata          \\
 PS1-0909010 & 02:28:28.368 & -04:04:44.08           & 01       &        55089 & $|x_1|>3.0$      \\
 PS1-0910012 & 03:29:53.232 & -28:05:12.12           & 02       &        55096 & Iax              \\
 PS1-0910016 & 02:21:13.752 & -03:54:35.82           & 01       &        55104 & $|x_1|>3.0$      \\
 PS1-0910017 & 02:21:06.600 & -05:01:23.84           & 01       &        55104 & \nodata          \\
 PS1-0910018 & 02:22:40.080 & -04:01:38.32           & 01       &        55104 & \nodata          \\
 PS1-0910020 & 03:38:23.400 & -28:15:11.88           & 02       &        55104 & LC incomplete    \\
 PS1-0910021 & 03:34:28.800 & -27:54:30.24           & 02       &        55104 & LC incomplete    \\
\enddata
\tablecomments{
Summary of the general properties of the \PS\ SN~Ia. The MDF fields
are defined in Table~\ref{tab:fields}.  The discovery date $t_{disc}$
is given in MJD.  The cuts are explained in \S\ref{sec:cuts} -- Cut
$7-10$ in \S\ref{sec:cuts} are referred to as ``LC incomplete''.
}
\end{deluxetable*}

\begin{deluxetable*}{lcc|cccccc}
\tabletypesize{\scriptsize}
\tablecaption{Contributions and Cuts of the Different Samples
\label{tab:fitoverview}}
\tablehead{
\colhead{Set} & \colhead{Total} & \colhead{Final} & \colhead{Initial} & \colhead{First Pass} & \colhead{Fit not} & \colhead{$P<0.001$} & \colhead{$|x_1|>3.0$} & \colhead{Chauveneut}\\
\colhead{} & \colhead{} & \colhead{} & \colhead{Cut} & \colhead{Cut} & \colhead{conv.} & \colhead{} &   \colhead{$|c|>0.25$} & \colhead{}
}
\startdata
 JRK07 & 133 & \phantom{1}49 & 30 & 34 & 1 & 12 &   7 & 0\\
  CFA3 & 185 & \phantom{1}85 & \phantom{1}8 & 70 & 0 & \phantom{1}7 &  15 & 0\\
  CFA4 & \phantom{1}94 & \phantom{1}43 & \phantom{1}8 & 26 & 3 & \phantom{1}8 &   5 & 1\\
   CSP & \phantom{1}85 & \phantom{1}45 & \phantom{1}5 & 13 & 0 & 14 &   7 & 1\\
   PS1 & 147 & 112 & 10 & 18 & 2 & \phantom{1}1 &   4 & 0\\
\enddata
\tablecomments{
Breakdown of the number of SNe~Ia that were cut from the different sets.
The different sets are defined in \S\ref{sec:cuts}.
}
\end{deluxetable*}

\begin{deluxetable*}{lccccccccc}
\tabletypesize{\scriptsize}
\tablecaption{Fit Parameters of low-z Sample
\label{tab:lowz}}
\tablehead{
\colhead{SN} & \colhead{Set} & \colhead{$z_{cmb}$} & \colhead{$t_{peak}$} & \colhead{$P_{fit}$} & \colhead{$x_1$} & \colhead{$c$} & \colhead{$m_B$} & \colhead{$\mu$} & \colhead{$E(B-V)$}
}
\startdata
   1990O & JRK07 & 0.031(0.001) & 48076.88(0.59) &  0.9960 & \phantom{-}0.476(0.231) & -0.054(0.036) & 15.993(0.065) & 35.587(0.166) & 0.075(0.012)\\
  1990af & JRK07 & 0.050(0.001) & 48196.03(0.27) &  1.0000 & -2.114(0.235) & -0.038(0.048) & 17.508(0.039) & 36.689(0.190) & 0.028(0.005)\\
   1992P & JRK07 & 0.026(0.001) & 48719.90(0.50) &  0.4810 & \phantom{-}0.069(0.352) & -0.083(0.045) & 15.811(0.036) & 35.441(0.195) & 0.017(0.003)\\
  1992ae & JRK07 & 0.075(0.001) & 48804.90(0.66) &  0.7470 & -0.708(0.186) & -0.029(0.055) & 18.216(0.052) & 37.565(0.199) & 0.029(0.005)\\
  1992aq & JRK07 & 0.101(0.001) & 48834.37(0.69) &  1.0000 & -1.326(0.264) & -0.071(0.063) & 19.065(0.053) & 38.460(0.215) & 0.010(0.002)\\
  1992al & JRK07 & 0.014(0.001) & 48838.84(0.17) &  1.0000 & -0.222(0.110) & -0.110(0.029) & 14.204(0.036) & 33.880(0.198) & 0.027(0.004)\\
  1992bc & JRK07 & 0.021(0.001) & 48913.16(0.08) &  0.3520 & \phantom{-}0.782(0.100) & -0.121(0.027) & 14.847(0.032) & 34.700(0.169) & 0.018(0.003)\\
  1992bh & JRK07 & 0.042(0.001) & 48921.06(0.57) &  0.9960 & -0.053(0.182) & \phantom{-}0.042(0.047) & 17.363(0.038) & 36.575(0.188) & 0.018(0.003)\\
\enddata
\tablecomments{
Fit parameters of all low-z SNe~Ia that passed all  cuts.
The redshift is corrected for coherent flow effects.
$t_{peak}$ is the fitted time of B-band maximum $m_B$.
$P_{fit}$ is the SALT2 Fit Probability.
$x_1$, $c$, $m_B$, and $\mu$ are the parameters as defined in Equation~\ref{eqn:salt2}.
The Milky Way Galaxy Reddening is given in the direction of the SN by \cite{SFD98}, and corrected by  6-14\% as described in \Scolnicsys.
}
\end{deluxetable*}
\pagebreak

\begin{deluxetable*}{lccccccccc}
\tabletypesize{\scriptsize}
\tablecaption{Fit Parameters of the Pan-STARRS1 Spectroscopic Sample
\label{tab:lcfitparams}}
\tablehead{
\colhead{PS1 name} & \colhead{$z_{cmb}$} & \colhead{$t_{peak}$} & \colhead{$P_{fit}$} & \colhead{$x_1$} & \colhead{$c$} & \colhead{$m_B$} & \colhead{$\mu$} & \colhead{$E(B-V)$} & \colhead{Host}\\
&&&&&&&&&Mass
}
\startdata
 PS1-0909006 & 0.283(0.001) & 55075.41(2.59) &   0.745 & -0.726(1.428) & \phantom{-}0.008(0.073) & 21.352(0.094) & 40.579(0.309) & 0.035(0.006) & \phantom{-}9.1620\\
 PS1-0910017 & 0.319(0.005) & 55109.65(0.57) &   0.902 & -0.704(0.473) & -0.115(0.037) & 21.412(0.045) & 41.035(0.129) & 0.018(0.003) & \nodata\\
 PS1-0910018 & 0.264(0.001) & 55109.23(0.59) &   0.246 & -2.253(0.625) & -0.174(0.050) & 20.975(0.055) & 40.573(0.135) & 0.020(0.003) & \nodata\\
 PS1-10c     & 0.153(0.001) & 55210.25(0.22) &   0.930 & -2.147(0.169) & \phantom{-}0.020(0.033) & 20.284(0.038) & 39.276(0.114) & 0.022(0.003) & \phantom{-}8.3920\\
 PS1-10d     & 0.230(0.001) & 55204.10(0.51) &   0.742 & -0.222(0.482) & \phantom{-}0.124(0.041) & 21.408(0.049) & 40.336(0.133) & 0.007(0.001) & 10.5900\\
 PS1-10e     & 0.246(0.001) & 55215.72(0.30) &   0.084 & \phantom{-}0.730(0.274) & \phantom{-}0.020(0.028) & 21.164(0.037) & 40.557(0.097) & 0.018(0.003) & 11.0200\\
 PS1-10g     & 0.136(0.001) & 55201.05(0.63) &   0.992 & \phantom{-}0.634(0.744) & -0.077(0.040) & 19.394(0.052) & 39.085(0.135) & 0.021(0.003) & 10.4100\\
 PS1-10h     & 0.249(0.005) & 55207.44(0.42) &   0.720 & -0.099(0.499) & -0.107(0.032) & 20.710(0.040) & 40.392(0.127) & 0.022(0.004) & \nodata\\
\enddata
\tablecomments{
Fit parameters of all SNe~Ia in the \PS\ spectroscopic sample that passed all  cuts.
The redshift is corrected for coherent flow effects.
$t_{peak}$ is the fitted time of B-band maximum $m_B$.
$P_{fit}$ is the SALT2 Fit Probability.
$x_1$, $c$, $m_B$, and $\mu$ are the parameters as defined in Equation~\ref{eqn:salt2}.
The Milky Way Galaxy Reddening is given in the direction of the SN by \cite{SFD98}, and corrected by  6-14\% as described in \Scolnicsys.
The host galaxy mass given as $\log_{10} M_{host}$. 
}
\end{deluxetable*}

{\it Facilities:}
\facility{PS1 (GPC1)},
\facility{Gemini:South (GMOS)},
\facility{Gemini:North (GMOS)},
\facility{MMT (Blue Channel spectrograph)},
\facility{MMT (Hectospec)},
\facility{Magellan:Baade (IMACS)},
\facility{Magellan:Clay (LDSS3)},
\facility{APO (DIS)}.

\acknowledgments

The Pan-STARRS1 Surveys (PS1) have been made possible through
contributions of the Institute for Astronomy, the University of
Hawaii, the Pan-STARRS Project Office, the Max-Planck Society and its
participating institutes, the Max Planck Institute for Astronomy,
Heidelberg and the Max Planck Institute for Extraterrestrial Physics,
Garching, The Johns Hopkins University, Durham University, the
University of Edinburgh, Queen's University Belfast, the
Harvard-Smithsonian Center for Astrophysics, the Las Cumbres
Observatory Global Telescope Network Incorporated, the National
Central University of Taiwan, the Space Telescope Science Institute,
the National Aeronautics and Space Administration under Grant
No. NNX08AR22G issued through the Planetary Science Division of the
NASA Science Mission Directorate, the National Science Foundation
under Grant No. AST-1238877, the University of Maryland, and Eotvos
Lorand University (ELTE).
% MMT
Some observations reported here were obtained at the MMT Observatory, a joint facility of the Smithsonian Institution and the University of Arizona. 
% Gemini
Based on observations obtained at the Gemini Observatory, which is operated by the 
Association of Universities for Research in Astronomy, Inc., under a cooperative agreement 
with the NSF on behalf of the Gemini partnership: the National Science Foundation 
(United States), the National Research Council (Canada), CONICYT (Chile), the Australian 
Research Council (Australia), Minist\'{e}rio da Ci\^{e}ncia, Tecnologia e Inova\c{c}\~{a}o 
(Brazil) and Ministerio de Ciencia, Tecnolog\'{i}a e Innovaci\'{o}n Productiva (Argentina).
% Magellan
This paper includes data gathered with the 6.5-m Magellan Telescopes located at Las Campanas Observatory, Chile. 
% APO
Based on observations obtained with the Apache Point Observatory 3.5-meter telescope, which is owned and operated by the Astrophysical Research Consortium.
% Stubbs grants
CWS and GN thank the DOE Office of Science for their support under grant ER41843.
% Tonry & RPK NSF grants
Partial support for this work was provided by National Science
Foundation grant AST-1009749.
% Photpipe
The ESSENCE/SuperMACHO data reduction pipeline {\it photpipe}
was developed with support from National Science
Foundation grant AST-0507574, and {\it HST} programs GO-10583 and GO-10903.
% RPK
RPKs supernova research is supported in part by NSF Grant AST-1211196 and {\it HST} program GO-13046.
% Odyssey
Some of the computations in this paper were run on the Odyssey cluster
supported by the FAS Science Division Research Computing Group at
Harvard University.
% CfA SN archive
This research has made use of the CfA Supernova Archive, which is
funded in part by the National Science Foundation through grant AST
0907903.
% ADS
This research has made use of NASA's Astrophysics Data System.
\clearpage

\bibliographystyle{fapj}
\bibliography{ms}

%%%%%%%%%%%%%%%%%%%%%%
% Appendix
%%%%%%%%%%%%%%%%%%%%%%
\appendix

\section{Astrometric Accuracy}
\label{sec:wcs}

Accurate astrometry is critical for the discovery and classification
of supernovae from a sequence of observations. In addition, forced
photometry requires accurate positions as input (see
\S\ref{sec:forcedphot}), therefore it is crucial that the
astrometry not exhibit any systematic biases. In this appendix, we
assses the accuracy of our PS1 astrometry, and quantify it's impact on
our overall systematic error budget.

Starting with the reported WCS from the PS1 IPP, we fine-tune the linear
terms that define the translation and rotation in 10'x10' cutouts
centered on the SN. 
The astrometric uncertainty $\sigma_a$ can be described by a
systematic floor $\sigma_{a1}$ and a poissonian term $\sigma_{a2}$
that scales with the FWHM and SNR:
\begin{eqnarray}
\sigma_a^2 & = & \sigma_{a1}^2 + \sigma_{a2}^2\left( \frac{\mathrm{FWHM}}{\mathrm{SNR}}\right)^2 \label{eq:astroerror}
\end{eqnarray}
The value of $\sigma_{a1}$ is sensitive to the accuracy of the
distortion terms, and the number of stars used to determine the WCS
solution.  If the PSF is undersampled, $\sigma_{a1}$ is limited by the
pixelation and can be as low as 0.05~pixels.  For a given
telescope/detector system, $\sigma_{a1}$ and $\sigma_{a2}$ are global
parameters, and our goal is to determine them for the PS1 system. This
allows us then to determine the astrometric uncertainty for a given
detection depending on its FWHM and SNR.
 
We compare the detections from a
nightly stack with the detections from a deep stack. The uncertainty
$\sigma_{\Delta}$ in the difference of position between the detections
in the nightly and the deep stack depends then on the SNR and FWHM of
both, the deep and the nightly stack. However, it can safely be
assumed that ${\rm SNR}_{\rm deep} \gg {\rm SNR}_{\rm nightly}$, and
also that its FWHM is similar or better.  Therefore the $\sigma_{a2}$
term from the deep stack is negligible, and we can write
\begin{eqnarray}
\sigma_{\Delta}^2 & = & \sigma_{a,{\rm nightly}}^2 + \sigma_{a,{\rm deep}}^2 \\
 & = & 2\sigma_{a1}^2 + \sigma_{a2}^2 \left( \mathrm{\frac{FWHM_{nightly}}{SNR_{nightly}}}\right)^2  + \sigma_{a2}^2\left( \mathrm{\frac{FWHM_{deep}}{SNR_{deep}}}\right)^2 \\
 & \approx & 2 \sigma_{a1}^2 +  \sigma_{a2}^2 \left( \mathrm{\frac{FWHM_{nightly}}{SNR_{nightly}}}\right)^2
\end{eqnarray}

\begin{figure}[t]
\includegraphics[width=250pt]{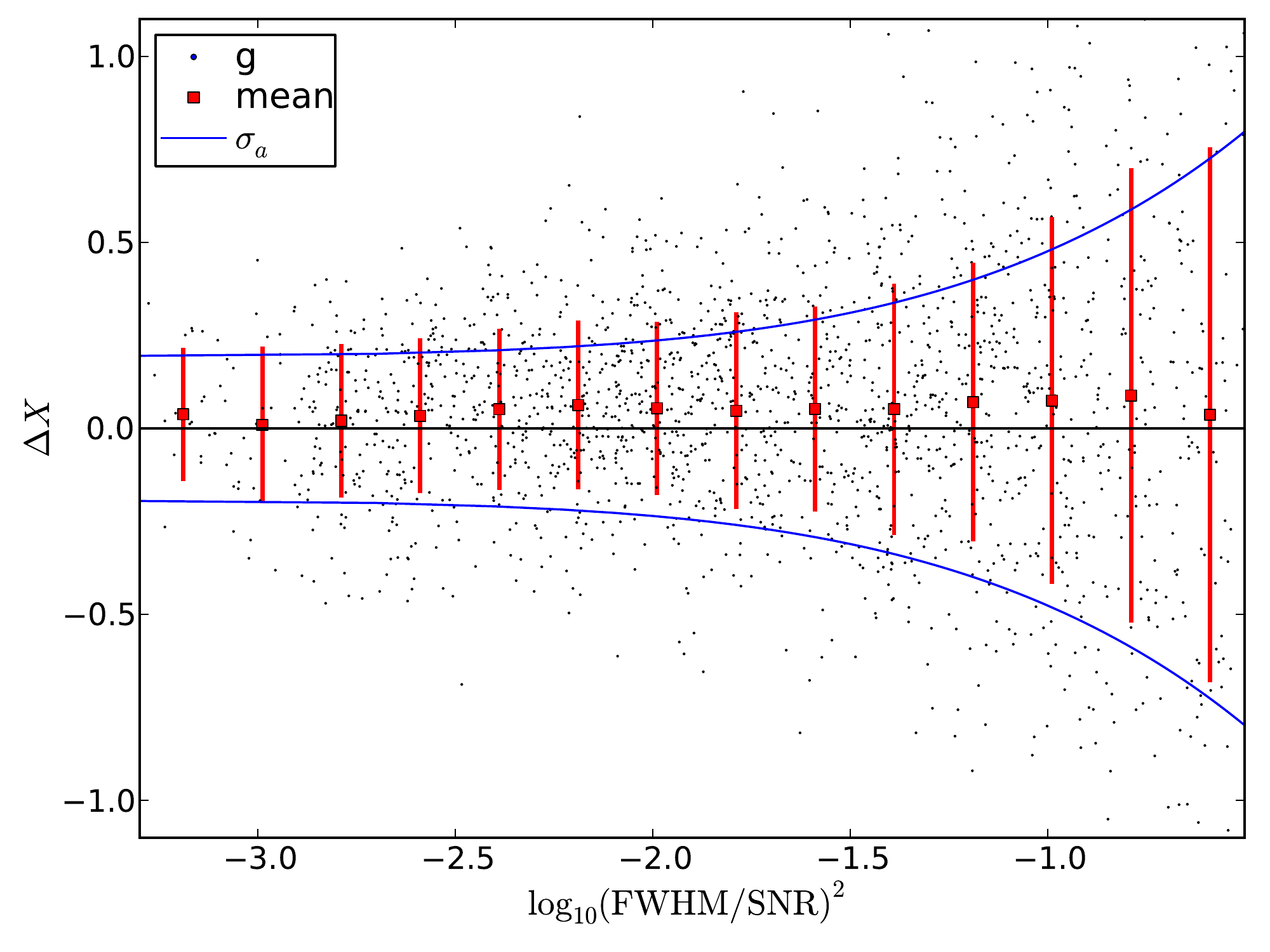} 
\includegraphics[width=250pt]{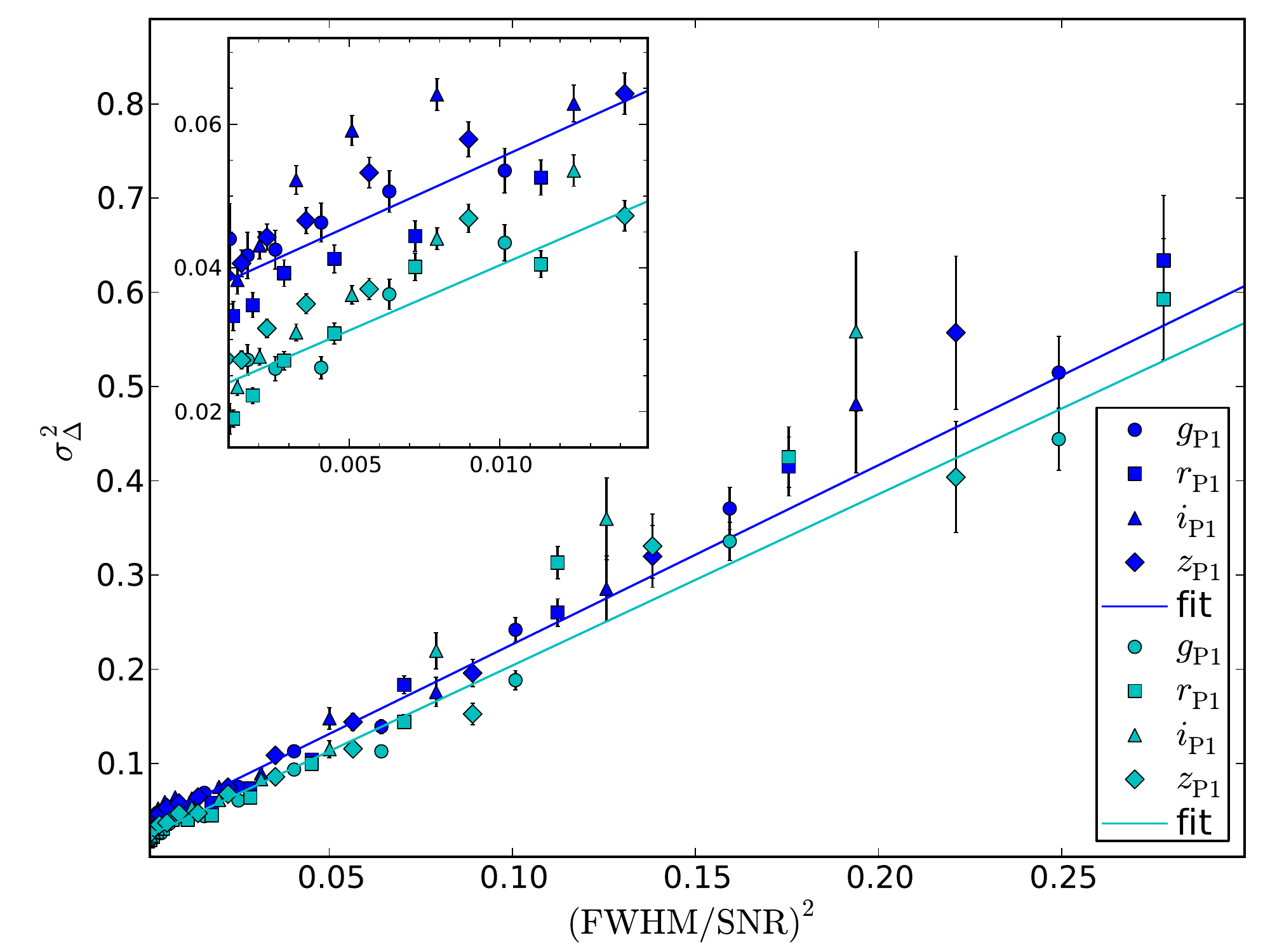} 
\caption[]{
{\it Left panel:} The  position difference $\Delta X$ in pixels of
nightly and deep stacks for stars from 100 randomly selected \gps\
band images (black dots) versus $\log_{10}(\mathrm{FWHM/SNR})^2$. The red
circles and its error bars indicate the average $\Delta X$ in
appropriate (FWHM/SNR)$^2$ bins and its standard deviation
$\sigma_{\Delta X}$, respectively. The blue line is the fitted
$\sigma_{\Delta X}$ from the fit shown in the right panel.
{\it Right panel:} The measured variance $\sigma_{\Delta
X}^2$ and $\sigma_{\Delta Y}^2$ in blue and cyan symbols, respectively,
for all 4 bands versus (FWHM/SNR)$^2$.  The straight line fits are
shown with the solid lines. The inset shows that for very small
(FWHM/SNR)$^2$, there is a significant difference in X- and
Y-direction.
\label{fig:dX}}
\end{figure}

Figure~\ref{fig:dX} shows the pixel position difference $\Delta X$ (RA
on the sky) and $\Delta Y$ (Dec. on the sky) between the nightly and
deep stacks for the stars in our photometric standard star catalog
(\Tonryphot) of 100 randomly selected images in the \gps\ band with a
plate scale of 0.2~arcsec per pixel. We have done a similar fit for
the other filters as well.  The mean and standard deviation of the
mean are shown with red symbols in appropriately spaced (FWHM/SNR)$^2$
bins.  We fit a straight line to $\sigma^{2}_{Delta}$ versus
(FWHM/SNR)$^{2}$ for all filters to determine $\sigma_{a1}$ and
$\sigma_{a2}$ from the slope and intercept of the fit. The fit is very
good, and we find
\begin{eqnarray}
\sigma_{a1,X}    & = & 0.135 \pm 0.010\ \mathrm{pixels} = 27 \pm 2\ \mathrm{mas} \label{eq:sigmaa1X}\\
\sigma_{a1,Y}    & = & 0.105 \pm 0.015\ \mathrm{pixels} = 21 \pm 3\ \mathrm{mas} \label{eq:sigmaa1Y}\\
\sigma_{a2,X}    & = & 1.379 \pm 0.012 \\
\sigma_{a2,Y}    & = & 1.348 \pm 0.011 \label{eq:sigmaa2Y}
\end{eqnarray}
We find no differences in a given direction for the different filters,
however we note that $\sigma_{a1}$ does differ at the 2$\sigma$-level
between the $X$ and $Y$ direction. This discrepancy might be due to the
Koppenhoefer effect, in which the positions of objects are biased in
the $X$-direction by an amount which increases for brighter objects.
This effect was present in half of the GPC1 chips from the start of
the mission until May 2011, at which point modifications to the camera
voltages successfully removed the bias.  For the affected chips, the
maximum displacement (for objects approaching saturation) is in the
range of 0.1 - 0.15 arcseconds (up to $\sim$ 0.6 pixels), depending on
the device.  Objects with fewer than $\sim$25,000 counts are
essentially unaffected.  As a result, the generally faint objects of
interest in this article are not directly affected.  The main effect
comes from the astrometric calibration, which uses the brighter stars,
and likely accounts for the enhanced astrometric scatter in the $X$
direction (Magnier et al., in prep.).

However, we find that this effect has only an insignificant impact on
our photometry ($<$1~mmag).  We find small average differences between
the positions measured in the nightly and deep stacked images on the
order or 10 mas, and we demonstrate that this also has negligible
impact on our forced photometry in
\S\ref{sec:forcedphot}. Conservatively, we adopt
\begin{eqnarray}
\sigma_{a1}&=&0.2\\
\sigma_{a2}&=&1.5 
\end{eqnarray}
when we use Equation~\ref{eq:astroerror} to determine the astrometric
uncertainty in a given detections, e.g. when we cluster detections
into obects.

%In the following, we apply Equation~\ref{eq:astroerror} using the
%parameters below whenever we need astrometric uncertainties, e.g. for
%clustering detections into objects or match catalogs to the
%detections.
%
%\begin{eqnarray}
%\sigma_{a1}    & = & 0.2 \label{eq:astroerrorvalues1} \\
%\sigma_{a2}    & = & 1.5 \label{eq:astroerrorvalues2}
%\end{eqnarray}
%

\subsection{SN centroids}

Accurate WCS is necessary, but not sufficient to ensure unbiased
forced photometry: difference image artificats from poor host-galaxy
subtraction or other background sources, can introduce additional
systematic biases.  We found that that a value of $\sigma_{a1}=0.2$
adequately takes into account these additional biases.  For a given
SN, we use all detections to re-determine the centroid, calculating
the average, weighted by the astrometric errors determined using
Equation~\ref{eq:astroerror}.

\section{PSF modeling}
\label{sec:psf}

\begin{figure}
\includegraphics[width=250pt]{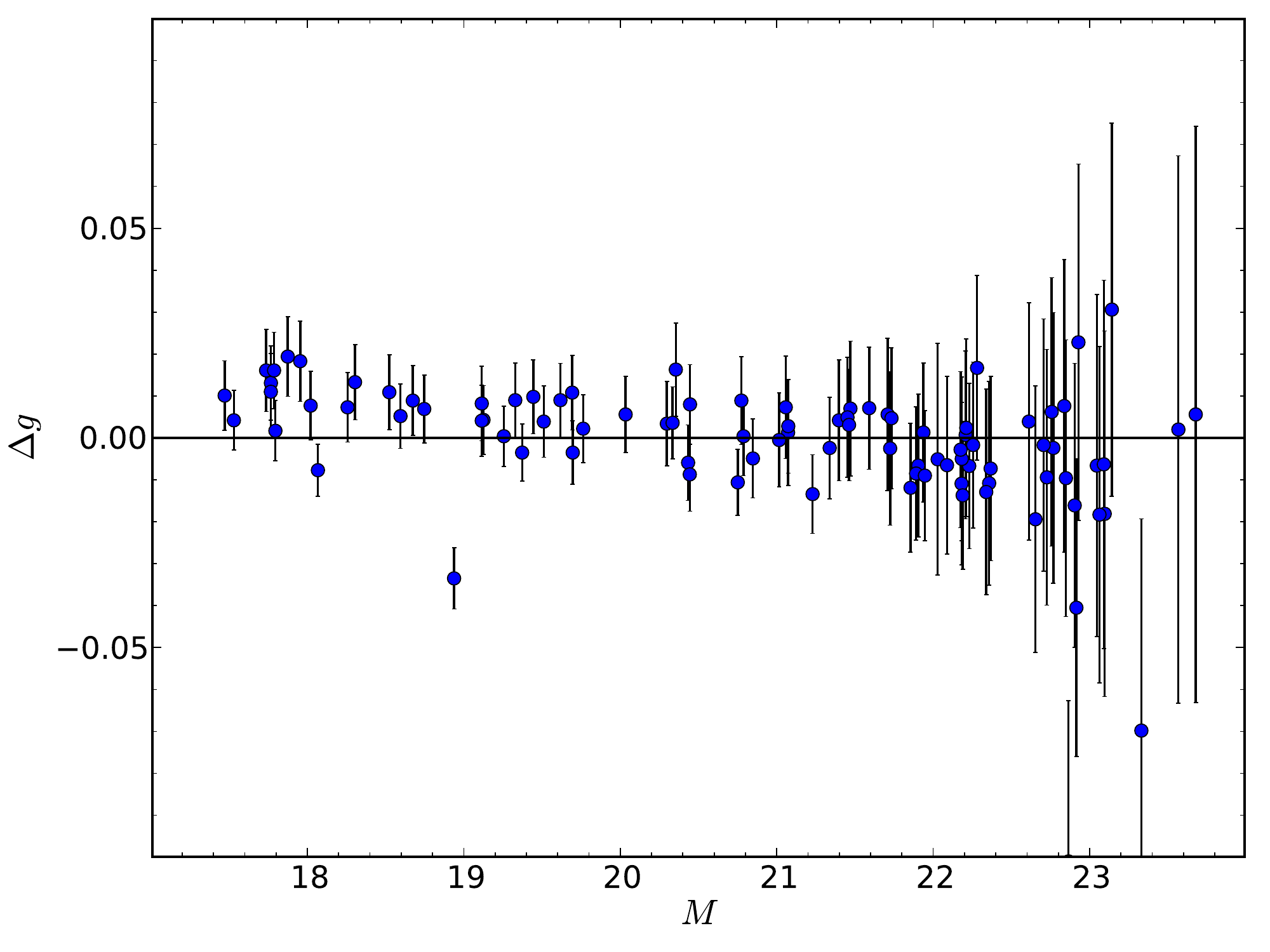}
\includegraphics[width=250pt]{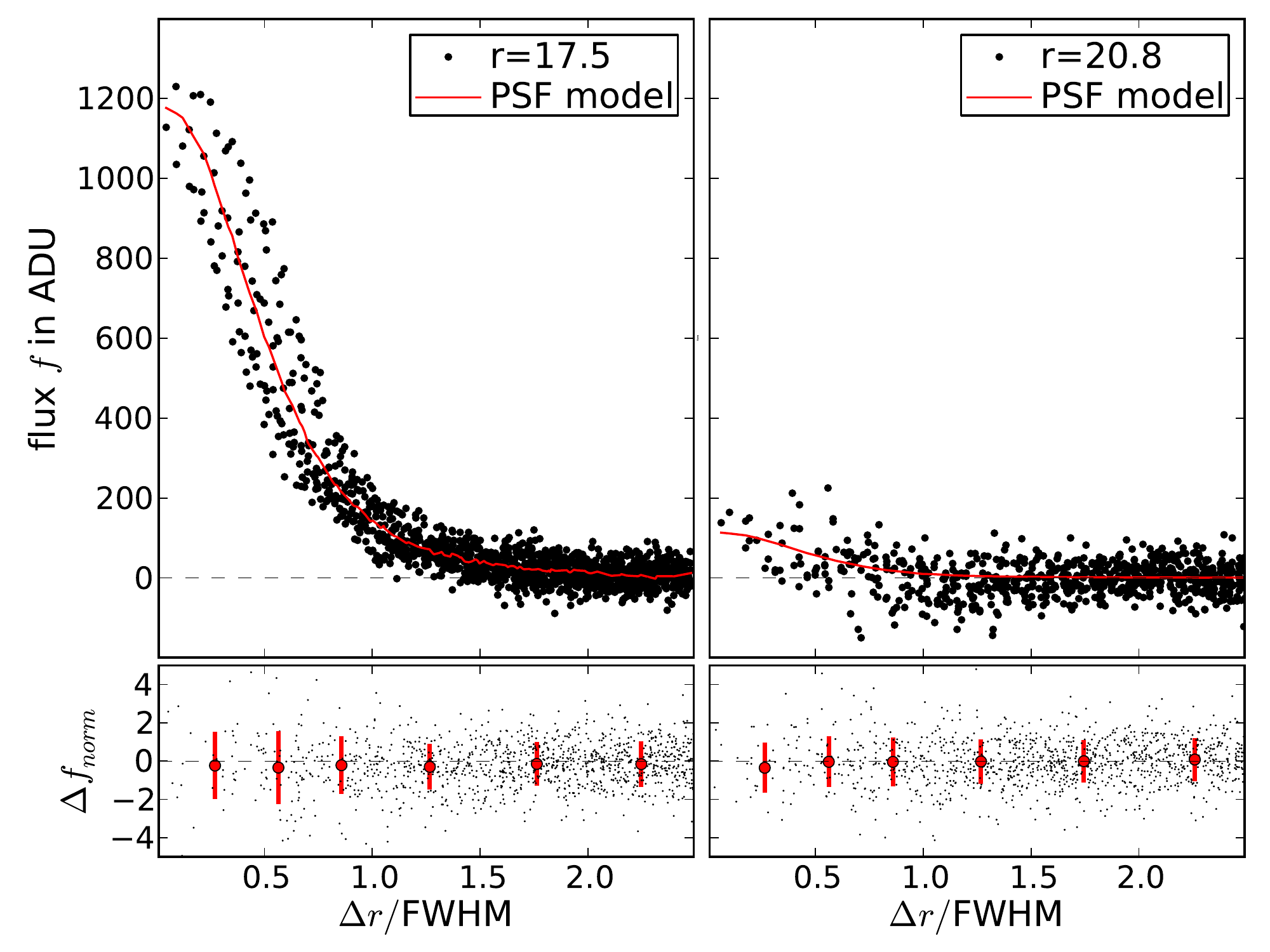}% 
\caption[]{
{\it Left:} Difference in \gps\ band magnitudes between {\it DoPHOT} and {\it DAOPHOT}
photometry. 
{\it Right:} {\it DAOPHOT} PSF for an  r=17.5 and r=20.8
magnitude star.  The red line shows the PSF model used,
and the black dots the observed flux. The x-axis is
the distance to the centroid position, normalized by the FWHM.
The lower panels show the difference between the measured flux and the
model normalized by the uncertainty  $\Delta f_{norm}$ for a subset of values randomly chosen
from 10 stars. The red circles indicate the average $\Delta
f_{norm}$ for appropriate radial bins, and the error bars indicate the
standard deviation.
%upper panels: The black symbols indicate the flux versus
%radius normalized by the FWHM for an r=17.5 (left) and r=20.8
%(right) magnitude stars.  The red line shows the PSF model used. In
%the lower panel, the difference between the measured flux and the
%model normalized by the uncertainty is shown (black dots) for 10
%randomly selected stars. The red circles indicate the average $\Delta
%f_{norm}$ for appropriate radial bins, and the error bars indicate the
%standard deviation.
\label{fig:psf}}
\end{figure}

For identifying transients with \photpipe, we use a customized
version of {\it DoPHOT}, which is quick, robust, and produces adequate
photometry. However, \PS\ has a PSF with structure that cannot
accurately fit with such an analytic model. This may cause systematic
biases in the photometry between faint and bright stars of up to 1\%.
The left of Fig.~\ref{fig:psf} shows an example of the difference
$\Delta g = g_{\mathrm{PS1,DoPHOT}}-g_{\mathrm{PS1,DAOPHOT}}$ between
{\it DoPHOT} and {\it DAOPHOT} photometry. There is a clear trend from
bright to faint magnitudes. {\it DAOPHOT} fits an empirical correction
in addition to the Gaussian model, which significantly better fits the
PSF in the image.

The right side of Fig.~\ref{fig:psf} shows the {\it DAOPHOT} flux
(black symbols) and PSF fit (red line) for $r=17.5$ and $r=20.8$
example stars in the upper left and right panel, respectively. In the
lower panels, the residual between the flux and the PSF model (black
symbols), normalized by the uncertainty, is shown for 10 randomly
selected stars in the magnitude bins $17.5<r<18.5$ and
$20.5<r<21.5$. The red circles indicate the average normalized
residual $\Delta f_{norm}$ for appropriate radial bins, and the error
bars indicate its standard deviation. The standard deviation is \about
1 for large radii, but increases to values on the order of up to 1.7
for radii closer to the core. This indicates that some of the PSF
structure is still not perfectly fitted for.  Most important though,
there is no significant bias toward higher or lower fit values for any
distance to the core for both the faint and bright stars.  We
therefore use {\it DAOPHOT} photometry for all the analysis in
\transphot.

\section{Empirical Adjustment of Uncertainties}
\label{sec:empadjust}

The propagated uncertainties are underestimates, as they do not
account for the pixel-pixel covariance introduced by warping,
sub-sampling, stacking, and convolution of the images.  In order to
empirically determine by how much the uncertainties are
underestimated, we measure the flux $f_{r}$ and its uncertainty
$\sigma_{r}$ at random positions in a given difference image in
exactly the same way we measure the SN flux.  We calculate the
weighted mean $\bar{f_r}$ of these flux measurements. In order to
guard against reduction and image artifacts, we apply a 3$\sigma$ cut
to the normalized flux distribution $(f_r - \bar{f_r})/(s_r\sigma_r)$,
rather than cutting on the underestimated errors, $\sigma_r$, for the
following reason: let's assume that all uncertainties are
underestimated by the same factor $s_r$. If we nominally apply a
N-sigma cut using these underestimated uncertainties, we effectively
apply an N/$s_r$-cut, e.g. for a nominal 3-sigma cut and $s_r=1.5$,
the real cut-off is at 2-sigma. In order to avoid this, we determine
the normalized flux distribution $(f_r - \bar{f_r})/\sigma_r$, which has
a standard deviation of $s_r$. The true 3-sigma outliers can then be
identified and removed by doing an 3-sigma cut on the normalized flux
distribution. Note that the standard deviation $s_r$ is equivalent to
the square-root of the chi-square distribution
\begin{equation}
s_r = \sqrt{\chi^2_r} = \frac{1}{N-2}\sum^N \left(\frac{{f_r} - \bar{f_r}}{\sigma_r} \right)^2
\end{equation}
We multiply all uncertainties by the factor $s_r$ in order to
empirically correct the uncertainties. We find that it is imperative
to employ this robust way of determining $s_r$ for the method to work
correctly. The fact that the reduced chi-square of the baseline flux
of the SN light curves peaks at 1.0 validates our method (see
\S\ref{sec:fluxoffset}).

In addition, for a given difference image, $\bar{f_r}$ is an estimate
of the bias in the flux measurements. The values of $\bar{f_r}$ are in
general very small, much smaller than the typical
uncertainties. Nevertheless, we adjust all fluxes by this value.

\section{Correction of Systematic Magnitude Biases due to Centroiding Errors}
\label{sec:centroiderrordmcorr}

We determine the position of a given SN as the weighted mean of all
its detections. This SN centroid $(RA_0, DEC_0)$ has a non-negligable astrometric
uncertainty $\sigma_{\mathrm{SN,cent}}$, which introduces a bias in our recovered photometry and
must be corrected. The recovered position of a detection can be different from 
the SN centroid for the following reasons:
\begin{enumerate} 
\item Poisson noise, in particular positive noise peaks, in the background sky.
\item Poisson noise in the SN flux
\item Difference image artifacts
\item The centroid accuracy of $(RA_0, DEC_0)$.
\item Accuracy of the WCS for a given image.
\end{enumerate} 
The first three items in the list above introduce a Malmquist bias for
regular photometry since the freedom in position will bias the fitted
PSF to be centered toward the positive noise peaks.  This effect is stronger for
detections with low SNR. However, using forced photometry eliminates
this bias (see also \S~\ref{sec:forcedphot} and
Figure~\ref{fig:comparison_forced_regdiff}). What is not corrected for
is if the position used for forced photometry of the SN is offset from
the true position of the SN, either because the uncertainty in SN
centroid or due to inaccuracies of the WCS for the given image.
In this section, our goal is to characterize how the photometric bias
depends on the astrometric offset, so that we can estimate the expectation value
of the photometric bias for a given centroid accuracy and correct
for it.

\begin{figure}
\includegraphics[width=500pt]{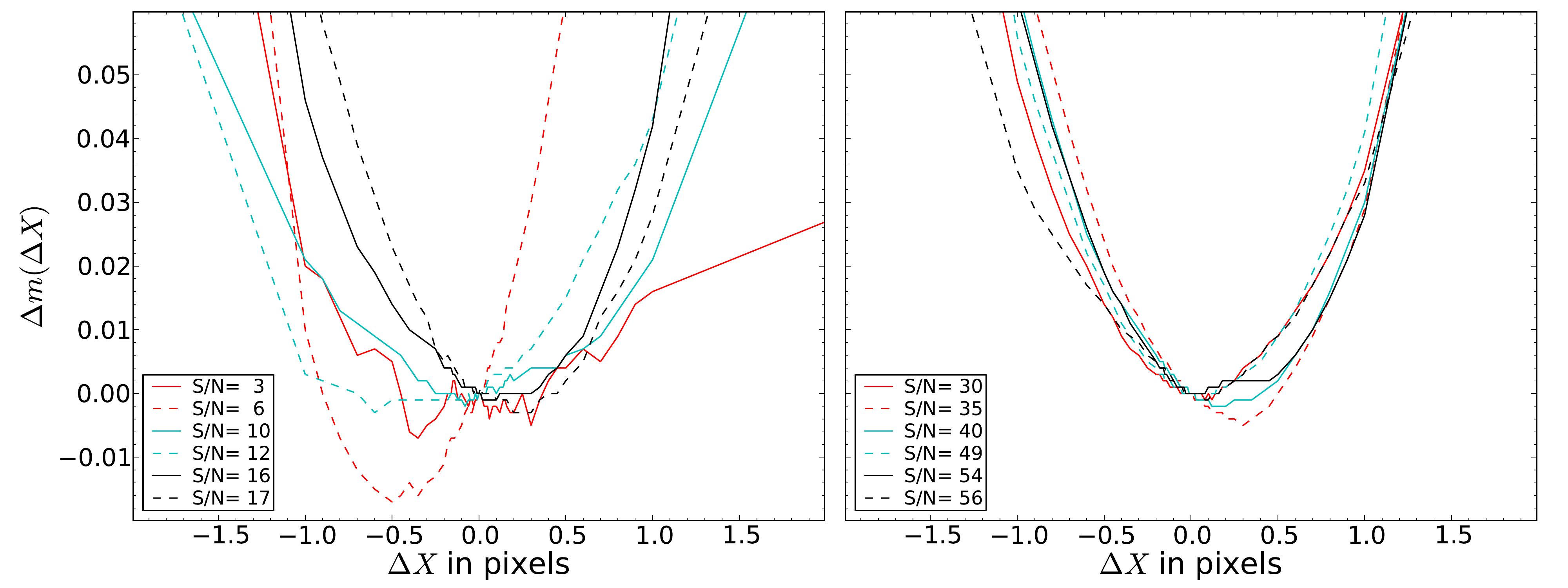}
\caption[]{
{\it Left:} The difference in forced photometry $\Delta m(\Delta X,\Delta Y) = m(X_0+\Delta X,Y_0+\Delta Y) - m(X_0,Y_0)$
between  $(X_0, Y_0)$ and $X_0+\Delta X,Y_0+\Delta Y$ for low SNR detections of PS1-10axx. 
{\it Right:} Same as the left panel, but for high SNR detections of PS1-10axx.
\label{fig:dmdx}}
\end{figure}
For our analysis, we use detections of PS1-10axx, a SN~Ia at $z=0.027$
with a good selection of detections with both high and low SNR. As
described in
\S~\ref{sec:centroids}, the position $(RA_0, DEC_0)$ was determined by
calculating the 3$\sigma$ clipped weighted mean of all
detections. With forced photometry, $(RA_0, DEC_0)$ is
translated into $(X_0, Y_0)$ for a given image, and the center of the
PSF is forced to be at $(X_0, Y_0)$.  We investigate how the
photometry depends on the centroid position by measuring the
photometry at position $(X,Y)=(X+\Delta X,Y+\Delta Y)$. We define the
change in photometry as
\begin{equation}
\Delta m(\Delta X,\Delta Y) = m(X_0+\Delta X,Y_0+\Delta Y) - m(X_0,Y_0).
\end{equation}
For simplicity, we concentrate on one dimension $(\Delta X)$, and
later on apply it to two dimensions.  In the right panel of
Figure~\ref{fig:dmdx}, we show $\Delta m(\Delta X)$ for detections of
PS1-10axx high SNR. Without noise, $\Delta m(\Delta X)$ would be a
parabola with a minimum at $\Delta X=0$, but due to a combination of
the reasons listed above, the parabola is shifted. We fit these parabolas with
\begin{equation}
\Delta m(\Delta X) = h (\Delta X - \Delta X_\mathrm{min})^2 + \Delta m_\mathrm{min} \label{eq:dmdxparabola}
\end{equation}
where $\Delta X_\mathrm{min},\Delta m_\mathrm{min}$ are the
coordinates of the mininum of the parabola for a given detection, and
$h$ defines the width of the parabola. 

If we assume that this shift in the position of the minimum is due to the error in the SN centroid, then the SN centroid used was off by $\Delta X_\mathrm{min}$ from the true
centroid, and the photometry was biased by $\Delta m_\mathrm{min}$.
Now we can turn around this argument: If we know $h$, the width of the
parabola, and the centroid accuracy of a given SN centroid
$\sigma_{\mathrm{SN,cent},X}$, we can calculate the expectation value
of the photometric bias $\Delta m_{\mathrm{SN,cent},X}$ as
\begin{eqnarray}
\Delta m_{\mathrm{SN,cent},X} & = & \int h t^2 \mathrm{PDF}(\sigma_{\mathrm{SN,cent},X},t)\,\mathrm{d}t \label{eq:dmSNcen_expectation}
\end{eqnarray}
where PDF($\sigma$,t) is the probability density function with sigma
$\sigma$, and assuming that $h$ is constant and independent of SNR.

\begin{figure}
\includegraphics[width=500pt]{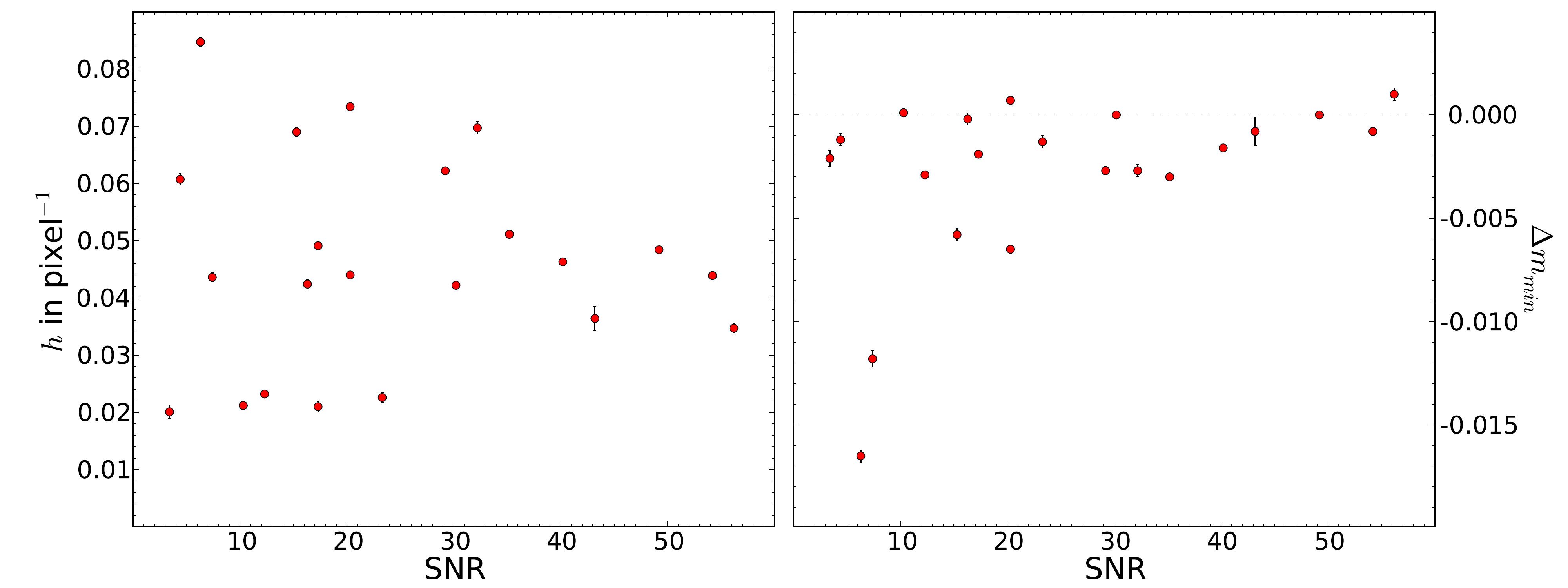}
\caption[]{
{\it Left:} Parameter $h$ from Equation~\ref{eq:dmdxparabola} derived by fitting $\Delta m(\Delta X)$ with a parabola.
{\it Right:} Parameter $\Delta m_\mathrm{min}$ from Equation~\ref{eq:dmdxparabola}. 
\label{fig:dmdx_fitparams}}
\end{figure}
The assumption that the width of the parabola is constant and independent of SNR
appears to valid to the first order.  For high SNR, the width
of the parabola seems to be constant (see right panel of
Figure~\ref{fig:dmdx}).  For low SNR detections, however, the width of
the parabola show a much larger variation (see left panel of
Figure~\ref{fig:dmdx}).
%The left panel of
Figure~\ref{fig:dmdx_fitparams} shows the fitted $h$ and $\Delta
m_\mathrm{min}$ for all detections of PS1-10axx in the left and right
panel, respectively.  For low SNR, the spread in $h$ significantly
increases because of poisson fluctuations in the flux and/or
difference image artifacts. We therefore use the $h$ derived only from
the SNR$>$35 detections:
\begin{eqnarray}
h&=&0.043 \pm 0.003\ \mathrm{pixel}^{-2}\label{eq:h}%\\
%&=& 2.2\times 10^{-3}\ \mathrm{mas}^{-1}
\end{eqnarray}
with a standard deviation of 0.007. Any effect of the spread in $h$ on
$\Delta m_{\mathrm{SN,cent},X}$ cancels out to first order, as long
the mean $h$ is independent of SNR.

\begin{figure}
\includegraphics[width=500pt]{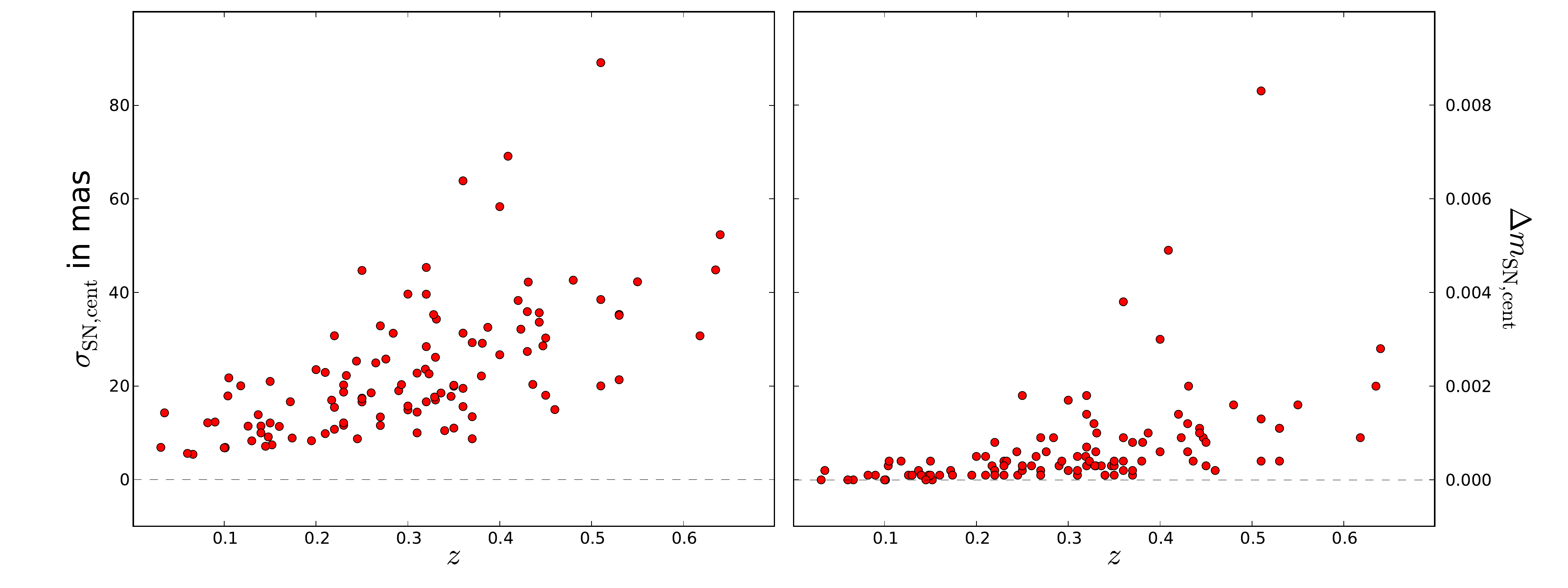}
\caption[]{
{\it Left:} The astrometric uncertainty in the SN centroid position determined by averaging the position of all detections
versus redshift. The uncertainty in the SN centroid increases with increasing redshift.
{\it Right:} The expectation value of the photometric bias due to the centroid uncertainty calculated with Equation~\ref{eq:dmcorrSN}.
\label{fig:dmcorrcentroid}}
\end{figure}

For a given SN, we estimate the uncertainty in the SN centroid
$\sigma_{\mathrm{SN,cent},X}$ and $\sigma_{\mathrm{SN,cent},Y}$ when
we calculate the SN position $(RA_0, DEC_0)$ as the weighted mean of
all the SN detections. Using the fitted $h$ and
Equation~\ref{eq:dmSNcen_expectation} we can now calculate the
expectaction value of the systematic bias $\Delta m_\mathrm{SN,cent}$
in our SN photometry:
\begin{eqnarray}
\Delta m_\mathrm{SN,cent} & = & \Delta m_{\mathrm{SN,cent},X} + \Delta m_{\mathrm{SN,cent},Y} \label{eq:dmcorrSN}
\end{eqnarray}
The left panel of Figure~\ref{fig:dmcorrcentroid} shows
$\sigma_{\mathrm{SN,cent}} = \sqrt{\sigma_{\mathrm{SN,cent},X}^2 +
\sigma_{\mathrm{SN,cent},X}^2}$ for all SNe versus the redshift, and
the right panel shows $\Delta m_\mathrm{SN,cent}$.
The photometric bias has a systematic trend with redshift on the order of 2mmag, and these corrections are added to our final photometry.

\end{document}